\documentclass[a4paper,fleqn,usenatbib,useAMS]{mnras}
\pdfoutput=1

\usepackage{graphicx}	
\usepackage{amsmath}	
\usepackage{amssymb}	
\usepackage{multicol}   
\usepackage{bm}		
\usepackage{pdflscape}	
\newcommand{\ct}{\citealt}
\renewcommand{\thesubsubsection}{\thesubsection.\alph{subsubsection}}

\usepackage[T1]{fontenc}
\usepackage{ae,aecompl}

\usepackage{txfonts}

\title[Disk Formation by Removing Small Grains]{Protostellar Disk Formation Enabled by Removal of Small Dust Grains}

\author[B. Zhao et al.]{Bo Zhao$^{1}$\thanks{Contact e-mail: \href{mailto:bo.zhao@mpe.mpg.de}{bo.zhao@mpe.mpg.de}}\thanks{Present address: Giessenbachstr. 1, D-85748, Garching, Germany},
Paola Caselli$^{1}$,
Zhi-Yun Li$^{2}$,
Ruben Krasnopolsky$^{3}$,
Hsien Shang$^{3}$,
\newauthor Fumitaka Nakamura$^{4}$\\
\\
$^{1}$Max-Planck-Institut f\"ur extraterrestrische Physik (MPE), Garching, Germany\\
$^{2}$University of Virginia, Astronomy Department, Charlottesville, USA\\
$^{3}$Academia Sinica Institute of Astronomy and Astrophysics, Taipei, Taiwan\\
$^{4}$National Astronomical Observatory of Japan, Tokyo, Japan}

\pubyear{2016}

\begin{document}
\label{firstpage}
\pagerange{\pageref{firstpage}--\pageref{lastpage}}
\maketitle

\begin{abstract}
It has been shown that a realistic level of magnetization of dense molecular 
cloud cores can suppress the formation of a rotationally supported disk (RSD) 
through catastrophic magnetic braking in the axisymmetric ideal MHD limit. 
In this study, we present conditions for the formation of RSDs through 
non-ideal MHD effects computed self-consistently from an equilibrium 
chemical network. We find that removing from the standard MRN distribution 
the large population of very small grains (VSGs) of 
$\sim$10~$\AA$ to few 100~$\AA$ that dominate the coupling of the bulk 
neutral matter to the magnetic field increases the ambipolar diffusivity 
by $\sim$1--2 orders of magnitude at densities below 10$^{10}$~cm$^{-3}$. 
The enhanced ambipolar diffusion (AD) in the envelope reduces the amount 
of magnetic flux dragged by the collapse into the 
circumstellar disk-forming region. Therefore, magnetic braking is weakened 
and more angular momentum can be retained. 
With continuous high angular momentum inflow, RSDs of tens of AU are able 
to form, survive, and even grow in size, depending on other parameters 
including cosmic-ray ionization rate, magnetic field strength, 
and rotation speed. Some disks become self-gravitating and evolve into rings 
in our 2D (axisymmetric) simulations, which have the potential to fragment 
into (close) multiple systems in 3D. 
We conclude that disk formation in magnetized cores is highly 
sensitive to chemistry, especially to grain sizes. 
A moderate grain coagulation/growth to remove the large population of VSGs, 
either in the prestellar phase or during free-fall collapse, 
can greatly promote AD and help formation of tens of AU RSDs.
\end{abstract}




\section{Introduction}
\label{Chap.Intro}

The formation of rotationally supported disk (RSD) from magnetized 
dense molecular cloud cores is a long-standing problem in star formation. 
Although RSDs are frequently observed around young stellar objects 
\citep{WilliamsCieza2011}, including deeply embedded sources 
\citep{Tobin+2012,Tobin+2013}, 
theoretical studies have found it difficult to form such disks due to 
strong magnetic braking which catastrophically removes most 
angular momentum of circumstellar gas. 
This is the so-called ``magnetic braking catastrophe'' in disk formation
\citep{MestelSpitzer1956,Allen+2003,MellonLi2008,Li+2011}. 

In the axisymmetric ideal MHD limit, the magnetic field strength required for 
such catastrophic magnetic braking is moderate \citep{Allen+2003}.
The observed magnetization of dense cores, quantified by the 
dimensionless mass-to-flux ratio $\lambda$, has a typical value of 
$\lambda \sim$ 2--3 from OH Zeeman survey \citep{TrolandCrutcher2008}. 
Although uncertainty exists due to OH depletion at high densities 
\citep{Tassis+2014}, the bulk of dense core should still have 
$\lambda \sim$ a few. Such a strong magnetic field, if not sufficiently 
decoupled from collapsing matter, can transport most angular momentum 
away from the circumstellar region via magnetic braking, 
and hence suppress the formation of RSDs 
\citep{MellonLi2008,HennebelleFromang2008}.

In reality, dense cores are only slightly ionized \citep{BerginTafalla2007}, 
and magnetic fields are expected to at least partially decouple from 
neutral matter through non-ideal MHD effects, including ambipolar diffusion 
(AD), Ohmic dissipation, and Hall effect \citep[e.g.,][]{Nakano+2002}. 
Their effects on disk formation have been studied previously 
\citep{MellonLi2009,DappBasu2010,Li+2011,Krasnopolsky+2011,Dapp+2012, 
Tomida+2013,Tomida+2015,Tsukamoto+2015a,Tsukamoto+2015b, 
Masson+2015,Wurst+2015}; however, whether a large (say 50~AU sized), 
long-lived RSD can form or not remains unclear. 
Most of these studies show that AD and Ohmic dissipation can only 
help to form a small RSD (<10~AU) early in the first core phase 
\citep{Larson1969}; but whether such disk can survive or grow in the 
subsequent evolution is unclear due to numerical difficulties. 
The RSD formed via Hall effect can indeed be larger ($\sim$30~AU), yet 
only when the direction of magnetic field is nearly anti-parallel to the 
net rotation axis of dense core \citep{Krasnopolsky+2011,BraidingWardle2012, 
Tsukamoto+2015b,Wurst+2015}. 
In a realistic turbulent core, it is unclear whether such a requirement 
in field geometry can be satisfied throughout the entire protostellar 
evolution phase. 
Therefore, the inconsistency between disks formed in theoretical 
calculations and those observed around young stellar objects remains 
unresolved. 
\citet{Shu+2006} and \citet{Krasnopolsky+2010} suggest that in order to 
bridge this gap, magnetic diffusivity over a large circumstellar region 
must be at least 1 order of magnitude above the classical value used 
in most studies. 

The magnetic diffusivity due to non-ideal MHD effects is essentially 
determined by chemistry and microscopic physical processes 
\citep{OppenheimerDalgarno1974,UmebayashiNakano1990,Nakano+2002}. 
The key is to calculate the fractional abundances of charged species 
(electron, ions, and charged grains) and their effective conductivities. 
However, uncertainty still exists in such calculations due to 
the lack of observational constraints on grain size distribution in 
dense molecular cloud cores \citep[e.g.,][]{Kim+1994}. 
Typical chemistry models either choose a single grain size 
\citep[e.g.,][]{UmebayashiNakano1990}, or adopt the standard MRN 
\citep[Mathis-Rumpl-Nordsieck;][]{Mathis+1977} size distribution from 
diffuse interstellar clouds \citep[e.g.,][]{Nakano+2002}. However, 
the choice of grain size can greatly affect the ionization fraction 
\citep{Nishi+1991} and the magnetic diffusivity \citep{Dapp+2012}; 
because grains not only provide surface for recombination of 
ions and electrons, but also couple to the magnetic field if 
they are charged (the smaller the grain, the stronger the coupling). 
Therefore, non-ideal MHD studies on disk formation should pay 
special attention to the effect of grain size on magnetic diffusivities. 

Recent studies have shown evidence of magnetic diffusivity enhancement 
by removing very small grains. 
\citet{WardleNg1999} find that removal of $\sim$10$^2~\AA$ grains greatly 
reduces the Hall conductivity and hence increases the diffusivity for 
Hall effect; however, its impact on ambipolar diffusivity has not been 
addressed. \citet{Dapp+2012} explore different single sized grains 
ranging from $\sim$0.01--0.1~$\mu$m, and reveal that the 
effective magnetic diffusivity (including both Ohmic dissipation 
and AD) with $\sim$0.1~$\mu$m grains is 1--2 orders of magnitude 
higher than that with a MRN size distribution (see their Fig.2). 
However, they did not elaborate on how this affects disk formation. 
\citet{Padovani+2014} apply post-processing chemistry models to the 
ideal MHD simulations of \citet{Joos+2012}, and find that the size of 
matter-field decoupling region increases significantly with a ``truncated'' 
MRN distribution ($a_{\rm min}=0.1~\mu$m) compared with the full 
MRN distribution. To verify their findings, a more complete 
chemical network and a self-consistent calculation are necessary. 
Inspired by the seminal work of \citet{Padovani+2014}, 
we improve their chemical network and revisit the non-ideal MHD
effects on disk formation in this paper, with a particular focus on 
enhancing AD by truncating very small grains 
(VSGs: $\sim$10~$\AA$ to few 100~$\AA$) off the MRN distribution. 
The truncation is a simple representation of the actual removal 
of very small grains in dense cores of molecular clouds. 

There is observational evidence for the removal of very small grains 
in dense cores. 
Extinction (optical and near-infrared) and polarimetric observations 
of dense molecular clouds indicate that dust grains have larger sizes  
than those in diffuse interstellar medium \citep{Cardelli+1989,Vrba+1993}. 
Recent discovery of ``coreshine'' (light scattered by large dust grain 
up to 1~$\mu$m; \ct{Pagani+2010,Steinacker+2010}) from nearby dense cores 
also implies that grains have grown substantially to larger sizes. 
Evidence of grain growth has also been collected using multi-wavelength 
studies \citep{Schnee+2014,Forbrich+2015}. 
Two grain growth mechanisms can operate in such cold dense environment: 
accretion onto grain mantles \citep[e.g.,][]{TielensHagen1982, Hasegawa+1992} 
and grain coagulation \citep[e.g.,][]{Chokshi+1993,DominikTielens1997}. 
The latter --- grain coagulation --- has been shown theoretically to be 
rather efficient in removing small grains (<0.1~$\mu$m) from the 
MRN size distribution within a few 10$^6$ years 
\citep{Rossi+1991,Ossenkopf1993,Ormel+2009,Hirashita2012}. 
Such a process may have already finished during the quiescent, 
slowly-evolving prestellar phase if the cloud is magnetically supported 
\citep{Shu+1987,MouschoviasCiolek1999}. 
Therefore, the exclusion of VSGs from MRN size distribution may in fact be 
more appropriate for evaluating the magnetic diffusivity in dense cores.

The rest of the paper is organized as follows. 
Section \ref{Chap.Chem} describes the chemistry model we adopted 
in our analytical and numerical calculations. 
The initial conditions of the simulation set 
are listed in Section \ref{Chap.IC}, together with an overview of the 
results. 
In Section \ref{Chap.ChemResult}, we first demonstrate the cause of 
AD enhancement by eliminating VSGs, 
using simple analytical results from our chemistry model. 
The simulation results are then presented in Section \ref{Chap.SimulResult}, 
with a comprehensive interpretation of the effects of various parameters 
on the formation of different types of RSDs. 
We summarize the main results and put them in context in 
Section \ref{Chap.Discuss}.

\section{Chemistry Model}
\label{Chap.Chem}

Magnetic diffusivities are determined by chemistry and 
microscopic physical processes, 
i.e., the degree of ionization and the thermal collision between different 
species. The ionization fraction of charged species can be obtained from 
standard chemistry models \citep{OppenheimerDalgarno1974, 
UmebayashiNakano1990}. To self-consistently couple the chemistry with 
MHD simulations during run-time, we adopt a simple equilibrium network 
based on \citet{Padovani+2014}, which is described as below.

\subsection{Ionization Rate}
\label{S.CRIRate}

The dominant source of ionization in dense molecular cloud cores is 
cosmic-rays. The interstellar UV-radiation becomes less important 
due to the relatively high column density in dense cores 
(visual extinction $A_{\rm v}$>4~mag, \ct{McKee1989}). 
The cosmic-ray ionization rate $\zeta_0^{\rm H_2}$ at cloud core scale 
ranges from $\sim$10$^{-17}$~s$^{-1}$ to $\sim$10$^{-16}$~s$^{-1}$, 
based on both theoretical models and observations 
\citep[e.g.,][]{SpitzerTomasko1968, Caselli+1998, vanTakDishoeck2000}.
The attenuation of cosmic-rays 
through energy loss of cosmic-ray particles can change the 
local ionization rate in dense cores. 
\cite{UmebayashiNakano1981} have shown that the cosmic-ray ionization rate
can be well described by the following relation:
\begin{equation}
\label{Eq:attenu}
\zeta^{\rm H_2}=\zeta_0^{\rm H_2}~{\rm exp}(-\Sigma_{\rm H_2}/\Sigma_0)
\end{equation}
where $\Sigma_0=96$~g~cm$^{-2}$ is the attenuation length, and the H$_2$
column density $\Sigma_{\rm H_2}$ can be estimated from the gas density 
$\rho$ as (\ct{Nakano+2002}):
\begin{equation}
\Sigma_{\rm H_2} \approx { 4kT\rho \over \pi G \mu m_{\rm H} }
\end{equation}
where $T$ is the gas temperature, $\mu=2.36$ is the mean molecular weight 
per hydrogen atom (assuming a mass fraction of 71$\%$ hydrogen, 
27$\%$ helium, and 2$\%$ metals), and $m_{\rm H}$ is the mass of 
a hydrogen atom. With this relation, the exponential decrease 
in Eq.~\ref{Eq:attenu} only becomes important at number density 
${\rm n(H_2)}\gtrsim$10$^{11}$~cm$^{-3}$. 

Above a few $10^{12}$~cm$^{-3}$, the ionization rate due to cosmic-ray 
decreases abruptly below $10^{-19}$~s$^{-1}$. However, at such densities, 
the radioactive decay of long-lived $^{40}$K or short-lived $^{26}$Al 
becomes the dominant source of ionization. They contribute to 
$\zeta^{\rm H_2}(^{40}{\rm K})=1.1 \times 10^{-22}$~s$^{-1}$ and 
$\zeta^{\rm H_2}(^{26}{\rm Al})=7.3 \times 10^{-19}$~s$^{-1}$, respectively, 
based on the newly measured abundances of radionuclides in 
primitive Solar nebula \citep{UmebayashiNakano2009, Cleeves+2013}. 
The short half-life $0.74$~Myr of $^{26}{\rm Al}$ is still 
much longer than the time-scale of protostellar collapse phase in this study. 
Therefore, we place a lower limit to the ionization rate at 
$7.3 \times 10^{-19}$~s$^{-1}$. 
\footnote{As will be shown in \S~\ref{Chap.ChemResult}, the main effects 
of this study occur at densities below 10$^{10}$~cm$^{-3}$, which are not 
affected by the limiting ionization rate.}

\subsection{Chemical Network}
\label{S.ChemNet}

The equilibrium chemical network used in this study is derived from 
\citet[Appendix A]{Padovani+2014}, with additional grain species 
(g$^+$ and g$^0$), instead of assuming all grains are negatively charged. 
The simplified network includes neutral species H$_2$, heavy molecules 
(denoted collectively as ``m''), and heavy metals (denoted collectively 
as ``M'') and charged species e$^-$, H$^+$, H$_3^+$, m$^+$ 
(typically HCO$^+$), and M$^+$, 
as well as neutral and singly charged dust grains g$^0$, g$^-$, and g$^+$. 
The abundance of neutral molecules and metals are fixed at 
$x(m) \approx 6 \times 10^{-4}$ and $x(M) \approx 4 \times 10^{-8}$ 
\citep{Caselli+2002a,Padovani+2014}.

Because of the relatively unconstrained grain size distribution in dense 
molecular clouds, we choose the standard $-3.5$ power law as in 
the MRN distribution \citep{Mathis+1977}, 
but with a varying minimum grain size $a_{\rm min}$ 
(the maximum grain size $a_{\rm max}$ is fixed at $0.25~\mu m$ 
for most models, although a higher value of 1~$\mu$m is also considered 
in some cases). 
We also fix the total grain mass at $q=1\%$ of the gas mass. 
The density of grain material is taken to be $\rho_{\rm g}=2.3$~g~cm$^{-3}$ 
\citep{KunzMouschovias2009}, i.e., the average density of silicates. 
Therefore, the size distribution function is given by, 
\begin{equation}
\label{Eq:MRN}
{{\rm d} n(a) \over {\rm d}a} = C a^{-3.5}~,
\end{equation}
where the normalization factor C can be determined as,
\begin{equation}
\label{Eq:MRNnorm}
C = {3 q m_{\rm H} \over 4\pi \rho_{\rm g} (\sqrt{a_{\rm max}}-\sqrt{a_{\rm min}})} n({\rm H}_2)~.
\end{equation}
Therefore, the total number density of grains can be written as a function 
of the minimum grain size $a_{\rm min}$, 
\begin{equation}
\label{Eq:xgr}
n(g) = {3 q m_{\rm H} \over 10 \pi \rho_{\rm g} (a_{\rm max}^{0.5}-a_{\rm amin}^{0.5})} (a_{\rm min}^{-2.5}-a_{\rm max}^{-2.5})~,
\end{equation}
which depends strongly on $a_{\rm min}$.

The abundances of all species are solved algebraically assuming a steady-state
in ionization, where creation and destruction of charged species balance 
each other (given the short time-scale of order a few 10$^1$ years for
the processes involved, \ct{Caselli+2002a}). 
The cosmic-ray ionization of H$_2$ is the primary source for 
H$^{+}$ and H$_3^{+}$ production in dense molecular clouds, 
which initiates the subsequent chemical reactions in the network through 
the relations: 
\begin{eqnarray}
& {\rm H}_2 + CR \rightarrow {\rm H}_2^+ + {\rm e}^-~,  \hspace{9mm}  & (1-\epsilon) \zeta^{\rm H_2} \label{Eq:CRionz1}\\
& {\rm H}_2 + CR \rightarrow {\rm H}^+ + {\rm H} + {\rm e}^-~,  \hspace{3mm} & \epsilon \zeta^{\rm H_2} \label{Eq:CRionz2}
\end{eqnarray}
in which $\epsilon \approx 0.05$ \citep{ShahGilbody1982}. 
H$_2^+$ reacts immediately with H$_2$ to form H$_3^+$, we thus consider
H$_3^+$ as a direct product of the ionization process given in 
Eq.~\ref{Eq:CRionz1}.

The steady-state chemical equations are summarized below, with 
rate coefficients derived from \citet{KunzMouschovias2009}. The 
charge transfer (CT) rate between atomic and molecular ions is 
$\beta \approx 2.5 \times 10^{-9}$
~cm$^3$~s$^{-1}$. 
The recombination rate of atomic ions and electrons is 
$\alpha_{\rm rec} \approx 2.8 \times 10^{-12} (300 {\rm K/T})^{0.86}$
~cm$^3$~s$^{-1}$.
The dissociative recombination (DR) rate of electrons and molecular ions is 
$\alpha_{\rm dr} \approx 2.0 \times 10^{-7} (300 {\rm K/T})^{0.75}$
~cm$^3$~s$^{-1}$.
The rate coefficients involving grains are calculated from 
\citet[Appendix A, see also \ct{DraineSutin1987}]{KunzMouschovias2009} 
and averaged over the MRN size distribution 
(ranging from $a_{\rm min}$ to $a_{\rm max}=0.25~\mu {\rm m}$): 
\begin{equation}
\label{Eq:avgMRN}
{\rm <}\alpha_{\rm xg} {\rm >}={\int_{a_{\rm min}}^{a_{\rm max}} \alpha_{\rm xg} {{\rm d}n \over {\rm d}a} {\rm d}a \over \int_{a_{\rm min}}^{a_{\rm max}} {{\rm d}n \over {\rm d}a} {\rm d}a}~,
\end{equation}
where $\alpha_{\rm xg}$ represents the recombination rate between a charged 
species ``x'' and a grain species ``g'' 
(neutral grain $g^0$, or charged grains $g^-$ and $g^+$). 
For simplicity, we denote the averaged quantity <$\alpha_{\rm xg}$> 
as $\alpha_{\rm xg}$ hereafter.

The protons H$^{+}$ produced by ionization of H$_2$ are mainly destroyed 
by CT with molecules and recombination on grains: 
\begin{equation}
\label{Eq:Hp}
\epsilon \zeta^{\rm H_2} n({\rm H}_2) = [\beta n(m)+\alpha_{\rm ig^-} n(g^-) + \alpha_{\rm ig^0} n(g^0)] n({\rm H}^+)~.
\end{equation}
The production of H$_3^+$ by ionization of H$_2$ is balanced by CT with 
heavy molecules, DR with electrons, and recombination on grains: 
\begin{equation}
\label{Eq:H3p}
(1-\epsilon) \zeta^{\rm H_2} n({\rm H}_2) = [\beta n(m)+\alpha_{\rm dr} n(e^-) + \alpha_{\rm ig^-} n(g^-) +\alpha_{\rm ig^0} n(g^0)] n({\rm H}_3^+)~.
\end{equation}
Molecular ions m$^+$ are mainly formed via CT of H$_3^+$ with heavy molecules, 
and destroyed by CT with metals, DR with electrons, and recombination 
on grains: 
\begin{equation}
\label{Eq:HCOp}
\beta n({\rm H}_3^+) n(m) = [\beta n(M)+\alpha_{\rm dr} n(e^-)+\alpha_{\rm ig^-} n(g^-) + \alpha_{\rm ig^0} n(g^0)] n(m^+)~.
\end{equation}
Metal ions M$^+$ are mainly formed via CT of H$_3^+$ and m$^+$ with metal 
atoms, and are destroyed by recombination with free electrons and on grains: 
\footnote{Even though the recombination rate $\alpha_{\rm rec}$ is 
$\gtrsim$10$^{4}$ smaller than $\alpha_{\rm ig^-}$, the recombination 
between metal ions and electrons can become very important when 
$n(g^-)\ll n(e^-)$ for large grain sizes.}
\begin{equation}
\label{Eq:Mp}
\beta [n({\rm H}_3^+)+n(m^+)] n(M) = [\alpha_{\rm rec} n(e^-)+\alpha_{\rm ig^-} n(g^-) + \alpha_{\rm ig^0} n(g^0)] n(M^+)~.
\end{equation}

The ionization equilibrium equations for charged grains are given as 
follows. 
Grains are negatively charged mainly by sticking electrons on the surface 
of neutral grains (assuming sticking probability as 1.0), while neutralized 
by recombination with ion species and CT with positively charged grains. 
So the equilibrium equation for $g^-$ is: 
\begin{equation}
\label{Eq:grm}
[\alpha_{\rm eg^0}n(e^-)] n(g^0) = [\alpha_{\rm ig^-} \sum n(i^+) + \alpha_{\rm g^+g^-} n(g^+)] n(g^-)~.
\end{equation}
Positively charged grains are formed by sticking ion species on the surface 
of neutral grains (sticking without fail), and are neutralized 
by recombination with electrons and CT with negatively charged grains. 
So the equilibrium equation for $g^+$ is: 
\begin{equation}
\label{Eq:grp}
[\alpha_{\rm ig^0} \sum n(i^+)]n(g^0) = [\alpha_{\rm eg^+} n(e^-) + \alpha_{\rm g^+g^-}n(g^-)] n(g^+)~,
\end{equation}
where we denote ion species collectively as 
$\sum n(i^+)=n({\rm H}^+)+n({\rm H}_3^+)+n(m^+)+n(M^+)$~. The CT between 
a charged grain and a neutral grain does not appear because 
it is already in steady-state and cancels out from both sides of the equation. 
Note that the grain-grain CT terms in right-hand side of 
Eq.~\ref{Eq:grm} and Eq.~\ref{Eq:grp} can be neglected in low density 
regimes (to simplify the solution) where the abundances of electrons and ions 
are orders of magnitude higher than that of grains, but not for 
higher densities when most electrons and ions have recombined and 
grains become the main charge carriers.

The set of steady-state equations Eq.~\ref{Eq:Hp}-\ref{Eq:grp}
are closed by charge neutrality, 
\begin{equation}
\label{Eq:charge}
n(g^-)+n(e^-) =  n({\rm H}^+)+n({\rm H}_3^+)+n(m^+)+n(M^+)+n(g^+)~,
\end{equation}
and the constraint on the total number of grains, 
\begin{equation}
\label{Eq:grsum}
n(g)=n(g^0)+n(g^-)+n(g^+)~.
\end{equation}
The above equations can be simplified algebraically (similar to 
\ct{Padovani+2014}). In the low density regime, 
where recombinations of electrons and ions are inefficient, 
we simplify the equations by ignoring 
the grain-grain CT terms in Eq.~\ref{Eq:grm} and Eq.~\ref{Eq:grp}.
When electrons are mostly recombined (solution of $n(e^-) \rightarrow 0$),
we approximate the abundance of electrons by a simple power law 
$x(e^-) \propto (n({\rm H}_2)/\zeta^{\rm H_2})^{-1}$ 
\citep{UmebayashiNakano1990}, hence to simplify the equations for 
the high density regime. Note that the boundary between 
low and high density regimes is grain-size dependent. 

The solution to our chemical network requires no iteration in low density 
regimes, and up to $\sim$20 iterations (depending on accuracy) in high density 
regimes. The low computational cost of the network makes it suitable for 
solving chemistry along with hydrodynamics during run-time 
(see the next section \S~\ref{S.MHDCoefs}).

\subsection{Non-ideal MHD Diffusivity}
\label{S.MHDCoefs}

The evolution of magnetic field in magnetohydrodynamics is governed by 
the induction equation, 
\begin{equation}
\label{Eq:inductB}
\begin{split}
{\partial \bmath{B} \over \partial t} = \nabla \times (\bmath{\rm v} \times \bmath{B}) - \nabla \times & \left\{ \eta_{\rm Ohm} \nabla \times \bmath{B} + \eta_{\rm Hall} (\nabla \times \bmath{B}) \times {\bmath{B} \over B} \right. \\
& \left. + \eta_{\rm AD} {\bmath{B} \over B} \times \left[ (\nabla \times \bmath{B}) \times {\bmath{B} \over B} \right] \right\}~,
\end{split}
\end{equation}
where {\bf v} is the fluid velocity, and $\eta_{\rm Ohm}$, 
$\eta_{\rm Hall}$, and $\eta_{\rm AD}$ are the Ohmic, Hall, and ambipolar 
diffusivities, respectively. 
The three non-ideal MHD coefficients can be expressed in terms of 
the components of the conductivity tensor $\sigma$ 
\citep[e.g.,][]{Wardle2007}: 
\begin{eqnarray}
\eta_{\rm AD} & = & {c^2 \over 4\pi} ({\sigma_{\rm P} \over \sigma_{\rm P}^2 + \sigma_{\rm H}^2} - {1\over \sigma_{\parallel}})~, \label{Eq:MHDcoef1}\\
\eta_{\rm Ohm} & = & {c^2 \over {4\pi \sigma_{\parallel}}}~, \label{Eq:MHDcoef2}\\
\eta_{\rm Hall} & = & {c^2 \over 4\pi} ({\sigma_{\rm H} \over \sigma_{\rm P}^2 + \sigma_{\rm H}^2})~; \label{Eq:MHDcoef3}
\end{eqnarray}
where the parallel $\sigma_{\parallel}$, Pedersen $\sigma_{\rm P}$, 
and Hall $\sigma_{\rm H}$ conductivities are related to the 
Hall parameter $\beta_{i,\rm H_2}$ as: 
\begin{eqnarray}
\sigma_{\parallel} & = & {{e c n({\rm H}_2)} \over B} \sum_i Z_i x_i \beta_{i,\rm H_2}~, \label{Eq:conduct1}\\
\sigma_{\rm P} & = & {{e c n({\rm H}_2)} \over B} \sum_i {{Z_i x_i \beta_{i, \rm H_2}} \over {1+\beta_{i,\rm H_2}^2}}~,\label{Eq:conduct2}\\
\sigma_{\rm H} & = & {{e c n({\rm H}_2)} \over B} \sum_i {{Z_i x_i} \over {1+\beta_{i,\rm H_2}^2}}~; \label{Eq:conduct3}
\end{eqnarray}
\citep{NormanHeyvaerts1985, WardleNg1999}, where $x_i$ is the abundance of 
charged species $i$.
The Hall parameter $\beta_{i,\rm H_2}$ is the key quantity 
that determines the relative importance of the Lorentz and drag forces 
in balancing the electric force for each charged species $i$ in a sea 
of neutral H$_2$ molecules. It is defined as: 
\begin{equation}
\label{Eq:mtr}
\beta_{i,\rm H_2} = ({{Z_i e B} \over {m_i c}}) {{m_i+m_{\rm H_2}} \over {\mu m_{\rm H} n({\rm H}_2) <\sigma v>_{i,\rm H_2}}}~,
\end{equation}
where $m_i$ and $Z_i e$ are the mass and the charge of charged species i, 
respectively, 
and <$\sigma v$>$_{\rm i,H_2}$ is the momentum transfer rate coefficient, 
parametrized as a function of temperature \citep{PintoGalli2008}, 
which quantifies the collisional coupling between neutral (H$_2$) and 
charged ($i$) species. Note that the momentum transfer rate coefficient 
$\beta_{g^-,\rm H_2}$ and $\beta_{g^+,\rm H_2}$
for charged grain species $g^-$ and $g^+$ are also averaged 
over the MRN distribution similar to the recombination rates 
(Eq.~\ref{Eq:avgMRN}).

The solution to our equilibrium chemistry network (\S~\ref{S.ChemNet}) 
provides the abundances for charged ion and grain species, 
which are used to update the non-ideal MHD coefficients 
(from the above equations Eq.~\ref{Eq:MHDcoef1}-\ref{Eq:mtr}) 
at each point in the computational domain. 
The set-up of numerical simulation is described next (\S~\ref{Chap.IC}).

\section{Initial Condition}
\label{Chap.IC}

We carry out two-dimensional (2D) axisymmetric numerical simulations 
using ZeusTW code \citep{Krasnopolsky+2010} 
-- a Zeus family MHD code including self-gravity and all three 
non-ideal MHD effects. The MHD is solved through constraint transport 
method to preserve the divergence-free condition for magnetic field. 
We implement a Lorentz force limiter similar to \cite{MillerStone2000} 
to avoid huge Alfv\'en speed in the low-density bipolar region. 
The chemistry network is solved at every hydrodynamic timestep, 
and at each spatial point to compute the non-ideal MHD coefficients. 
In this study, we consider only ambipolar diffusion 
and Ohmic dissipation for diffusion of magnetic field 
(i.e., the second term inside the curly braces in Eq.~\ref{Eq:inductB} 
is neglected). The Hall effect will be addressed in future studies.
To avoid extremely small timesteps, we cap the Ohmic diffusivity at 
$1.0\times10^{20}$~cm$^2$~s$^{-1}$. 

We initialize a uniform, isolated spherical core with total mass 
$M_{\rm c}=1.0~M_{\sun}$, 
and radius $R_{\rm c}=10^{17}$~cm~$\approx 6684$~AU. This corresponds
to an initial mass density $\rho_0=4.77 \times 10^{-19}$~g~cm$^{-3}$ and 
a volume density for molecular hydrogen $n({\rm H}_2)=10^5$~cm$^{-3}$ 
(assuming mean molecular weight $\mu=2.36$). The free-fall time of the core
is thus $t_{\rm ff} =3 \times 10^{12}$~s~$\approx 9.6 \times 10^4$~yr.
We assume an isothermal equation of state below a critical density 
$\rho_{\rm cr}=10^{-13}$~g~cm$^{-3}$, and $P \propto \rho^{5/3}$ at 
densities above. \footnote{In general, the adiabatic index varies with density.
For the density range considered in this study 
($n(\rm H_2)\lesssim$ 10$^{13}$~cm$^{-3}$ or 
$\rho\lesssim$ 4$\times$10$^{-11}$~g~cm$^{-3}$), 
gas temperature is well below $\sim$200-300~K almost everywhere. 
Therefore, molecular hydrogen still behaves like monotomic gas with an 
adiabatic index of 5/3 \citep{MasunagaInutsuka2000,Tomida+2013}.}
The ratio of thermal to gravitational energy is 
$\alpha_{\rm thm}={5R_{\rm c} c_{\rm s} ^2 \over 2GM_{\rm c}}=0.75$, in which 
$c_{\rm s}=0.2$~km~s$^{-1}$ is the isothermal sound speed.
The core is rotating initially as a solid-body with angular speed 
$\omega_0=1 \times 10^{-13}$~s$^{-1}$ for slow rotating case, and 
$2 \times 10^{-13}$~s$^{-1}$ for fast rotating case, 
which corresponds to a ratio of rotational to gravitational energy 
$\beta_{\rm rot}={R_{\rm c}^3 \omega_0^2 \over 3 G M_{\rm c}}=0.025$ 
~and~$0.1$, respectively (within range of the typical 
$\beta_{\rm rot}$ estimated by \citet{Goodman+1993,Caselli+2002b}). 
The initial core is threaded by a uniform magnetic field along the 
rotation axis with a constant strength of 
$B_0 \approx 2.13 \times 10^{-5}$~G for weak field case 
and $4.25\times 10^{-5}$~G for strong field case, which corresponds to 
a mass-to-flux ratio of $\lambda=4.8$ and $\lambda=2.4$, respectively. 
The strong field case is particularly consistent with the 
mean value of $\lambda$ inferred from the OH Zeeman observations by 
\citet{TrolandCrutcher2008}.

We adopt the spherical coordinate system (r, $\theta$) and non-uniform grid  
to provide high resolution towards the innermost region of simulation domain. 
The inner boundary has a radius $r_{\rm in}=3 \times 10^{13}$~cm~$=2$~AU 
and the outer has $r_{\rm out}=10^{\rm 17}$~cm. At both boundaries, we 
impose a standard outflow boundary conditions to allow matter to leave 
the computational domain. The mass accreated across the inner boundary 
is collected at the center as the stellar object. We use a total of 
$120 \times 96$ grid points. The grid is uniform in the $\theta$-direction, 
and is non-uniform in the $r$-direction with a spacing $\delta r=0.1$~AU 
next to the inner boundary. The spacing increases geometrically outward
by a constant factor of $\sim$1.0647. 

In this study, we investigate two typical cosmic-ray ionization rate
$\zeta_0^{\rm H_2}=1.0\times 10^{-17}$~s$^{-1}$ 
~and~$5.0\times 10^{-17}$~s$^{-1}$, along with different levels of 
initial rotation and magnetization, and focus on the effect of dust 
grain size on the formation of RSDs. The simulation models are summarized 
in Table~\ref{Tab:model1}--\ref{Tab:model2}.
\begin{table}
\caption{Model Parameters for strong B-field $B_0=42.5~\mu$G ($\lambda$=2.4)}
\label{Tab:model1}
\begin{tabular}{lccllc}
\hline\hline
Model & Grain Size & $\zeta_0^{\rm H_2}$ & $\beta_{\rm rot}$ & RSD & Disk Radius \\
 & & $10^{-17}$~s$^{-1}$ & & & AU \\
\hline
Slw-MRN5 & MRN & 5 & .025 & N & -- \\
Slw-trMRN5 & tr-MRN & 5 & .025 & N$^{\rm Trans}$ & $\lesssim$12$\rightarrow$0\\
Slw-LG5 & LG & 5 & .025 & N$^{\rm Trans}$ & <5$\rightarrow$0 \\
Slw-MRN1$^{\rm R}$  & MRN & 1 & .025 & N$^{\rm Trans}$ & $\lesssim$9$\rightarrow$0 \\
Slw-trMRN1 & tr-MRN & 1 & .025 & {\bf Y} & {\bf $\sim$20 ID} \\
Slw-LG1 & LG & 1 & .025 & N$^{\rm Trans}$ & $\lesssim$11$\rightarrow$0 \\
\hline
Fst-MRN5 & MRN & 5 & .1 & N$^{\rm Trans}$ & $\lesssim$10$\rightarrow$0 \\
Fst-trMRN5 & tr-MRN & 5 & .1 & Y$^{\rm Shrink}_{\rm 4.7~kyr}$ & $\lesssim$18$\rightarrow$12\\
Fst-LG5 & LG & 5 & .1 & N$^{\rm Trans}$ & $\lesssim$13$\rightarrow$0 \\
Fst-MRN1 & MRN & 1 & .1 & N$^{\rm Trans}$ & $\lesssim$11$\rightarrow$0 \\
Fst-trMRN1 & tr-MRN & 1 & .1 & {\bf Y} & {\bf $\sim$40 ID+OR} \\
Fst-LG1 & LG & 1 & .1 & Y$^{\rm Shrink}_{\rm 5.7~kyr}$ & $\lesssim$18$\rightarrow$7 \\
\hline
\end{tabular}
$^*$Slw-MRN1$^{\rm R}$ is the reference model; 
other symbols are explained in Table~\ref{Tab:model2}. 
\end{table}

\begin{table}
\caption{Model Parameters for weak B-field $B_0=21.3~\mu$G 
($\lambda$=4.8)}
\label{Tab:model2}
\begin{tabular}{lccllc}
\hline\hline
Model & Grain Size & $\zeta_0^{\rm H_2}$ & $\beta_{\rm rot}$ & RSD & Disk Radius \\
 & & $10^{-17}$~s$^{-1}$ & & & AU \\
\hline
Slw-MRN5 & MRN & 5 & .025 & N$^{\rm Trans}$ & $\lesssim$14$\rightarrow$0 \\
Slw-trMRN5 & tr-MRN & 5 & .025 & Y$^{\rm Shrink}_{\rm 6.7~kyr}$ & $\lesssim$21$\rightarrow$15 \\
Slw-LG5 & LG & 5 & .025 & N$^{\rm Trans}$ & $\lesssim$14$\rightarrow$0 \\
Slw-MRN1 & MRN & 1 & .025 & Y$^{\rm Shrink}_{\rm 4.1~kyr}$ & $\lesssim$16$\rightarrow$9 \\
Slw-trMRN1 & tr-MRN & 1 & .025 & {\bf Y} & {\bf $\sim$35 ID+OR} \\
Slw-LG1 & LG & 1 & .025 & Y$^{\rm Shrink}_{\rm 6.0~kyr}$ & $\lesssim$19$\rightarrow$13\\
\hline
Fst-MRN5 & MRN & 5 & .1 & N$^{\rm Trans}$ & $\lesssim$20$\rightarrow$0 \\
Fst-trMRN5 & tr-MRN & 5 & .1 & {\bf Y$^*$} & $\lesssim$35$\rightarrow${\bf 16 ID}\\
Fst-LG5 & LG & 5 & .1 & N$^{\rm Trans}$ & $\lesssim$15$\rightarrow$0 \\
Fst-MRN1 & MRN & 1 & .1 & Y$^{\rm Shrink}_{\rm 6.3~kyr}$ & $\lesssim$25$\rightarrow$10 \\
Fst-trMRN1 & tr-MRN & 1 & .1 & {\bf Y} & {\bf $\sim$40--60 OR} \\
Fst-LG1 & LG & 1 & .1 & Y$^{\rm Shrink}_{\rm 7.5~kyr}$ & $\lesssim$30$\rightarrow$13 \\
\hline
\end{tabular}
$\dagger$~MRN: full MRN distribution with $a_{\rm min}=0.005~\mu$m, 
$a_{\rm max}=0.25~\mu$m \\
$\dagger$~tr-MRN: truncated MRN with $a_{\rm min}=0.1~\mu$m, 
$a_{\rm max}=0.25~\mu$m \\
$\dagger$~LG: large grain of single size $a=1.0~\mu$m \\
$\dagger$~N$^{\rm Trans}$: a transient disk forms from ``first core''-like 
structure, not rotationally supported for most of its lifetime, 
and disappears quickly in $\sim$few $10^2$ years.\\
$\dagger$~Y$^{\rm Shrink}$: initially forms a RSD but shrinks in size
over time (estimated lifetime $\sim$few $10^3$ years ). \\
$\dagger$~Y$^*$: an intermediate type of RSD, shrinking in size over 
$16$~kyrs. The final disk radius holds around $\sim$16~AU. \\
$\dagger$~ID: inner rotationally supported disk. \\
$\dagger$~OR: outer self-gravitating ring.
\end{table}

\section{Analytical Results of Chemistry Model}
\label{Chap.ChemResult}

In this section, we discuss the results from our analytic chemistry model, 
with a particular focus on the enhancement of AD by changing dust grain size. 
The numerical results from ZeusTW simulations are presented in the 
next section.

To illustrate the effect of dust grain size on magnetic diffusivities, 
we use the charge abundances solved by our chemistry network 
(\S~\ref{S.ChemNet}) plus a simple field strength-density relation 
\begin{equation}
\label{Eq:Brelation}
|\bmath{B}|=0.143~\left[{n({\rm H}_2) \over {\rm cm}^{-3}}\right]^{0.5}~\mu{\rm G}~,
\end{equation}
\citep{Nakano+2002}, to estimate analytically 
the three non-ideal MHD coefficients, 
$\eta_{\rm AD}$, $\eta_{\rm Ohm}$, and $\eta_{\rm Hall}$. 
Note that the magnetic field strength will be computed 
self-consistently in our simulations.

We here compare three cases in terms of chemical abundances and 
magnetic diffusivities: 
a) full MRN distribution with $a_{\rm min}=0.005~\mu$m, 
b) truncated MRN (tr-MRN) distribution with $a_{\rm min}=0.1~\mu$m, 
and c) large grains (LG) of single size $a=1.0~\mu$m. 
The cosmic-ray ionization rate is set at $\zeta_0^{\rm H_2}=1.0\times10^{-17}$
or $5.0\times10^{-17}$~s$^{-1}$, 
with attenuation prescribed by Eq.~\ref{Eq:attenu}.

\subsection{Fractional Abundance}
\label{S.Abund}

\begin{figure*}
\includegraphics[width=1.03\textwidth]{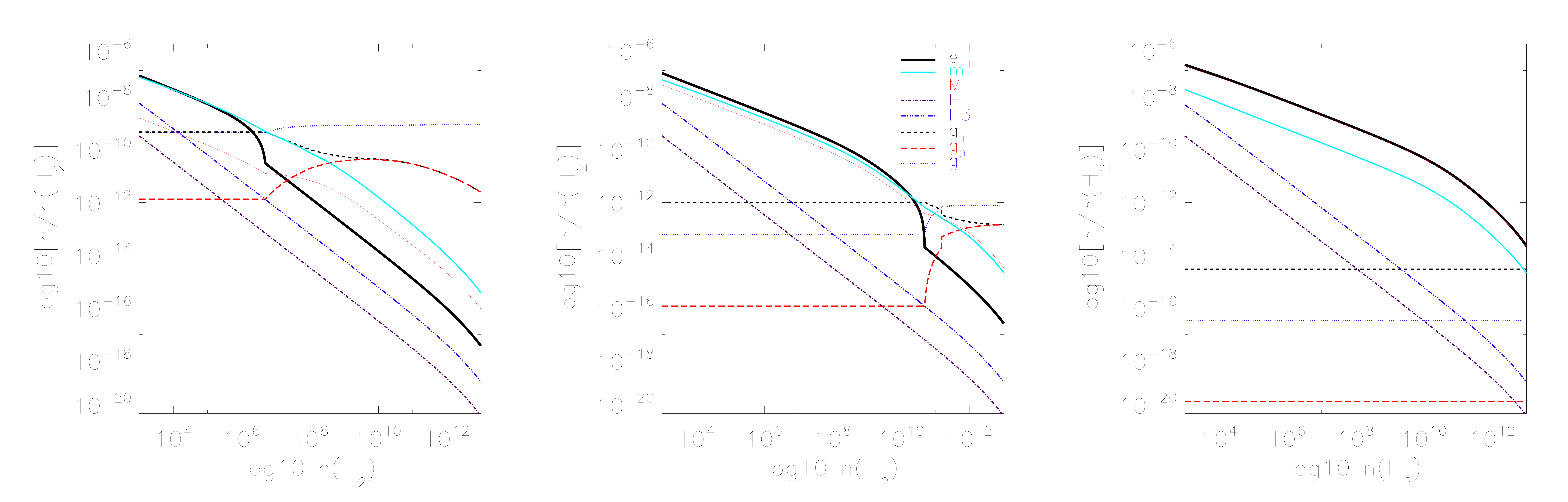}
\caption{Fractional abundances for MRN grain size distribution (left panel),
tr-MRN size distribution (middle panel), and LG of single size 1~$\mu$m 
(right panel). Labels of species are marked in the middle panel.}
\label{Fig:abund}
\end{figure*}

Fig.~\ref{Fig:abund} shows the fractional abundances of charged gas-phase
species and dust grains for the three grain size cases (MRN, tr-MRN, and LG), 
from which the following trends are evident.

{\it First}, in the MRN case, charges are mainly dominated 
by $e^-$, $m^+$, and to a lesser extent $M^+$ in the low density regime 
($\lesssim$10$^{7}$~cm$^{-3}$), 
and by charged grains $g^-$ and $g^+$ in the high density regime 
($\gtrsim$10$^{9}$~cm$^{-3}$), 
with a transition region in between when recombination of $e^-$ starts to 
become very efficient. In the tr-MRN case, such transition occurs much later
only when number density reaches above $\gtrsim$10$^{10}$~cm$^{-3}$. In the LG
case, the transition shifts so much beyond $\gtrsim$10$^{14}$~cm$^{-3}$ that
in the density range of interest to us, only $e^-$ and $M^+$ are the dominant
charges. Beyond the transition region in both MRN and tr-MRN cases, 
the abundances of positively and negatively charged grains become nearly 
equal; this will affect the Hall conductivity $\sigma_{\rm H}$ 
in high density regimes (see \S~\ref{S.Conduct}).

{\it Second}, as the grain size increases, the abundances of 
$e^-$ and $M^+$ also increase, due to the decrease of
total grain surface area. Note that such change affects the $e^-$ abundance 
by less than a factor of $10$, and has a stronger effect on the abundance 
of $M^+$ (roughly proportional to grain size $a$). The reason is that the 
destruction of metal ions $M^+$ is mainly through recombination 
on grains ($\alpha_{\rm ig^-}$ and $\alpha_{\rm ig^0}$, both are orders 
of magnitude larger than $\alpha_{\rm rec}$), 
which is sensitive to total grain surface area. 
However, for electrons, the $\propto a$ dependence 
from recombination on grains is weakened by the recombination 
with excessive metal and molecular ions (Eq.~\ref{Eq:HCOp}-\ref{Eq:Mp}), 
as total grain surface area decreases.
Nevertheless, the competing mechanism makes the 
overall gas-phase charge density $x(m^+)+x(M^+)+x(e^-)$ 
depend only weakly on total grain surface area.

{\it Third}, the abundance of grains decreases drastically when $a_{\rm min}$ 
increases (the dependence is slightly weaker than $a_{\rm min}^{-2.5}$; 
see Eq.~\ref{Eq:xgr}), 
because we assume a constant grain mass fraction --- $1\%$ of gas mass. 
Although this is a very rough representation for the actual grain growth 
process in dense cores, we find the change in grain abundances has 
a great impact on the ambipolar diffusivity 
(along with changes in the Hall parameter of grains).

\subsection{Enhanced Ambipolar-Diffusion}
\label{S.EnhancAD}

\begin{figure}
\includegraphics[width=\columnwidth]{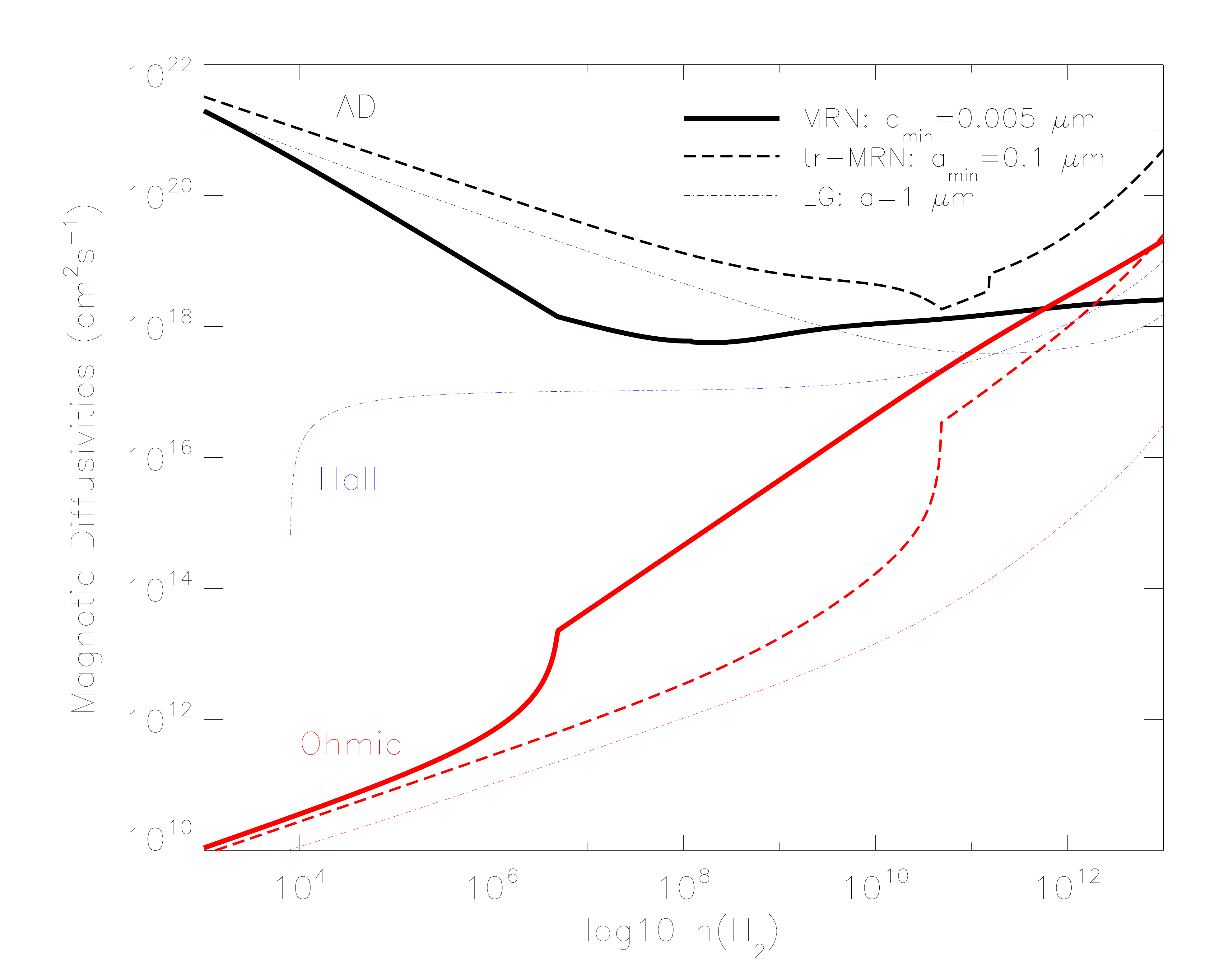}
\caption{Magnetic diffusivities for different cases of grain size with  
$\zeta_0^{\rm H_2}=1.0 \times 10^{-17}$~s$^{-1}$, for the illustrative 
magnetic field-density relation given in Eq.~\ref{Eq:Brelation}. 
Black: $\eta_{\rm AD}$; red: $\eta_{\rm Ohmic}$; blue: $\eta_{\rm Hall}$. 
The Hall diffusivity in the MRN and tr-MRN cases are negative 
within the density range. }
\label{Fig:MHDcoefs}
\end{figure}

The magnetic diffusivities 
$\eta_{\rm AD}$, $\eta_{\rm Ohm}$, and $\eta_{\rm Hall}$, 
computed via Eq.~\ref{Eq:MHDcoef1}-\ref{Eq:MHDcoef3}, are shown in 
Fig.~\ref{Fig:MHDcoefs} for all three grain size cases (MRN, tr-MRN, and LG). 
Note that the Hall diffusivity $\eta_{\rm Hall}$ is always negative for 
both MRN and tr-MRN cases. 

The most remarkable effect is the enhancement of ambipolar diffusivity
$\eta_{\rm AD}$ in the tr-MRN case compared with that of the MRN case 
($\sim$1--2 orders of magnitude),
particularly at lower densities ($n({\rm H}_2)\lesssim$10$^{10}$~cm$^{-3}$) 
where magnetic diffusivity matters most for saving gas angular momentum 
from catastrophic magnetic braking. The enhancement of AD is essentially 
caused by the lack of a large population of VSGs 
(with size from $\sim$10~$\AA$ to few 100~$\AA$), 
which are well-coupled to the magnetic field and dominate the 
fluid conductivity (see \S~\ref{S.Conduct} for detailed analysis). 
Recall the analysis in \S~\ref{S.Abund} (see Fig.~\ref{Fig:abund}), 
the decrease of total surface area from 
$a_{\rm min}=0.005~\mu$m to $a_{\rm min}=0.1~\mu$m 
only increases the overall gas-phase charge density by a factor of a few, 
whose effect on conductivity is at least 1 orders of magnitude smaller 
than that of the large population of VSGs. 
Therefore, truncating the lower end of MRN distribution (both 
a lower grain fractional abundance and a lack of highly conductive VSGs) 
boosts the ambipolar diffusivity $\eta_{\rm AD}$ by $\sim$1--2 
orders of magnitude, while the increased gas-phase charge density 
only lowers $\eta_{\rm AD}$ by a factor of a few.

We have explored other values of $a_{\rm min}$, while fixing 
$a_{\rm max}=0.25~\mu$m, and report the following trends 
(not plotted). \\
$\bullet$ The peak of the AD enhancement occurs just near 
$a_{\rm min}\approx 0.1~\mu$m, above which $\eta_{\rm AD}$ 
starts to fall off but slowly due to the weak counteracting effect 
from the increasing gas-phase charges. For example, 
the magnitude of $\eta_{\rm AD}$ in the LG case 
(shown in Fig~\ref{Fig:MHDcoefs}, in between the other two cases) 
is roughly comparable to that of $a_{\rm min}\approx 0.04~\mu$m 
in the low density regime ($\lesssim$10$^{10}$~cm$^{-3}$); but it is 
$\sim$10--100 times smaller at high densities ($\gtrsim$10$^{10}$~cm$^{-3}$) 
than that of the small grain cases due to insufficient recombination of 
ions and electrons on grain surfaces. For even larger grains (of single size 
$\gtrsim$10$^1~\mu$m), their $\eta_{\rm AD}$ are roughly comparable to 
that of $a_{\rm min}\approx 0.03\mu$m, which are still bigger than that 
of the MRN case in low density regimes. \\
$\bullet$ From $a_{\rm min}\approx 0.1~\mu$m down to 
$a_{\rm min}\approx 0.01~\mu$m, the $\eta_{\rm AD}$ curve drops down 
because of the rapid increase in the Hall conductivity $\sigma_{\rm H}$ 
(Eq.~\ref{Eq:MHDcoef1}), which is determined by negatively charged grains 
(see \S~\ref{S.Conduct}). Here, the increase in $\sigma_{\rm H}$ is mainly 
caused by the increase in the abundance of negatively charged grains 
$x(g^-)$. When $a_{\rm min}\lesssim0.04~\mu$m, 
$\sigma_{\rm H}$ rises above $\sigma_{\rm P}$ between 
$n({\rm H}_2)\sim$10$^{6}$-10$^{11}$~cm$^{-3}$. 
Below $a_{\rm min}\approx 0.02~\mu$m, negatively charged grains 
start to dominate the Pedersen conductivity $\sigma_{\rm P}$ as well 
at low densities. 
These small size grains (10--200~$\AA$), relatively well-coupled to 
the magnetic field (|$\beta_{g^-,\rm H_2}$| around unity), 
can exert stronger drag to ${\rm H}_2$ molecules than ions and electrons 
do \citep{Pinto+2008}. A large number of such grains 
(by reducing $a_{\rm min}$) therefore dominate the conductivity 
$\sigma_{\rm P}$ (e.g., see Fig.~\ref{Fig:conds50} in \S~\ref{S.Conduct}). \\
$\bullet$ There exists a ``worst grain size'' $a_{\rm min}\approx 0.01~\mu$m 
that produces an overall lowest ambipolar diffusivity in the low density 
regime, with a minimum of $\eta_{\rm AD}\sim$10$^{17}$~cm$^2$~s$^{-1}$ 
between $n({\rm H}_2)\sim$10$^8$-10$^9$~cm$^{-3}$. This is because the 
Hall conductivity (Eq.~\ref{Eq:MHDcoef1}) reaches a maximum, which is 
$\gtrsim$1 order of magnitude higher than the Pedersen conductivity 
at densities between $\sim$10$^8$-10$^9$~cm$^{-3}$. \\
$\bullet$ From $a_{\rm min}\approx 0.01~\mu$m down to 
$a_{\rm min}\approx 0.003~\mu$m=30~$\AA$, the $\eta_{\rm AD}$ curve 
lifts up slightly between $n({\rm H}_2) \sim$10$^7$-10$^9$~cm$^{-3}$ 
as Pedersen conductivity $\sigma_{\rm P}$ starts to increase 
(both $x(g^-)$ and |$\beta_{g^-,\rm H_2}$| increases) and Hall conductivity 
$\sigma_{\rm H}$ remains relatively unchanged (|$\beta_{g^-,\rm H_2}$| 
becomes larger than 1 and offsets the increase in $x(g^-)$). Note that 
in this size range $\sigma_{\rm P}\ll\sigma_{\rm H}$ at low densities. \\
$\bullet$ Below $a_{\rm min}\approx 30$~$\AA$, $\sigma_{\rm P}$ 
finally becomes larger than $\sigma_{\rm H}$ again, and 
the $\eta_{\rm AD}$ curve slowly rises as both Pedersen and 
Hall conductivities gradually decrease due to the increasing 
|$\beta_{g^-,\rm H_2}$|. Note that the PAH-type grains with size 
10~$\AA$ or less have Hall parameter comparable to that of ions. 
However, even with $a_{\rm min}\approx 4~\AA$ 
\citep[smallest possible grain size][]{WeingartnerDraine2001}, 
the $\eta_{\rm AD}$ curve only returns to the level of 
$a_{\rm min}\approx 0.03~\mu$m in low density regimes.

We also investigate other maximum grain sizes $a_{\rm max}$. 
For example, when $a_{\rm max}=1~\mu$m, the optimal $a_{\rm min}$ 
that yields the largest AD enhancement is reduced 
to $\approx$0.055~$\mu$m. Besides, the overall $\eta_{\rm AD}$ curve 
with [0.055$\mu$m, 1$\mu$m] is slightly lower than the [0.1$\mu$m, 0.25$\mu$m]
curve. The reason is that, large $a_{\rm max}$ reduces the total grain 
surface area for recombination; thus to suppress the excessive ions and 
electrons, one needs to add back more small grains to restore enough grain 
surface area. But too many small grains of few 100~$\AA$ will instead 
reduce $\eta_{\rm AD}$ because they dominates the fluid conductivity, 
especially when $a_{\rm min}\lesssim0.02~\mu$m. 

The differences in Ohmic diffusivity $\eta_{\rm Ohm}$ 
are also prominent among all cases. Fig.~\ref{Fig:MHDcoefs} shows 
that grain size anti-correlates with Ohmic diffusivity $\eta_{\rm Ohm}$. 
The main current carrier --- electrons --- are responsible for the strength of 
Ohmic dissipation. 
As grain size increases, the recombination of electrons becomes 
less efficient due to a smaller total grain surface area, which boosts 
the fluid conductivity (see \S~\ref{S.Conduct} for detailed analysis) 
and suppresses the Ohmic diffusivity.
Note that the Ohmic dissipation in Fig.~\ref{Fig:MHDcoefs} 
only starts to strongly affect the diffusion of magnetic fields
at high densities $n({\rm H}_2) \gtrsim$10$^{11}$~cm$^{-3}$, 
below which AD and Hall effect dominate 
(consistent with existing literature, e.g., \ct{KunzMouschovias2010,Li+2011}).

Our result also shows that the relative importance between 
Ohmic dissipation and AD varies with grain size distribution. 
In the standard MRN case, Ohmic dissipation dominates over AD when 
$n({\rm H}_2) \gtrsim$few $10^{11}$~cm$^{-3}$. In contrast, 
in both tr-MRN and LG cases, ambipolar diffusivity $\eta_{\rm AD}$ is 
always orders of magnitude larger than Ohmic diffusivity $\eta_{\rm Ohm}$ 
for the density range in this study ($\lesssim$10$^{13}$~cm$^{-3}$). 
Our chemistry model indicates that such a result also holds for other 
values of $a_{\rm min}$ as long as $a_{\rm min} \gtrsim$0.02~$\mu$m. 
Particularly, the strong AD in the tr-MRN case at high densities 
($\sim$10$^{11}$-10$^{13}$~cm$^{-3}$) is caused by the lack of highly 
conductive VSGs which dominate the Pedersen conductivity $\sigma_{\rm P}$ 
in the MRN case \citep[\S.~\ref{S.Conduct}, see also][]{DeschMouschovias2001,
Dapp+2012}.

\subsection{Effect of Cosmic-ray Ionization Rate}
\label{S.CRrate}

\begin{figure}
\includegraphics[width=\columnwidth]{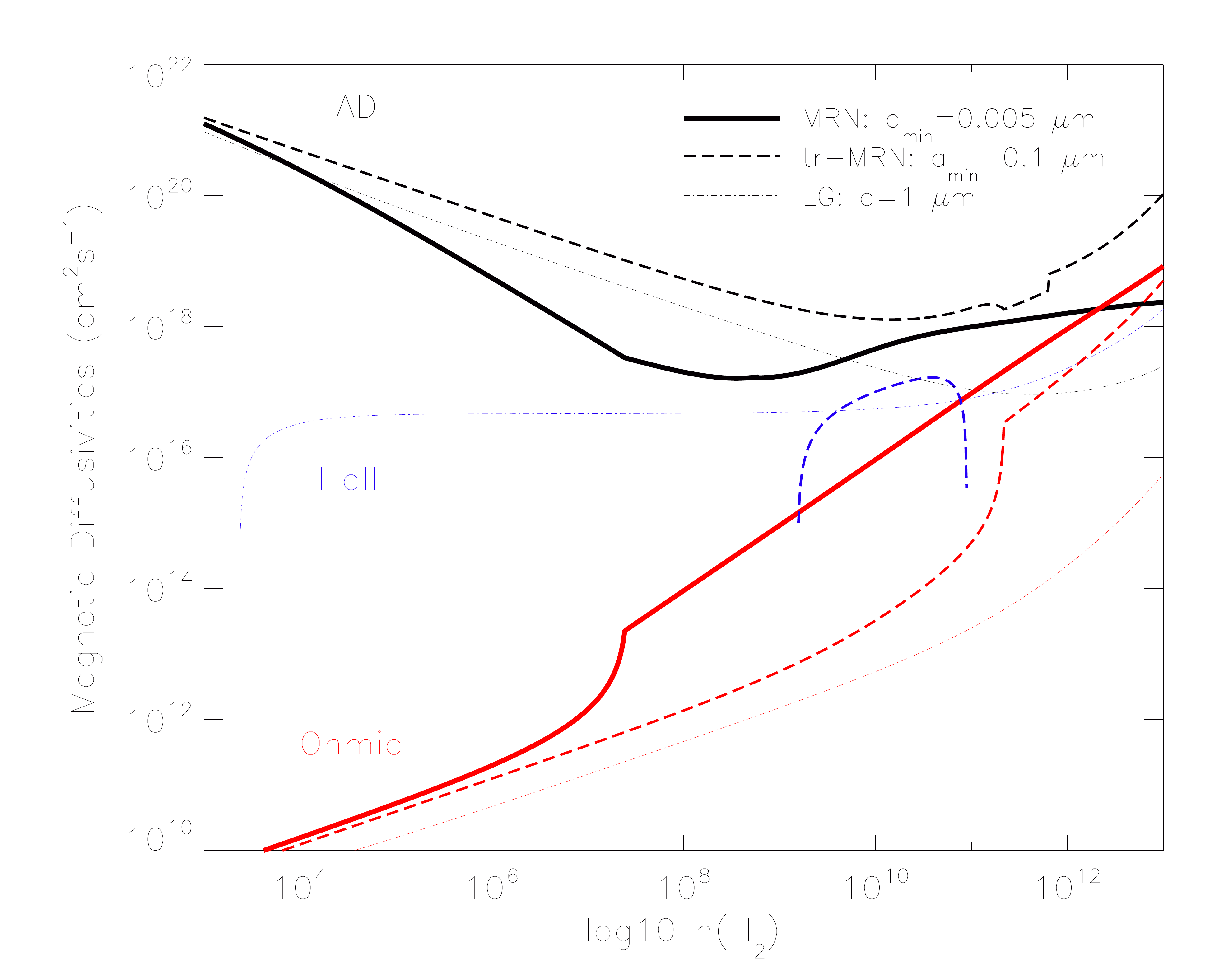}
\caption{Magnetic diffusivities for different cases of grain size with 
$\zeta_0^{\rm H_2}=5.0 \times 10^{-17}$~s$^{-1}$, for the illustrative 
magnetic field-density relation given in Eq.~\ref{Eq:Brelation}. 
The Hall diffusivity in the MRN case are negative within the density 
range.}
\label{Fig:MHDcoefs5}
\end{figure}

The charge abundances and hence magnetic diffusivities also depend on 
the cosmic-ray ionization rate. 
Our chemistry model shows that the magnetic diffusivity scales with 
cosmic-ray ionization rate roughly as $\propto \sqrt{\zeta^{\rm H_2}}$ 
(see Fig.~\ref{Fig:MHDcoefs5}). Though the effect varies somewhat 
with density, it largely agrees with theoretical prediction \citep{Shu1991}.
The factor of $\sim$2 difference in magnetic diffusivities, 
by changing $\zeta_0^{\rm H_2}$ from $1.0 \times 10^{-17}$~s$^{-1}$ to 
$5.0 \times 10^{-17}$~s$^{-1}$, 
will also play a role in the formation of RSDs
(see Table~\ref{Tab:model1}-\ref{Tab:model2} and \S~\ref{S.ShrinkDisk}).
Its role is not as decisive as that of the grain size, 
but can greatly affect the age and morphology of disks 
and is more important than parameters other than grain size.

\subsection{Analysis of Conductivity}
\label{S.Conduct}

In this section, we will elaborate on the conductivity of charged 
species to provide a more thorough explanation for the results above.
The bottom line is that: 
1) both the fractional abundances of charged species (\S~\ref{S.Abund})
and their Hall parameters (Eq.~\ref{Eq:mtr}) are important to 
fluid conductivity; 
and 2) a large population of conductive VSGs in MRN distribution 
greatly reduces the ambipolar diffusivity $\eta_{\rm AD}$.

\begin{figure}
\includegraphics[width=\columnwidth]{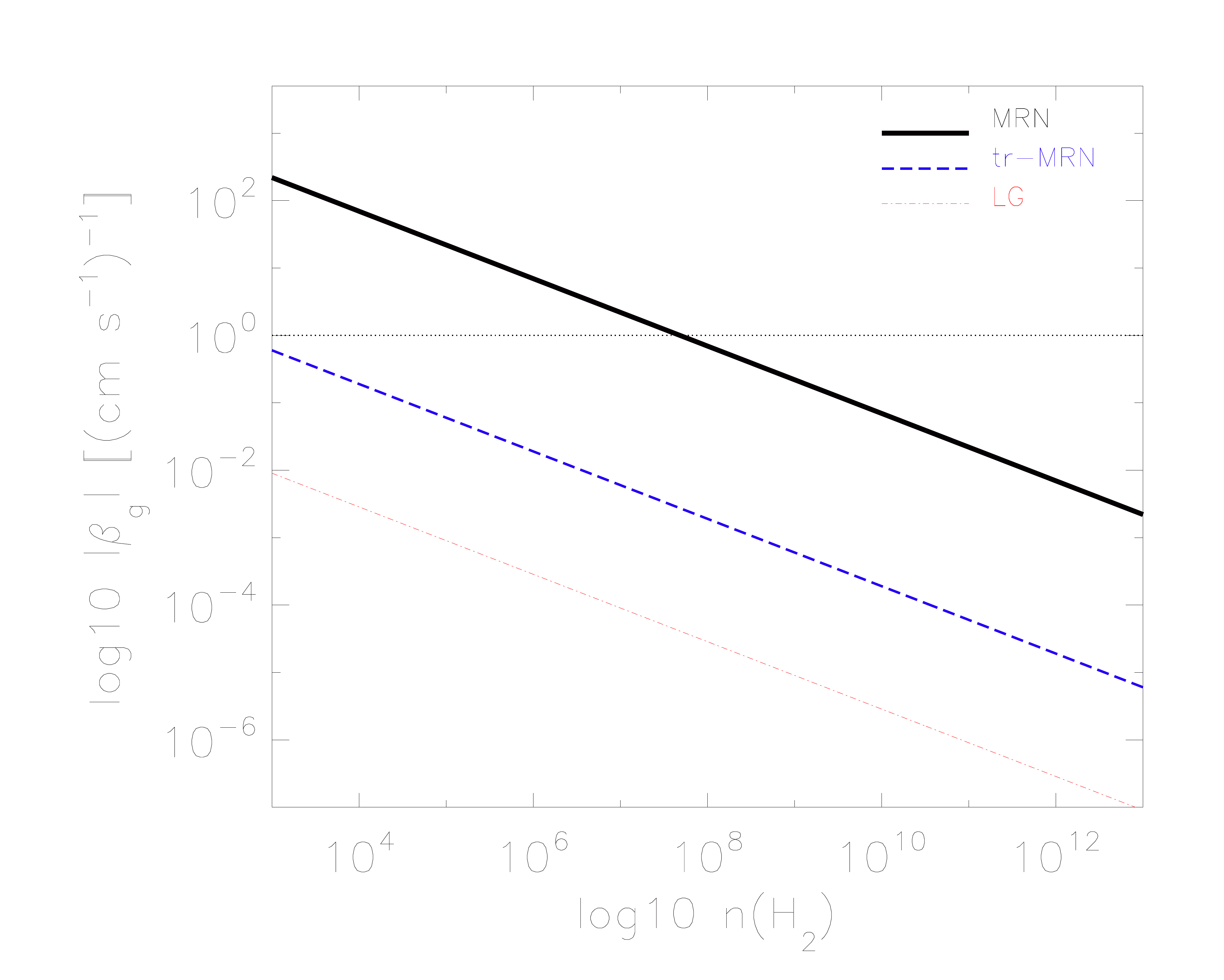}
\caption{Hall parameter |$\beta_{\rm g^-,H_2}$| for different cases of grain
size, the values in the MRN and tr-MRN cases are averaged over the size 
distribution. Horizontal dotted line indicates Hall parameter of unity.}
\label{Fig:betagr}
\end{figure}

In general, each charged species is coupled to the magnetic field 
by the Lorentz force to certain extent; the degree of decoupling is 
determined by the drag force exerted through collision with neutrals. 
The relative importance of the Lorentz force versus drag force 
is quantified through the Hall parameter $\beta_{i,\rm H_2}$. 
Ions and electrons couple more strongly to the 
magnetic field owing to their light weight and small collisional 
cross-section. For charged grains, the coupling to the magnetic field 
depends on the grain size; smaller size grains are better-coupled 
than larger ones. Although the grain-field coupling is generally 
weaker than the ion-(electron-)field coupling, however, grains exert much 
stronger drag to their surrounding neutral molecules than 
ions and electrons do \citep{Pinto+2008}. 
Nevertheless, the grain's Hall parameter |$\beta_{\rm g^-,H_2}$| is a 
good indicator of the two competing effects, which is plotted in 
Fig.~\ref{Fig:betagr} for the three grain size cases MRN, tr-MRN, and LG 
(averaged over the size distribution).

\begin{figure}
\includegraphics[width=\columnwidth]{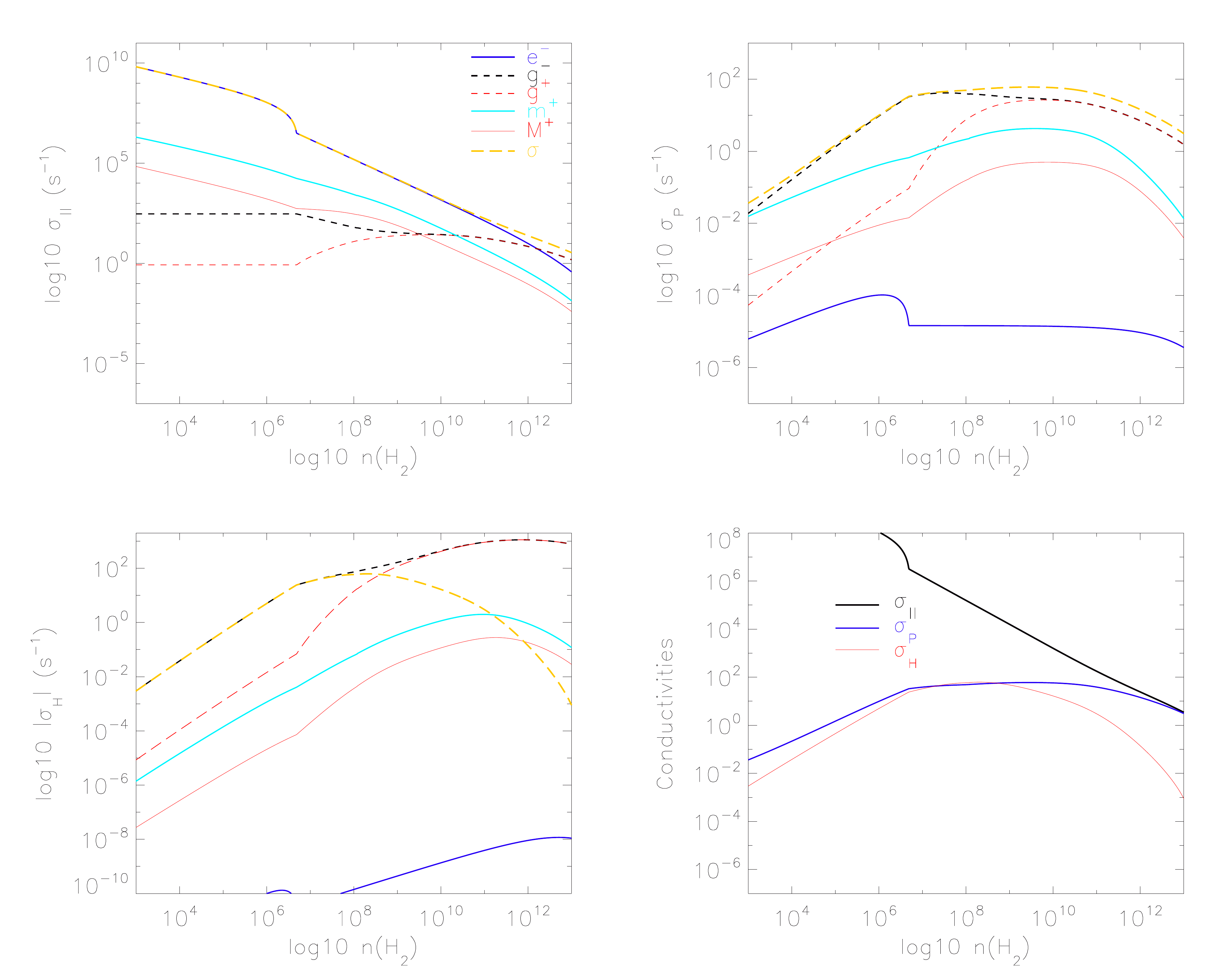}
\caption{Parallel $\sigma_{\parallel}$ (top-left), 
Pedersen $\sigma_{\rm P}$ (top-right), 
and Hall $\sigma_{\rm H}$ (bottom-left) components of the conductivity tensor, 
and contributions made by different species to each conductivity component 
for the MRN grain size distribution. Bottom-right panel plots the three 
conductivity components together, which are shown as $\sigma$ 
({\it yellow dashed} lines) in the other three panels respectively.
The computation adopts the illustrative magnetic field-density relation 
given in Eq.~\ref{Eq:Brelation}.}
\label{Fig:conds50}
\end{figure}

\begin{figure}
\includegraphics[width=\columnwidth]{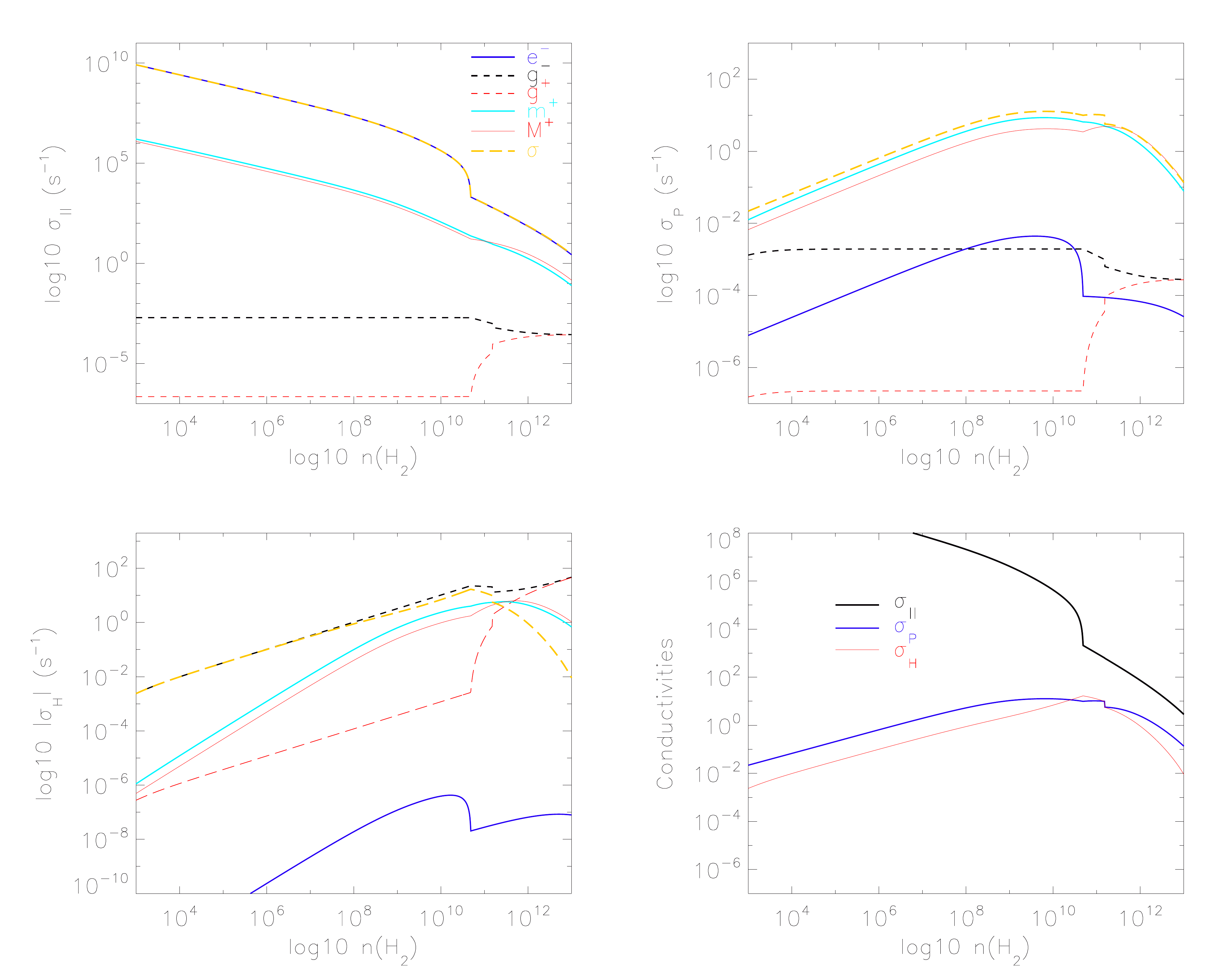}
\caption{As for Fig.~\ref{Fig:conds50}, but for tr-MRN grain size 
distribution.}
\label{Fig:conds1k}
\end{figure}

\begin{figure}
\includegraphics[width=\columnwidth]{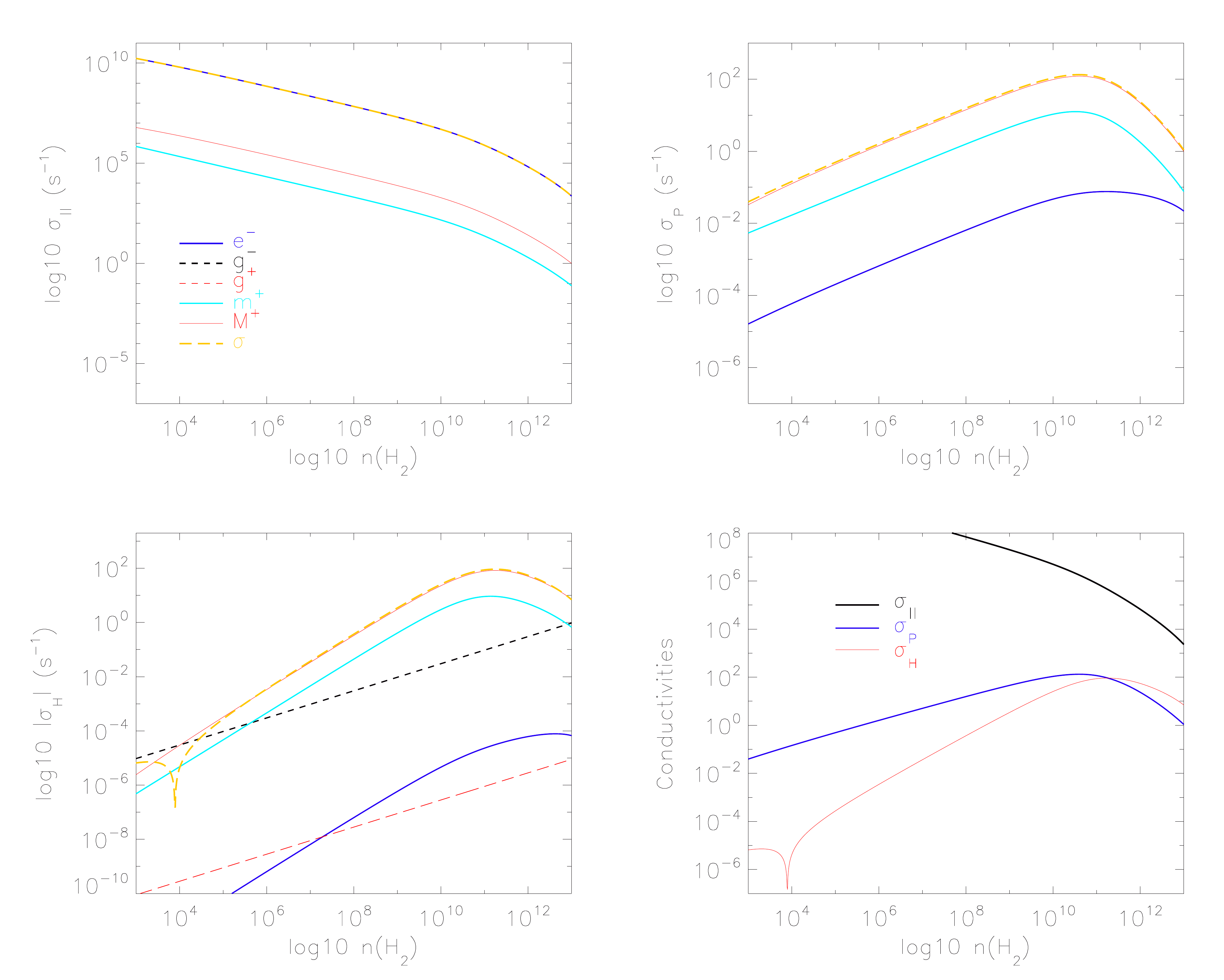}
\caption{As for Fig.~\ref{Fig:conds50}, but for LG of single size 1~$\mu$m.}
\label{Fig:conds1um}
\end{figure}

The comparison of |$\beta_{\rm g^-,H_2}$| for different grain sizes 
clearly shows the $\propto a^{-2}$ dependence due to the inverse of 
momentum transfer rate <$\sigma v$>$^{-1}$. The overall magnitude of 
|$\beta_{\rm g^-,H_2}$| in the MRN case is $\sim$20$^2$ times of that in 
tr-MRN case; and the value in tr-MRN case is $\sim$10$^2$ times of that in 
LG case. 
In the MRN case, |$\beta_{\rm g^-,H_2}$| is well above $1$ at low 
densities ($n({\rm H}_2)\lesssim$10$^7$~cm$^{-3}$), 
implying a strong coupling to the magnetic field from the large 
population of small grains. The coupling only weakens at high 
densities after |$\beta_{\rm g^-,H_2}$| drops below unity. 
For the tr-MRN and LG cases, however, the Hall parameter is well 
below $1$ for all densities, suggesting that the drag force exerted 
on grains by grain-neutral collision always dominates the Lorentz force. 
The large difference in the Hall parameter of charged grains 
strongly affects the conductivities, especially the Pedersen conductivity 
$\sigma_{\rm P}$ (Fig.~\ref{Fig:conds50}-\ref{Fig:conds1um});
it is one of the key origins for the AD enhancement (\S~\ref{S.EnhancAD}).

Comparing Fig.~\ref{Fig:conds50} and Fig.~\ref{Fig:conds1k}, 
the most notable difference is the g$^-$'s contribution 
to the Pedersen conductivity $\sigma_{\rm P}$. \\
$\bullet$ In the MRN case, the $\sigma_{\rm P}$ is entirely determined by 
the component from negatively charged grains $\sigma_{\rm P}(g^-)$,
whose value is over 1--2 orders of magnitude higher than that of 
molecular ions $\sigma_{\rm P}(m^+)$, and about 3 orders of 
magnitude higher than that of metal ions $\sigma_{\rm P}(M^+)$; 
because the grains, even of small sizes in the MRN case, have 
significantly smaller Hall parameter than typical ions or electrons. 
The collisional drag between MRN grains and neutrals is generally stronger 
than that between ions (electrons) and neutrals. Or mathematically, 
$\sigma_{\rm P} \propto {x_i \beta_{i,\rm H_2} \over (1+\beta_{i,\rm H_2}^2)}$,
and $\propto {x_i \over \beta_{i, \rm H_2}}$ for $\beta_{i, \rm H_2} \gg 1$
(note that the abundances $x_i$ of ions and grains do not differ much 
in the intermediate densities $10^6$-$10^{10}$~cm$^{-3}$ 
in Fig.\ref{Fig:abund}). \\
$\bullet$ In the tr-MRN case, in contrast, $\sigma_{\rm P}(g^-)$ 
is about 10$^4$ times smaller than that in the MRN case, 
and the total Pedersen conductivity $\sigma_{\rm P}$ is dominated by 
molecular and metal ions instead ($\sim$1--2 orders of magnitude lower 
than the MRN value). This is mainly for two reasons. 
First, the Hall parameter $\beta_{\rm g^-,H_2}$ decreases 
from $\gg 1$ in the MRN case to <$1$ in the tr-MRN case 
(Fig.~\ref{Fig:betagr}), which completely changes $\sigma_{\rm P}(g^-)$'s 
dependence on Hall parameter into 
$\sigma_{\rm P} \propto x(g^-) \beta_{\rm g^-,H_2}$. 
At density $n({\rm H}_2) \sim$10$^7$~cm$^{-3}$ for example, 
$\propto {1 \over \beta_{\rm g^-,H_2}}$ in the MRN case 
gives a value of a few $10^{-1}$, and $\propto \beta_{\rm g^-,H_2}$ in the 
tr-MRN case gives a value of $\sim$10$^{-2}$; 
the difference is a factor of a few 10$^1$. 
Second, the abundance of negatively charged grains $x(g^-)$ reduces by 
$\approx$500 times (Fig.~\ref{Fig:abund}) when $a_{\rm min}$ changes 
from $0.005~\mu$m (MRN) to $0.1~\mu$m (tr-MRN). 
Therefore, both a difference in $\beta_{\rm g^-,H_2}$ (by order of $10^1$)
and a difference in grain abundance $x(g^-)$ (by order of $10^2$) 
together lead to the large difference in the grains' Pedersen conductivity 
$\sigma_{\rm P}(g^-)$. 
Note that in the LG case, $\sigma_{\rm P}(g^-)$ drops below 
$10^{-7}$ that it is not shown in Fig.~\ref{Fig:conds1um}.

Due to the smaller Hall parameter of grains than that of ions and electrons, 
the Hall conductivity |$\sigma_{\rm H}$| is normally dominated by the 
$g^-$ component |$\sigma_{\rm H}(g^-)$| as long as grain size 
is below $\lesssim$0.5~$\mu$m (large size grains become much less abundant) 
and above $\gtrsim$10~$\AA$ (PAH-type grains have Hall parameter 
comparable to that of ions). 
Comparing Fig.~\ref{Fig:conds50} and Fig.~\ref{Fig:conds1k},
|$\sigma_{\rm H}(g^-)$| is larger in the MRN case at densities below 
$\sim$10$^{10}$~cm$^{-3}$, and is larger in the tr-MRN case 
at higher densities. 
As shown in Eq.~\ref{Eq:conduct3}, the Hall conductivity 
of negatively charged grains depends on their 
abundance $x(g^-)$ and Hall parameter $\beta_{\rm g^-,H_2}$ as 
|$\sigma_{\rm H}(g^-)$| $\propto {x(g^-) \over 1+\beta_{\rm g^-,H_2}^2}$.
When $n({\rm H}_2)\lesssim$10$^6$~cm${^3}$, 
the abundance $x(g^-)$ differs by $\approx$500 times between 
the two cases; however, the effect is partly alleviated by the inverse
dependence on $1+\beta_{\rm g^-,H_2}^2$, because |$\beta_{\rm g^-,H_2}$| 
$\gg 1$ for MRN case but $\ll 1$ for tr-MRN case.
For higher densities, |$\beta_{\rm g^-,H_2}$| drops below unity for both cases 
and |$\sigma_{\rm H}(g^-)$| becomes roughly proportional to $x(g^-)$.
The difference in |$\sigma_{\rm H}(g^-)$| between the two cases peaks
around number density $10^7$~cm${^3}$, with the MRN case 
|$\sigma_{\rm H}(g^-)$| being $\sim$100 
times larger than that of the tr-MRN case. 
Between $10^7$ and $10^{10}$~cm$^{-3}$, such difference dwindles 
because the abundance $x(g^-)$ in the MRN case starts to decline 
due to efficient grain-grain neutralization, while not for the tr-MRN case yet.
At even higher densities $\gtrsim$10$^{10}$~cm$^{-3}$, 
the Hall conductivity in the tr-MRN case overtakes that 
in the MRN case, because the canceling effect from positively-charged grains 
($\sigma_{\rm H}(g^+) \propto x(g^+)$) is weaker in the former. 
Note that in the LG case, $\sigma_{\rm H}$ is mostly determined by 
metal ions, which changes its sign to positive.

The parallel component of the conductivity tensor --- $\sigma_{\parallel}$ 
--- is almost always determined by electrons in all three cases. 
Recall that $\sigma_{\parallel} \propto x_i \beta_{i, \rm H_2}$. 
Electron is so light-weighted that its Hall parameter 
$\beta_{e^-,\rm H_2}$ is more than 3 orders of magnitude larger than 
that of ions, and more than $10^5$ times than MRN grains 
($10^7$ times than tr-MRN grains). 
Therefore, electron generally contributes the most to parallel conductivity, 
as long as its abundance $x(e^-)$ is not significantly lower 
than the other charged species.
Exception only occurs at very high densities ($>10^{12}$~cm$^{-3}$) 
in the MRN case, where the abundance of electrons $x(e^-)$ 
is $10^6$-$10^7$ times lower than the abundances of charged grains; 
so that both $g^-$ and $g^+$ start to dominate $\sigma_{\parallel}$ 
near the high density tail (see Fig.~\ref{Fig:conds50}).

We conclude this section by comparing the three 
components of the conductivity tensor: 
$\sigma_{\parallel}$, $\sigma_{\rm P}$, and $\sigma_{\rm H}$.
By definition of Eq.~\ref{Eq:MHDcoef1}--\ref{Eq:MHDcoef3}, 
the strength of each type of magnetic diffusivity is determined 
by the absolute magnitude as well as the relative importance of 
the three conductivities, summarized as below \citep[see also][]{WardleNg1999}:
\begin{enumerate}
\item strong $\eta_{\rm AD}$ requires: small $\sigma_{\rm P}$ while 
  $\sigma_{\parallel} \gg \sigma_{\rm P} \gg$ |$\sigma_{\rm H}$|.
\item strong $\eta_{\rm Ohm}$ requires: small $\sigma_{\parallel}$;
  (note that, for Ohmic dissipation to dominate over the other two effects, we 
  need $\sigma_{\parallel}$ slightly > $\sigma_{\rm P} \gg$ |$\sigma_{\rm H}$|.
  )
\item strong $\eta_{\rm Hall}$ requires: small |$\sigma_{\rm H}$| while 
  |$\sigma_{\rm H}$| $\gg \sigma_{\rm P}$; (note that, for Hall effect to 
  dominate, we need $\sigma_{\parallel} \gg$ |$\sigma_{\rm H}$|.)
\end{enumerate}

The AD enhancement seen in the tr-MRN case exactly matches the criteria in 
(i). As discussed above, the small Pedersen conductivity $\sigma_{\rm P}$ 
owes to the removal of a large amount of conductive VSGs, so that 
both the abundance and Hall parameter of grains become low. 
Hence only ions (m$^+$ and M$^+$) dominate $\sigma_{\rm P}$. 
The relation $\sigma_{\parallel} \gg \sigma_{\rm P} \gg$ |$\sigma_{\rm H}$|
is also satisfied for most densities 
because |$\sigma_{\rm H}$| --- primarily controlled by grains --- plunged
with the low grain abundance, 
and $\sigma_{\parallel}$ --- controlled by electrons --- 
are always orders of magnitude larger than the other two conductivities.

In the MRN case, the conductivities satisfy the criteria in (ii) 
at high densities ($\gtrsim$few $10^{11}$~cm$^{-3}$); however, 
at lower densities, although AD dominates the magnetic diffusion, 
the strength of $\eta_{\rm AD}$ is weaker by $\sim$1--2 orders of magnitude 
than that of the tr-MRN case (Fig.~\ref{Fig:MHDcoefs}), 
owing to the large Pedersen and Hall conductivities here.
At densities below $\sim$10$^7$~cm$^{-3}$, 
the Pedersen conductivity $\sigma_{\rm P}$ is larger (only slightly) 
than the Hall conductivity |$\sigma_{\rm H}$|; 
but the latter increases with density faster than the former, 
which undermines $\eta_{\rm AD}$ (Eq.~\ref{Eq:MHDcoef1}). 
Between $\sim$10$^7$ and $\sim$10$^9$~cm$^{-3}$,
$\sigma_{\rm P}\approx$|$\sigma_{\rm H}$|, 
so that the $\eta_{\rm AD}$ curve reaches a minimum.
From $\sim$10$^9$ to $\sim$10$^{11}$~cm$^{-3}$, $\eta_{\rm AD}$ starts to 
increase slightly because $\sigma_{\rm H}$ decreases rapidly 
as a result of efficient grain-grain neutralization, 
while $\sigma_{\rm P}$ does not change much.
At high densities $n({\rm H}_2)\gtrsim$10$^{11}$~cm$^{-3}$, 
Ohmic dissipation becomes the dominant diffusion process.
It is mainly caused by the large $\sigma_{\rm P}$ that nearly equals
$\sigma_{\parallel}$, hence the criteria in (ii) are satisfied.

In the LG case, the conductivities satisfy criteria in (i) at low densities 
($\lesssim$10$^{11}$~cm$^{-3}$) and criteria in (iii) at high densities 
($\gtrsim$10$^{11}$~cm$^{-3}$). However, the overall strength of 
$\eta_{\rm AD}$ is $\sim$few times smaller than that in the tr-MRN case, 
yet still $\sim$few $10^1$ times bigger than the MRN case in the low density 
regimes. Here, both Pedersen and Hall conductivities are controlled mostly 
by metal ions. The Hall conductivity $\sigma_{\rm H}$ changes sign at 
density $\sim$10$^4$~cm$^{-3}$ from $g^-$ dominated regime into 
$M^+$ dominated regime. It increases faster with density than 
Pedersen conductivity $\sigma_{\rm P}$ due to 
the $\propto {1 \over \beta_{i,\rm H_2}^2}$ dependence 
compared with the $\propto {1 \over \beta_{i,\rm H_2}}$ dependence 
for $\sigma_{\rm P}$ ($\beta_{M^+,\rm H_2} \gg 1$ and is 
decreasing with density). 
When density $n({\rm H}_2)\gtrsim$10$^{11}$~cm$^{-3}$, 
$\sigma_{\rm H}$ overtakes $\sigma_{\rm P}$, therefore Hall diffusivity 
$\eta_{\rm Hall}$ becomes the strongest among the three magnetic diffusivities.

\section{Simulation Results}
\label{Chap.SimulResult}

With a better understanding of analytical results from the chemistry model, 
we present the simulation results in this section, which largely follow 
the analytical expectations above. In comparison, the numerical simulation 
offers more realistic strength and geometry of the magnetic field than the 
simple relation Eq.~\ref{Eq:Brelation} used in the previous section.
It also allows us to study the interplay of different physical processes, 
and to determine the relative importance of different parameters 
on disk formation.

As summarized in Table~\ref{Tab:model1}-\ref{Tab:model2}, 
a 4-parameter space including
magnetic field strength (in terms of mass-to-flux ratio $\lambda$), 
grain size, cosmic-ray ionization rate, and initial rotation speed
(in terms of $\beta_{\rm rot}$),
is explored with a total of 24 calculations.
We find that each parameter has an impact on the formation of RSDs 
and the lifetime of disks.
Weaker magnetic field strength, lower cosmic-ray ionization rate, or 
faster rotation speed all promote disk formation to certain degree 
(e.g., increase disk size and/or lifetime); 
however, the most effective is by changing the grain size distribution.
Formation of long-lived stable RSDs (even self-gravitating disks and rings) 
is only possible in the tr-MRN cases, or with similar grain size 
distributions that are free of a large population of VSGs.

\subsection{Disk Morphology}
\label{S.DiskMorph}

The whole parameter space produces 3 major types of disks, categorized 
based on their lifetime and size as follows: 
\begin{enumerate}
\item N$^{\rm Trans}$ (or N): a small transient disk 
(not rotationally supported for most of its lifetime) 
of radius $\sim$10~AU forms from ``first core''-like structure 
and disappears quickly within $\sim$few $10^2$ to $10^3$ years 
due to accretion of low angular momentum gas. 
\item Y$_{t_{\rm d}}^{\rm Shrink}$: a RSD of radius $\sim$20~AU 
is able to form from initial collapse; but its size shrinks over time, 
with a lifetime $t_{\rm d} \sim$few $10^3$ to $10^4$ years. 
\item Y: large RSDs of radius $\sim$20--50~AU are able to form and survive 
(at least few $10^4$ years). 
The disk maintains a relatively stable size by accreting 
high angular momentum gas; its mass can grow up to $30\%$--$40\%$ of 
the total core mass.
\end{enumerate}
Type (iii) disk only forms when the grain size distribution is a truncated MRN.
Depending on different combinations with other parameters, such disks in 2D 
can have substructures including 
{\it ID} --- small Keplerian inner disk around the central star, 
and {\it OR} --- massive rotationally supported outer ring at the 
centrifugal barrier, 
which is an evolutionary outcome of a single disk that forms first.

The categorization criteria above are not strict. 
Some type (i) transient disks (N$^{\rm Trans}$) 
in the weaker B-field ($\lambda=4.8$) cases 
do partly become rotationally supported for a small fraction 
of their lifetime. 
Another example is the intermediate type of RSD 
formed in the $\lambda=4.8$ Fst-trMRN5 case which, 
despite shrinking in size, can survive a much longer time (>$16$~kyr); 
and the disk radius holds steady around $16$~AU at the end of 
the simulation. It is possible that, as more parameter space 
is explored, the category boundary can become less sharp.

\subsection{AD-Enabled Disk Formation: from MRN to tr-MRN}
\label{S.AD-RSDs}

The formation of long-lived RSDs in the tr-MRN runs is essentially caused by 
the enhanced AD as shown in the analytical result 
\S~\ref{S.EnhancAD}. 
The strong AD weakens the coupling of neutral gas to magnetic field 
and reduces the amount of magnetic flux being dragged in by the 
collapsing flow. 
Therefore, magnetic braking is weakened, especially in the 
circumstellar region, and sufficient angular momentum can be preserved for 
a rotationally supported disk. 
To illustrate this, we compare the reference model Slw-MRN1$^{\rm R}$ 
(Fig.~\ref{Fig:2.4Slw-MRN1}) with model Slw-trMRN1 
(Fig.~\ref{Fig:2.4Slw-trMRN1}), both for the strong initial B-field 
$\lambda=2.4$.

\subsubsection{MRN Grain Reference Model: $\lambda2.4$ Slw-MRN1$^{\rm R}$}
\label{S.ModelMRN}

The reference model Slw-MRN1$^{\rm R}$ is representative of the type (i) 
transient disks in this study; it is also considered as a standard case 
in other disk formation studies
\citep[e.g.,][]{Li+2011, Tomida+2013, Masson+2015}.
Although a RSD is frequently claimed to form from Larson's first core
\citep[e.g.,][]{MachidaMatsumoto2011, Tomida+2013, Tsukamoto+2015a}, 
these studies do not follow the disk evolution very long.
Hence, there is no direct evidence that the small disk formed from 
the first core can survive or grow in size by accreting infalling 
gas with enough angular momentum.

\begin{figure*}
\includegraphics[width=\textwidth]{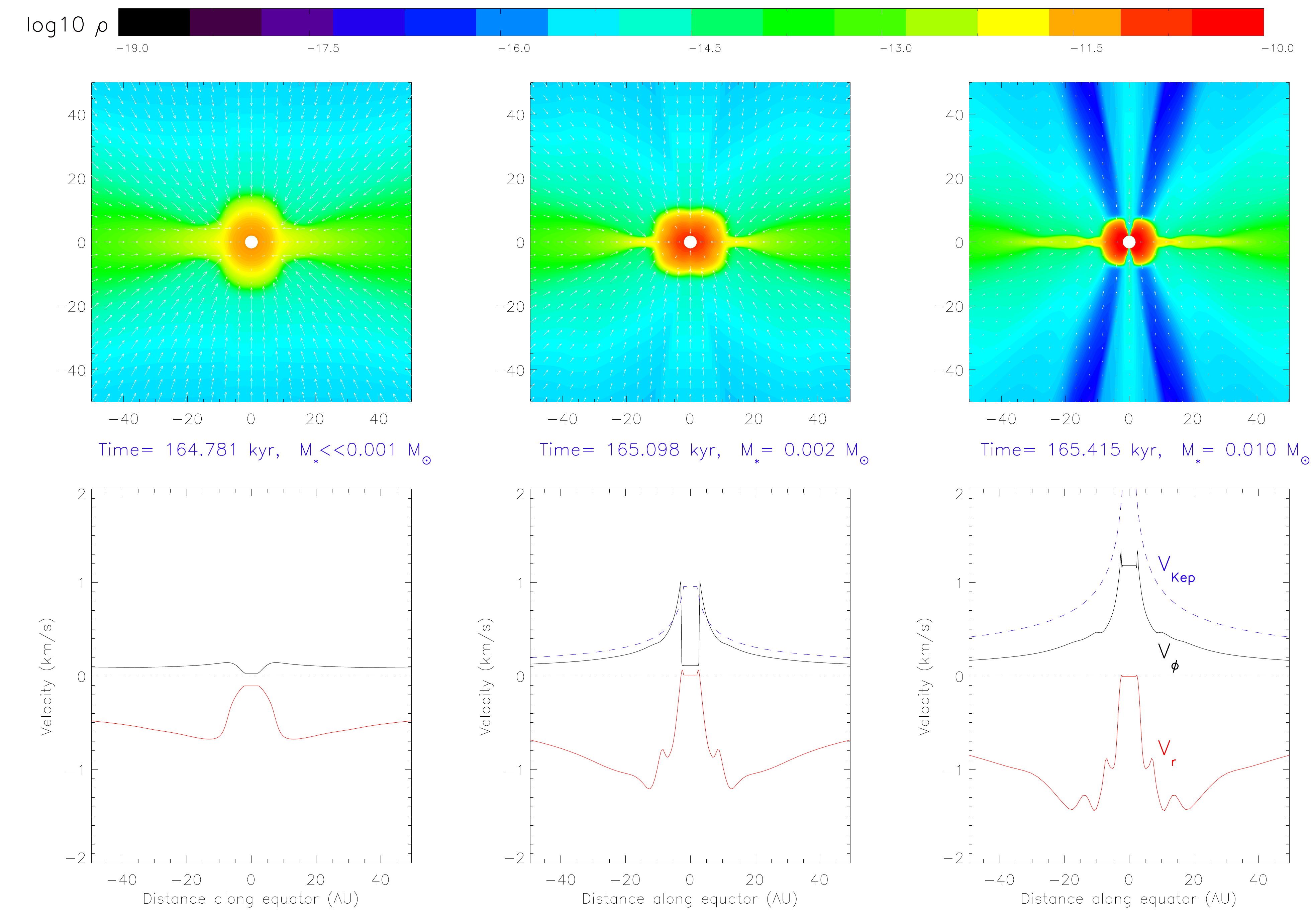}
\caption{Early evolution of density distribution (color map) and 
velocity field (white arrows) for the $\lambda2.4$ Slw-MRN1$^{\rm R}$ 
reference model inside 50~AU radius. The corresponding velocity profiles 
along the equator are plotted in the bottom panels. 
Note that the inner hole has a radius of 2~AU.}
\label{Fig:earlyMRN1}
\end{figure*}
\begin{figure*}
\includegraphics[width=\textwidth]{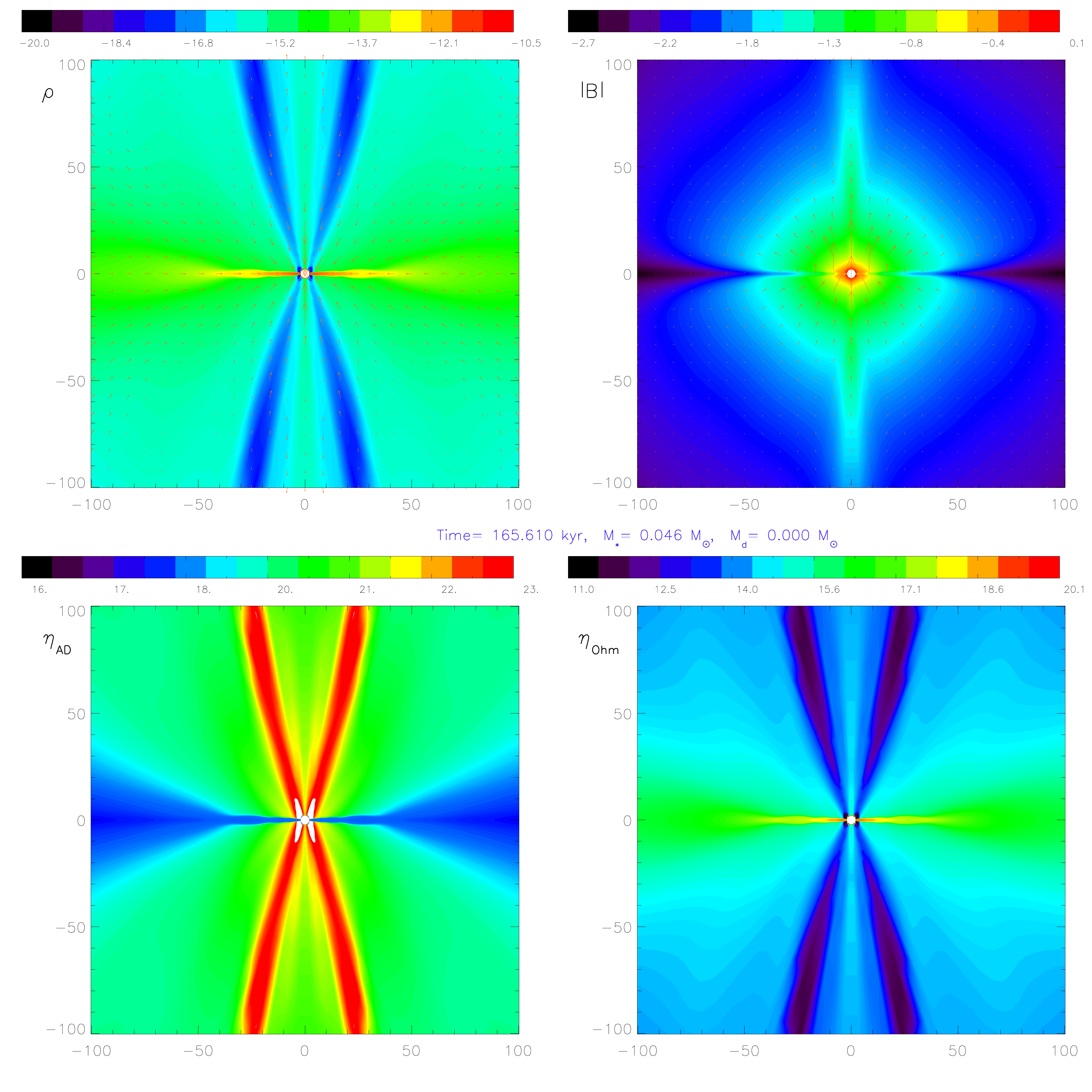}
\caption{Distributions of density $\rho$, magnetic field strength |$B$|, 
ambipolar diffusivity $\eta_{\rm AD}$ and Ohmic diffusivity $\eta_{\rm Ohm}$,
all in logarithmic scale, for the $\lambda=2.4$ Slw-MRN1$^{\rm R}$ 
reference model at a later time $t \approx 165.6$~kyr.  
The poloidal velocity field (top-left) and magnetic field (top-right) 
are shown as orange arrows. Length unit of the axes is in AU.}
\label{Fig:2.4Slw-MRN1}
\end{figure*}

As shown in the middle panel of Fig.~\ref{Fig:earlyMRN1}, 
a spherical structure ($\lesssim$10~AU) similar to the Larson's first core 
do appear at $t \approx 165.1$~kyr ($\sim$1.7 $t_{\rm ff}$, middle panel), 
with a $0.002$~M$_{\sun}$ ($\sim$2 Jupiter mass) stellar object in its center.
Because we do not treat the full radiative transfer in our study, 
formation of this structure owes to the stiffening of the equation of state, 
i.e., transitioning from isothermal to adiabatic regime. 
In the central part of the structure, the temperature reaches a few 10$^2$~K. 
The inner $\sim$8~AU is rotating temporarily faster than Keplerian speed as 
the star mass is still small. On the other hand, the gas at radius beyond 8~AU 
all rotates with sub-Keplerian speed, indicating a low specific 
angular momentum in circumstellar regions (see Fig.~\ref{Fig:2.4SlwAMRcent}).
As these materials fall towards the center, the ``first core'' structure 
quickly becomes non-rotationally supported in less than $300$~years 
at $t \approx 165.4$~kyr. The stellar mass grows to $0.01$~M$_{\sun}$, 
and the density rarefaction caused by infall carves out the bipolar 
regions, which is shaping the core into a more disk-like structure. 
Up to this time, the inner $\sim$10~AU is dominated by the thermal pressure 
$P_{\rm th}$, where the plasma $\beta$ ($\equiv {P_{\rm th} \over P_{\rm B}}$) 
can reach a few 10$^2$ to 10$^3$.
Unfortunately, this ``first core''-like structure lacks further 
angular momentum supply, which quickly leads to its disappearance in 
less than $200$ years. At $t\approx 165.6$~kyr (Fig.~\ref{Fig:2.4Slw-MRN1}), 
the whole structure has been accreted by the central star, leaving only a 
magnetically-dominated pseudo-disk around a $0.046$~M$_{\sun}$ star.

In Fig.~\ref{Fig:2.4Slw-MRN1}, the magnetic field lines along the equator 
are strongly pinched towards the central star, with field strength up to
$\sim$1~G; however, diffusion of magnetic field is inefficient. 
The Ohmic diffusivity $\eta_{\rm Ohm}$ only becomes large 
($\gtrsim$10$^{18}$~cm$^2$~s$^{-1}$) near the surface of the $2$~AU 
inner hole and in the central thin layer of the pseudo-disk, 
where densities are above few 10$^{-12}$~g~cm$^{-3}$. 
The ambipolar diffusivity $\eta_{\rm AD}$ is overall low along the equator,
implying an inefficient AD. Therefore in this case, decoupling of matter 
from magnetic field occurs only near the inner hole through Ohmic dissipation, 
which allows the accretion flow to land onto the central star.
Note that the high $\eta_{\rm AD}$ in the outflow cavity is because 
of the low density in these regions.

\begin{figure*}
\begin{tabular}{ll}
\includegraphics[width=\columnwidth]{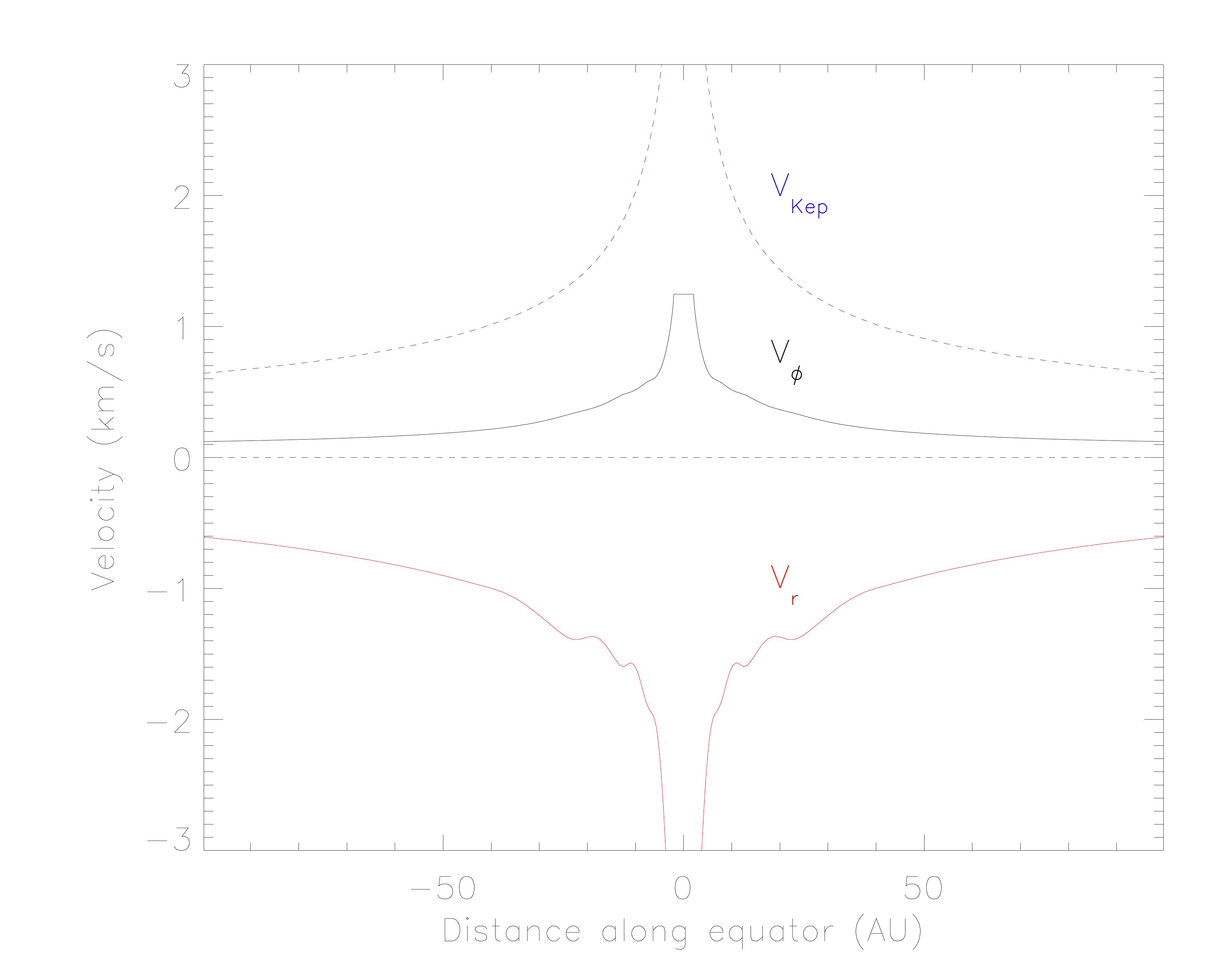}
\includegraphics[width=\columnwidth]{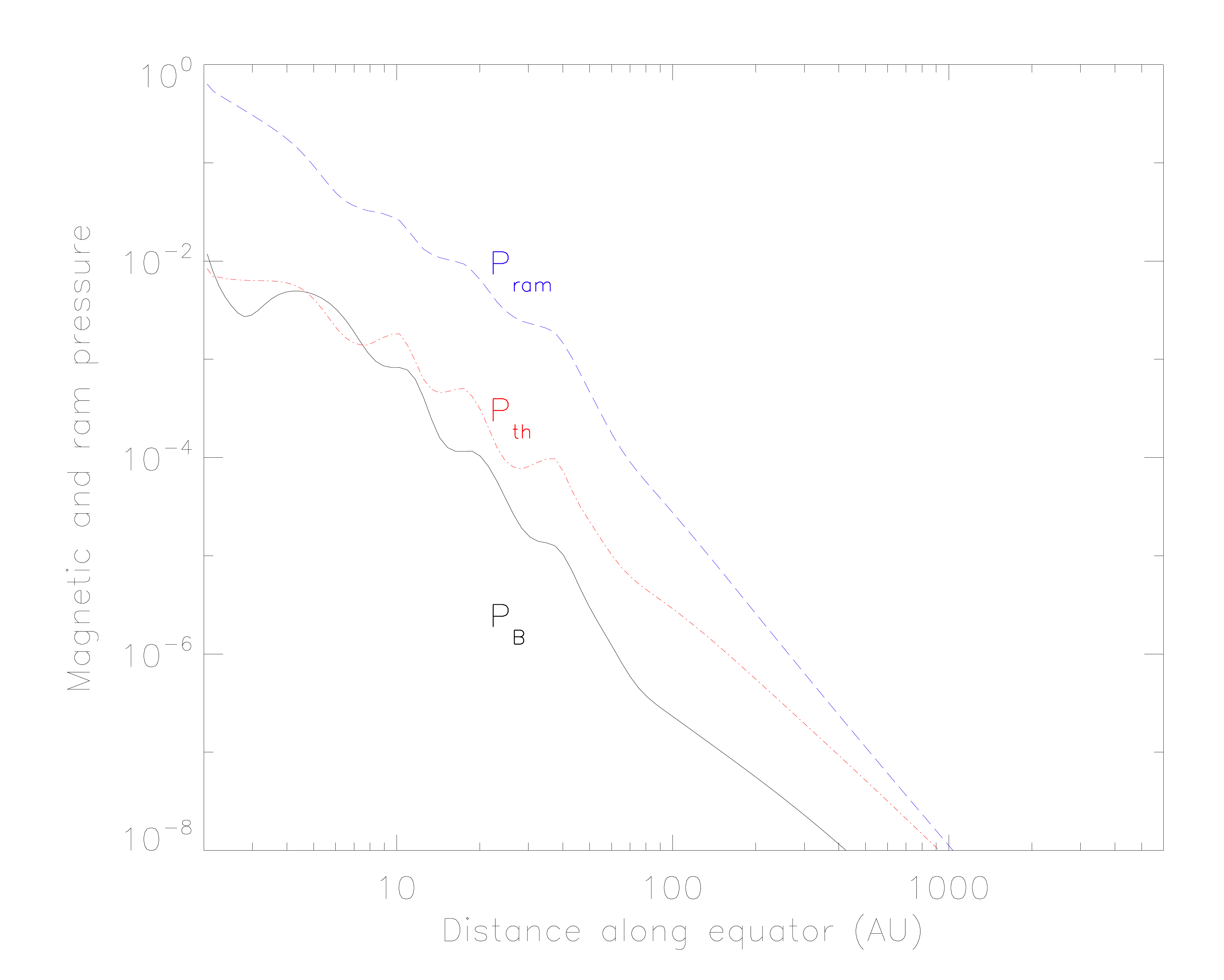}
\end{tabular}
\caption{Left panel: profile of equatorial infall and rotation speed 
in the $\lambda2.4$ Slw-MRN1$^{\rm R}$ reference model at 
$t \approx 165.6$~kyr (same time frame as Fig.~\ref{Fig:2.4Slw-MRN1}). 
The Keplerian speed is also plotted based on the central mass. 
Right panel: profile of thermal $P_{\rm th}$, magnetic $P_{\rm B}$, 
and ram $P_{\rm ram}$ pressures along the equator.}
\label{Fig:VP2.4SlwMRN1}
\end{figure*}

The non-rotationally-supported nature of the pseudo-disk at $t=165.6$~kyr
is also obvious from the velocity profile in Fig.~\ref{Fig:VP2.4SlwMRN1}.
The gas on the pseudo-disk is infalling supersonically towards 
the central star with increasing ${\rm v_r}$, 
while the rotation speed ${\rm v}_\phi$ is highly sub-Keplerian. 
No obvious bump in the infall velocity due to AD-shock 
\citep{LiMcKee1996, Li+2011} is present in this case, 
because the stellar mass is still quite low and not much magnetic 
flux has been decoupled from the accreted matter. 
The magnetic pressure $P_{\rm B}$ (=|B|$^2$/(8$\pi$)), 
ram pressure $P_{\rm ram}$ (=$\rho {\rm v_r}^2$), 
and thermal pressure $P_{\rm th}$ are all increasing towards the center. 
The thermal pressure no longer dominates the inner $\sim$10~AU, where 
the plasma $\beta$ drops to unity.
At this stage, gas on the pseudo-disk has already lost most of its
angular momentum through magnetic braking; this results in 
the fast infall motion of gas and a rapid increase in stellar mass.

\subsubsection{Truncated-MRN Grain Model: $\lambda2.4$ Slw-trMRN1}
\label{S.ModeltrMRN}

\begin{figure*}
\includegraphics[width=\textwidth]{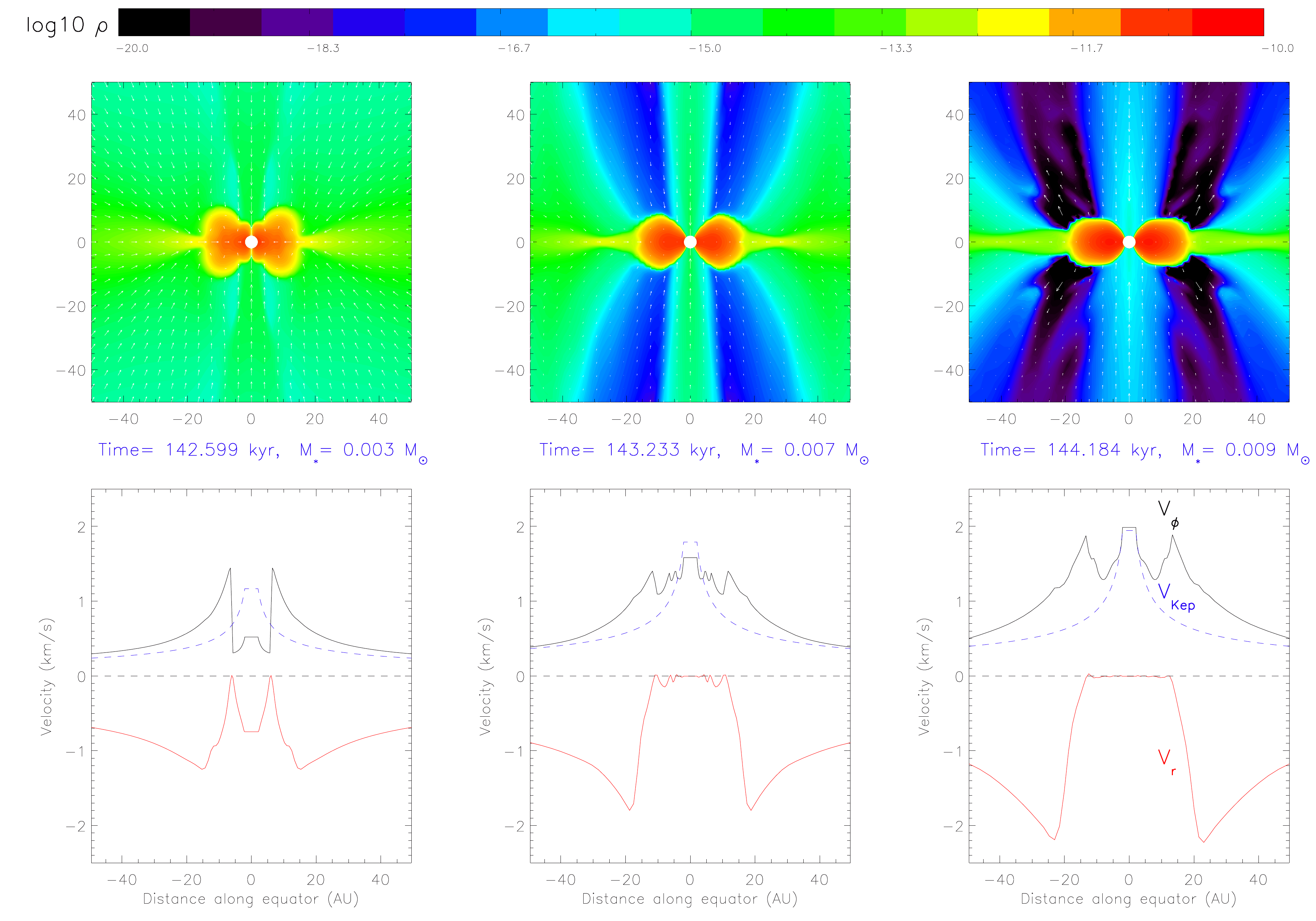}
\caption{Early evolution of density distribution (color map) and 
velocity field (white arrows) for the $\lambda2.4$ Slw-trMRN1 model 
inside 50~AU radius. The corresponding velocity profiles along the equator 
are plotted in the bottom panels.}
\label{Fig:earlytrMRN1}
\end{figure*}

In contrast, the $\lambda2.4$ Slw-trMRN1 model clearly demonstrates 
formation of large sustainable RSDs; it also represents the type (iii) disks 
in this study. As shown in Fig.~\ref{Fig:earlytrMRN1}, the central dense 
structure is persistently super-Keplerian, from the early 
``first core''-like structure to the well defined disk later on. 
The ``first core''-like structure starts as an elongated shape with radius
$\sim$15~AU,  at $t \approx 142.6$~kyr ($\sim$1.5 $t_{\rm ff}$). 
Its innermost $\sim$10~AU is close to the shock front created by the 
centrifugal barrier (see detailed discussion in \S~\ref{S.CentriShock}), 
therefore both infall and rotation motions are suppressed. 
However, at larger radius ($\gtrsim$10~AU) along the equator, 
gas rotates with super-Keplerian speed, contrary to that of the 
reference model. It implies that the infalling gas
still retains a relatively high angular momentum 
(see Fig.~\ref{Fig:2.4SlwAMRcent}). As collapse continues, this 
angular momentum influx is able to spin up the inner $\sim$15--20~AU
and to enable the formation of a RSD in $\sim$600 years 
($t \approx 143.2$~kyr) around a 0.007~M$_{\sun}$ ($\sim$7 Jupiter mass) 
protostar.
Meanwhile, density cavities appear in the bipolar regions as a result of 
rapid infall onto the flattened equatorial region. 
However, magnetically driven outflows only appear shortly afterwards 
(at $\approx 143.6$~kyr, not shown), which continuously carves the cavity 
wall as the disk grows in size. The third panel of Fig.~\ref{Fig:earlytrMRN1} 
captures a representative moment when such magnetically driven outflow is 
pushing out more materials near the cavity wall and widening the cavity angle. 
The launching point of the outflow locates at the disk outer edge 
$\sim$20~AU, where gas flow is being accreted onto the disk and field lines 
are strongly pinched. 
As evolution continues, the RSD in this model survives beyond $14$~kyr 
with mass growing above $0.22$~M$_{\sun}$, and shows no sign of disappearing.
\footnote{We stopped the 2D axisymmetric simulation because of the lack 
of 3D gravitational and MHD instabilities to redistribute angular momentum 
in the disk, which is needed for disk accretion onto the central object. 
As a result, the disk mass is much larger than the stellar mass at the 
end of the simulation.}

\begin{figure*}
\includegraphics[width=\textwidth]{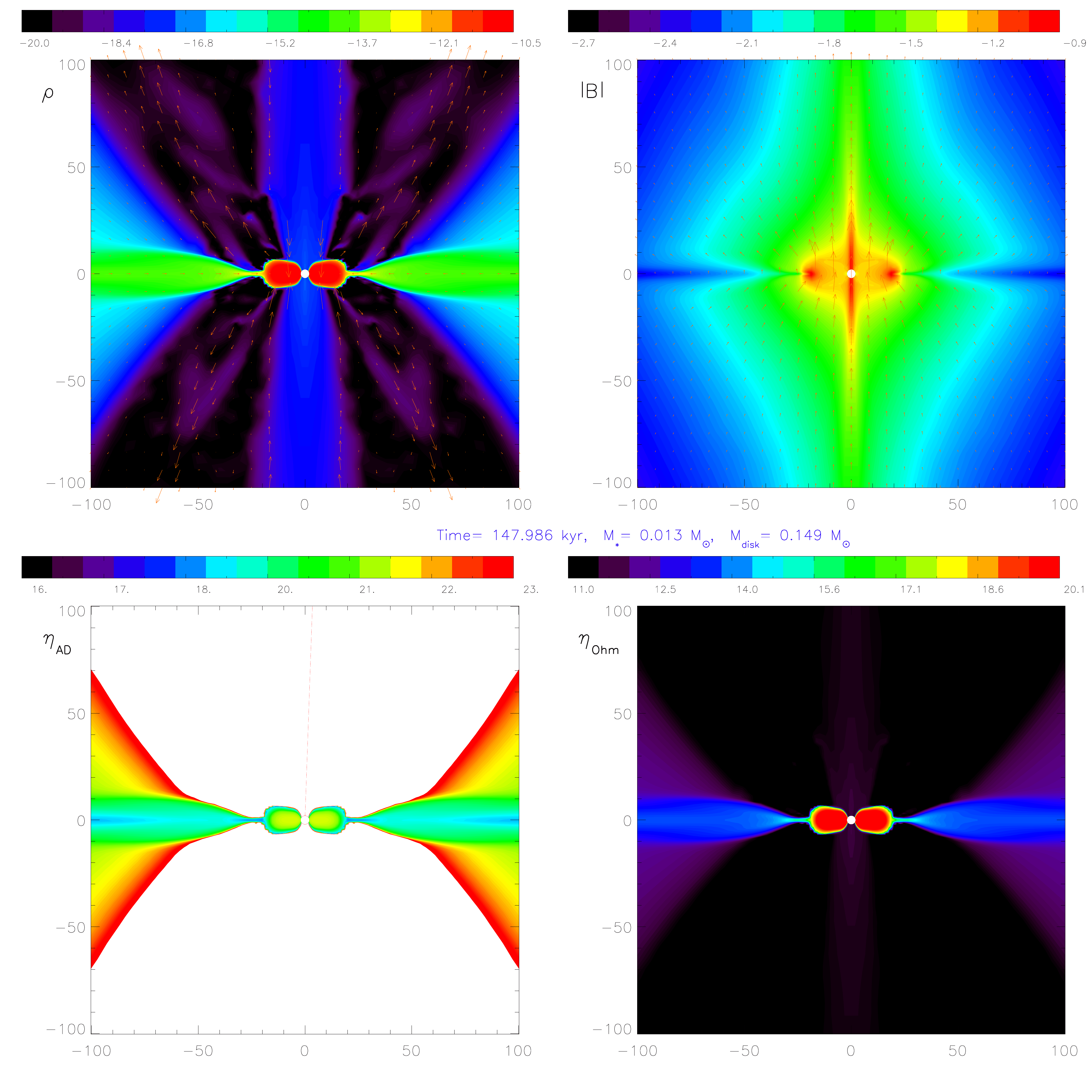}
\caption{Distributions of density $\rho$, magnetic field strength |$B$|, 
ambipolar diffusivity $\eta_{\rm AD}$ and Ohmic diffusivity $\eta_{\rm Ohm}$, 
all in logarithmic scale, for the $\lambda=2.4$ Slw-trMRN1 model 
at a later time $t \approx 148.0$~kyr. 
The poloidal velocity field (top-left) and magnetic field (top-right) 
are shown as orange arrows. The white bipolar regions in the bottom-left 
panel are of high $\eta_{\rm AD}$ above 10$^{23}$~cm$^2$~s$^{-1}$. 
Length unit of the axes is in AU.}
\label{Fig:2.4Slw-trMRN1}
\end{figure*}
\begin{figure*}
\includegraphics[width=\textwidth]{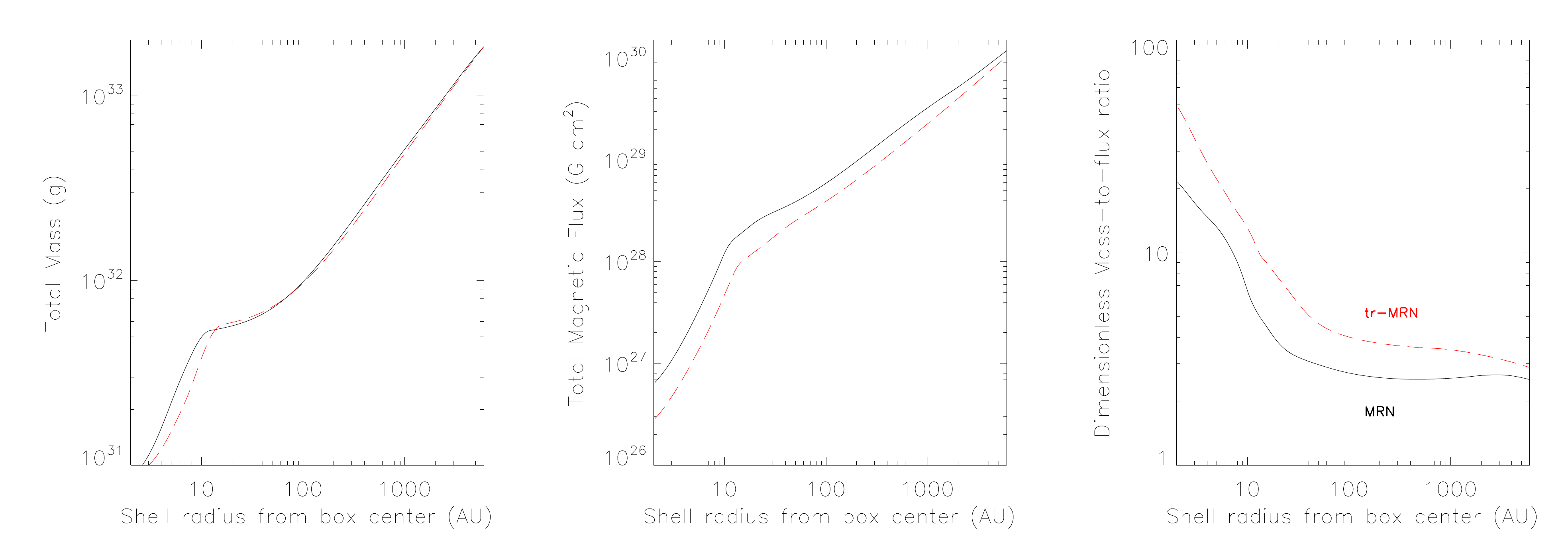}
\caption{Distribution of total mass (left panel), 
total magnetic flux (middle panel) and the corresponding 
mass-to-flux ratio (right panel) inside spheres of different radii for 
the $\lambda2.4$ Slw-MRN1$^{\rm R}$ reference model (black solid) at 
$t \approx 165.1$~kyr and the $\lambda2.4$ Slw-trMRN1 model (red dashed) at 
$t \approx 143.0$~kyr. The time frames are chosen at roughly the 
``first-core'' phase when central structures are relatively simple.}
\label{Fig:2.4Slw1mtf}
\end{figure*}
\begin{figure*}
\begin{tabular}{ll}
\includegraphics[width=\columnwidth]{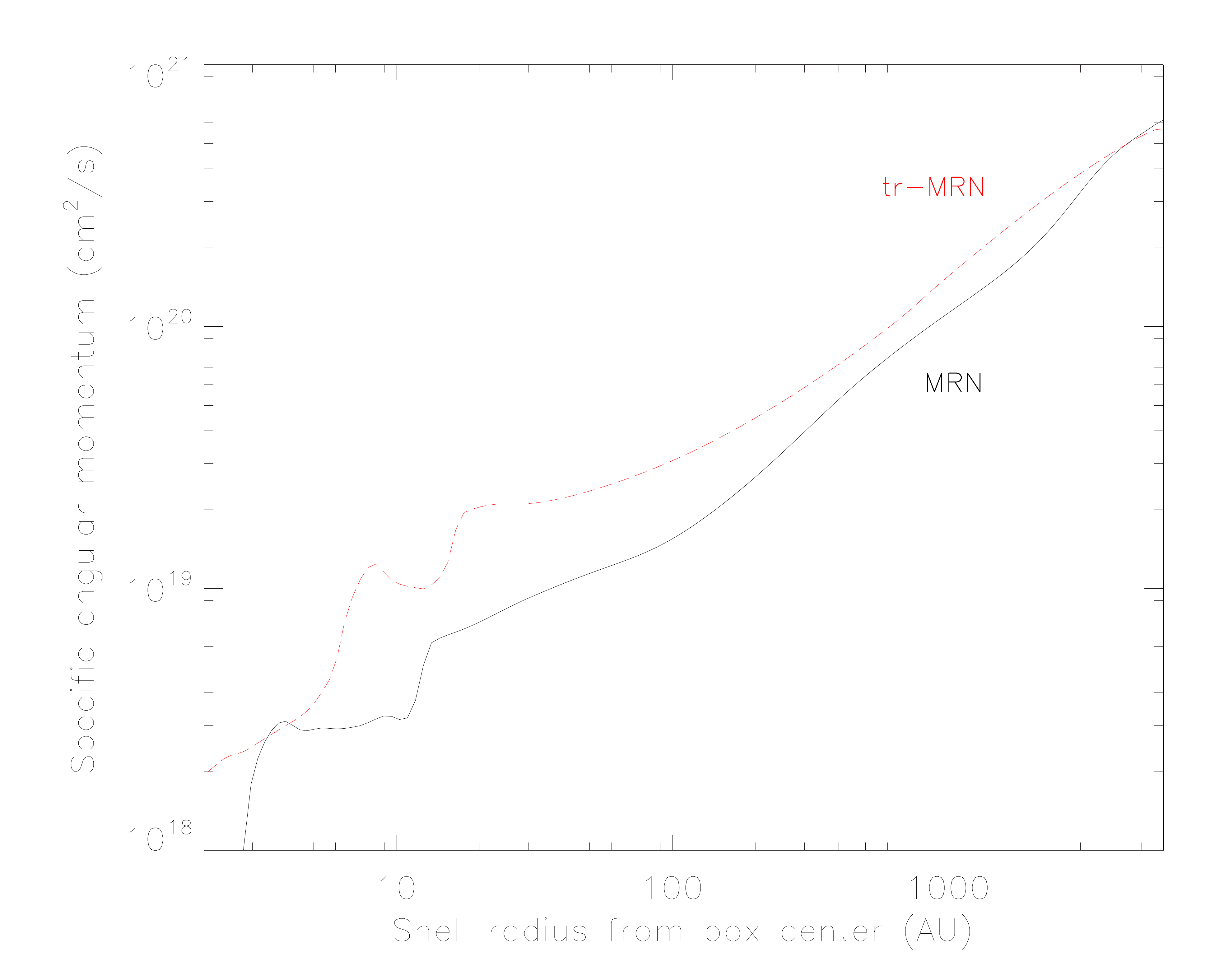}
\includegraphics[width=\columnwidth]{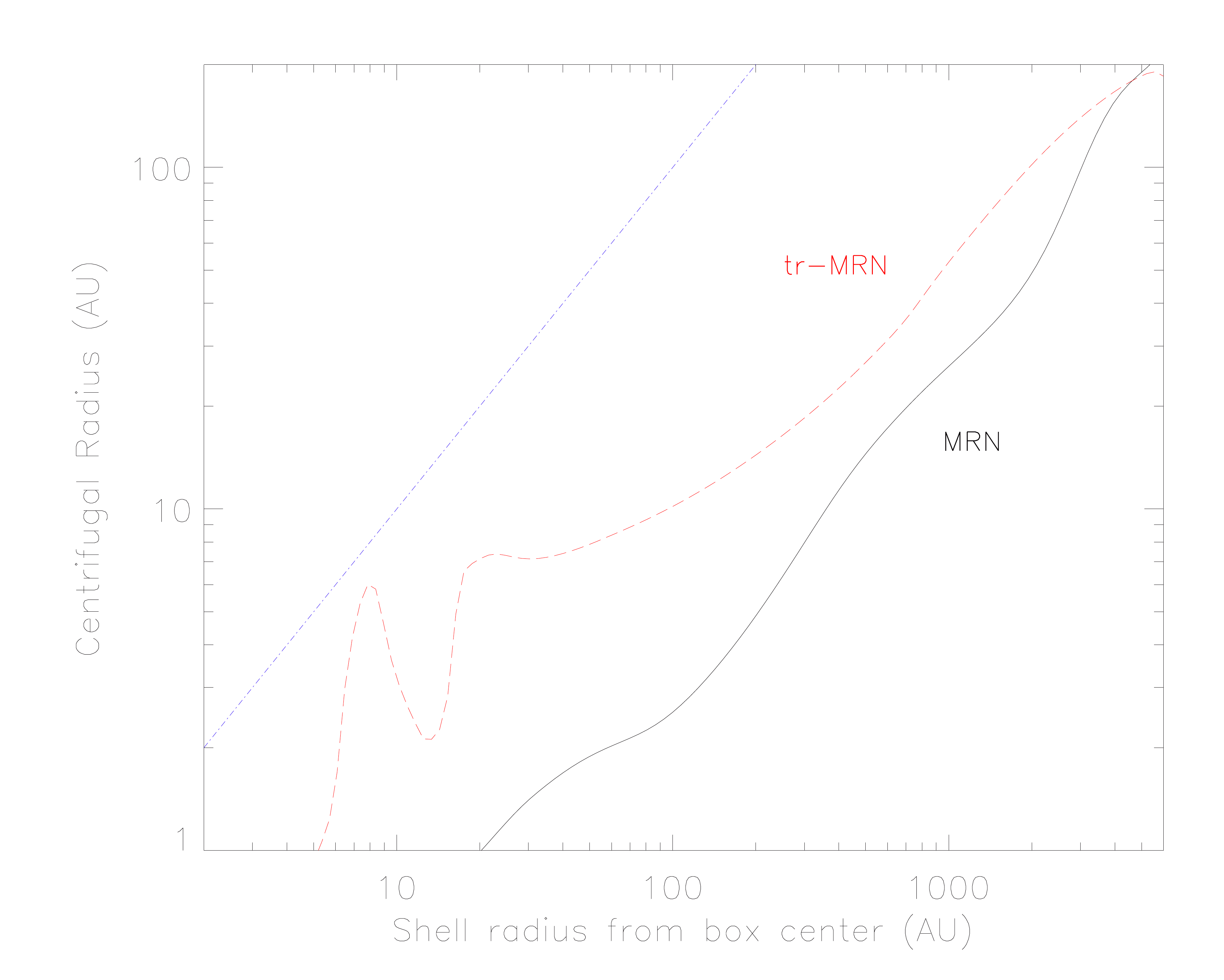}
\end{tabular}
\caption{Left panel: distribution of specific angular momentum inside shells 
at different radii for the $\lambda2.4$ Slw-MRN1$^{\rm R}$ reference model 
at $t \approx 165.1$~kyr (black solid) and the $\lambda2.4$ Slw-trMRN1 model 
at $t \approx 143.0$~kyr (red dashed). 
Right panel: the estimated centrifugal radius from Eq.~\ref{Eq:Rcent} 
for the two models in the left panel. Blue dash-dotted line represents 
a ``break-even'' relation $R_{\rm cent}(r_{\rm sh})=r_{\rm sh}$.}
\label{Fig:2.4SlwAMRcent}
\end{figure*}

The formation and growth of the RSD in this tr-MRN model owes to 
the enhanced AD, which weakens the coupling of neutral gas 
to the magnetic field and allows the gas to collapse or rotate 
without dragging as much magnetic flux as in the MRN reference model.
As shown in Fig.~\ref{Fig:2.4Slw-trMRN1} 
at a typical time $t \approx 148.0$~kyr, the ambipolar diffusivity 
$\eta_{\rm AD}$ along the equatorial region is generally above 
10$^{18}$~cm$^2$~s$^{-1}$, at least a few 10$^1$ times 
higher than that of the MRN reference model; this provides a sufficient 
gas-field decoupling for the infalling envelope material 
\citep{Krasnopolsky+2010}.
In the disk, $\eta_{\rm AD}$ reaches $\gtrsim$10$^{21}$~cm$^2$~s$^{-1}$, 
even $\sim$10 times higher than the Ohmic diffusivity $\eta_{\rm Ohm}$, 
consistent with the analytical result above (Fig.~\ref{Fig:MHDcoefs}). 
The high ambipolar diffusivity in this tr-MRN model ensures a low influx 
of magnetic flux throughout the entire envelope, as shown in 
the middle panel of Fig.~\ref{Fig:2.4Slw1mtf}. 
Comparing the ``first core'' stage, 
the total magnetic flux inside any given radius (cylinder) is a few times 
smaller in the tr-MRN model than that of the MRN reference model. 
While having a similar total mass, the resulting mass-to-flux ratio 
at any given radius is also larger in the tr-MRN model by a few times.

A direct consequence of the reduced magnetic flux everywhere is a higher 
angular momentum in the gas, because of a weaker magnetic braking 
(assuming a similar magnetic field geometry). 
The left panel of Fig.~\ref{Fig:2.4SlwAMRcent} shows the  
specific angular momentum in each shell of gas from the 
inner 2~AU hole to the edge of cloud at the ``first core'' stage. 
Except for the cloud edge where collapse is still slow, 
gas in the tr-MRN model clearly retains higher specific 
angular momentum in any given shell than the MRN reference model.
This also holds over the subsequent collapse; as more materials that 
have experienced less magnetic braking bring in more angular momentum, 
a RSD is able to survive and even grow in size. 
Notice that there is a ``plateau'' near $\gtrsim$10--30~AU 
in the tr-MRN model and $\sim$4--10~AU in the 
MRN reference model, indicating the centrifugal barrier where the 
free-falling gas starts to feel the centrifugal force inside. 
The centrifugal barrier in the MRN reference model will disappear 
quickly in less than 300 years as gas with low specific angular momentum 
keeps falling in, which again implies the unsustainable nature of the small 
transient RSD formed in these cases. In contrast, the centrifugal 
barrier remains in the tr-MRN model as more angular momentum flows in 
with gas (not plotted). Note that the rapid decrease of specific angular 
momentum just inside the plateau is caused by the strong magnetic braking 
at this location (see the next section \S~\ref{S.CentriShock} 
for detailed analysis).

The location of centrifugal barrier, i.e. centrifugal radius 
$R_{\rm cent}(r_{\rm sh})$ for gas in any give shell r$_{\rm sh}$, 
can be approximated by the following formula, 
\begin{equation}
\label{Eq:Rcent}
R_{\rm cent}(r_{\rm sh})={j(r_{\rm sh})^2 \over G (M_*+\int_{\rm r<r_{sh}} M(r))}~,
\end{equation}
where $M_*$ is the star mass and $j(r_{\rm sh})$ is the averaged 
specific angular momentum in this shell; 
and we assume all the mass interior to this shell collapses 
self-similarly towards the origin, hence contributing to the gravitational 
force for this shell. The formula is only an approximation, in which 
we ignore the persistent magnetic braking during collapse, 
and the non-spherical distribution of mass in the circumstellar region, 
especially the bipolar outflow which will decrease the mass 
inside $r_{\rm sh}$.

The distribution of centrifugal radii for the ``first core'' stage is shown 
in the right panel of Fig.~\ref{Fig:2.4SlwAMRcent}. 
The tr-MRN model has a much larger centrifugal radius for each given shell 
than the MRN reference model, consistent with the dependence on 
the square of specific angular momentum, $j^2$. In the tr-MRN model, 
circumstellar materials in a few 10$^1$ to a few 10$^2$~AU 
are expected to settle down at a centrifugal radius of $\sim$8--20~AU; 
while materials in a similar region in the MRN reference model are likely to 
land within a few AU radius or even enter the inner 2~AU hole 
(considered as accreted by the star). These expectations are indeed 
realized at later times, when the disk in the tr-MRN model reaches 
a radius of $\sim$15--20~AU and the MRN reference model shows no 
disk at all. Therefore, the centrifugal radius $R_{\rm cent}(r_{\rm sh})$ 
estimated from Eq.~\ref{Eq:Rcent} is a good indicator of the 
expected disk size. It reinforces the notion that a long-lived RSD with 
a reasonable size requires continuous supply of high angular momentum 
materials from the envelope, which have reasonably large centrifugal 
radii.

\subsubsection{Centrifugal Shock and Magnetic Braking}
\label{S.CentriShock}

The presence of a centrifugal barrier naturally creates a shock by 
slowing down and piling up infalling materials. 
As a result, magnetic field geometry changes abruptly from 
1) pinched field lines outside centrifugal radius 
--- dragged by rapid accretion flow onto the disk --- to 
2) straight vertical field lines inside the centrifugal radius 
where infall motions are largely halted. 
At the centrifugal radius where the field lines pile up, 
magnetic pressure strengthens, and magnetic braking becomes efficient 
(see below). The centrifugal radius is also the launching site of the 
outflow via the magneto-centrifugal acceleration (``slingshot'') mechanism 
\citep{BlandfordPayne1982,PudritzNorman1986}. Note that part of the outflow 
region is affected by the force limiter technique.

\begin{figure*}
\begin{tabular}{ll}
\includegraphics[width=\columnwidth]{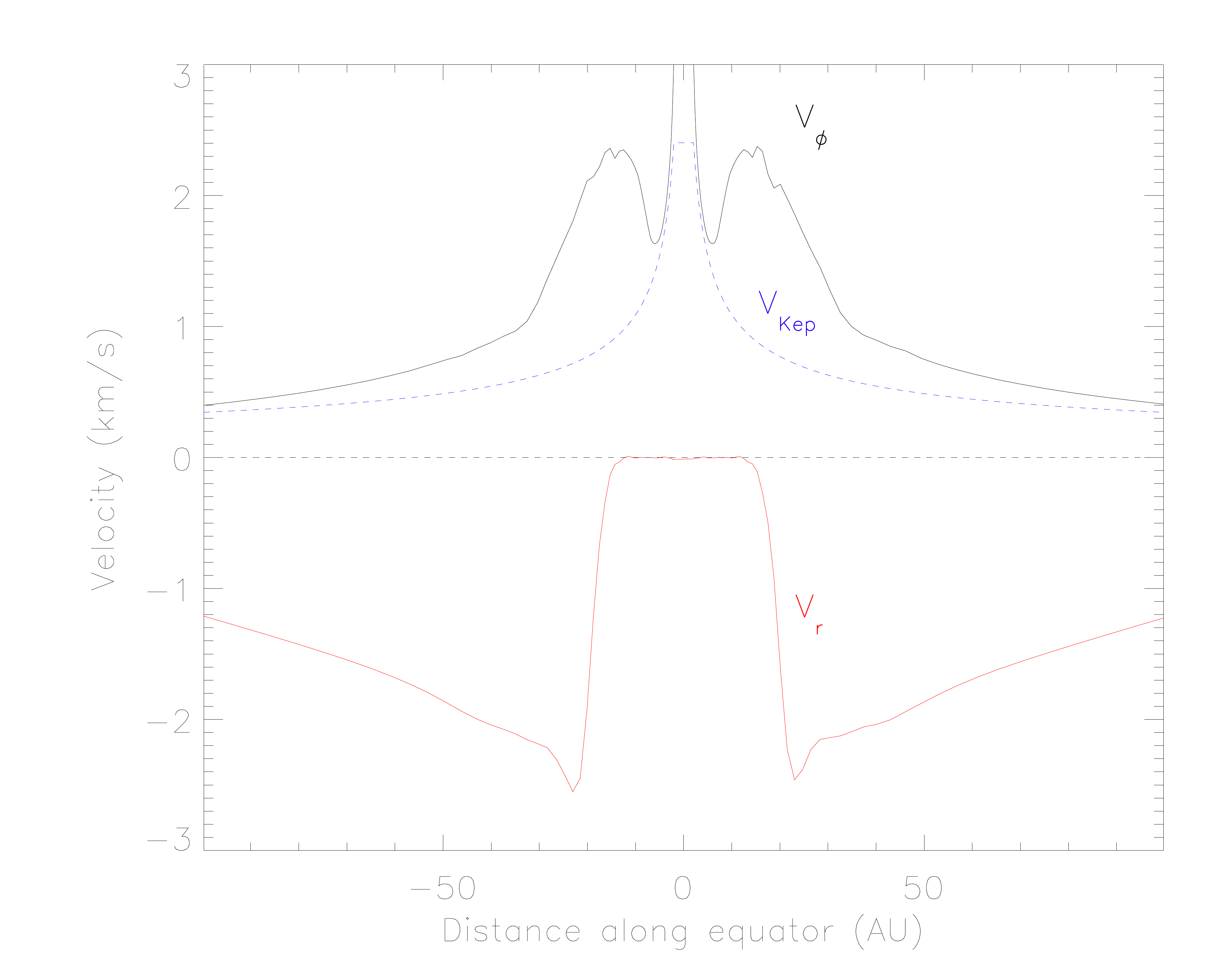}
\includegraphics[width=\columnwidth]{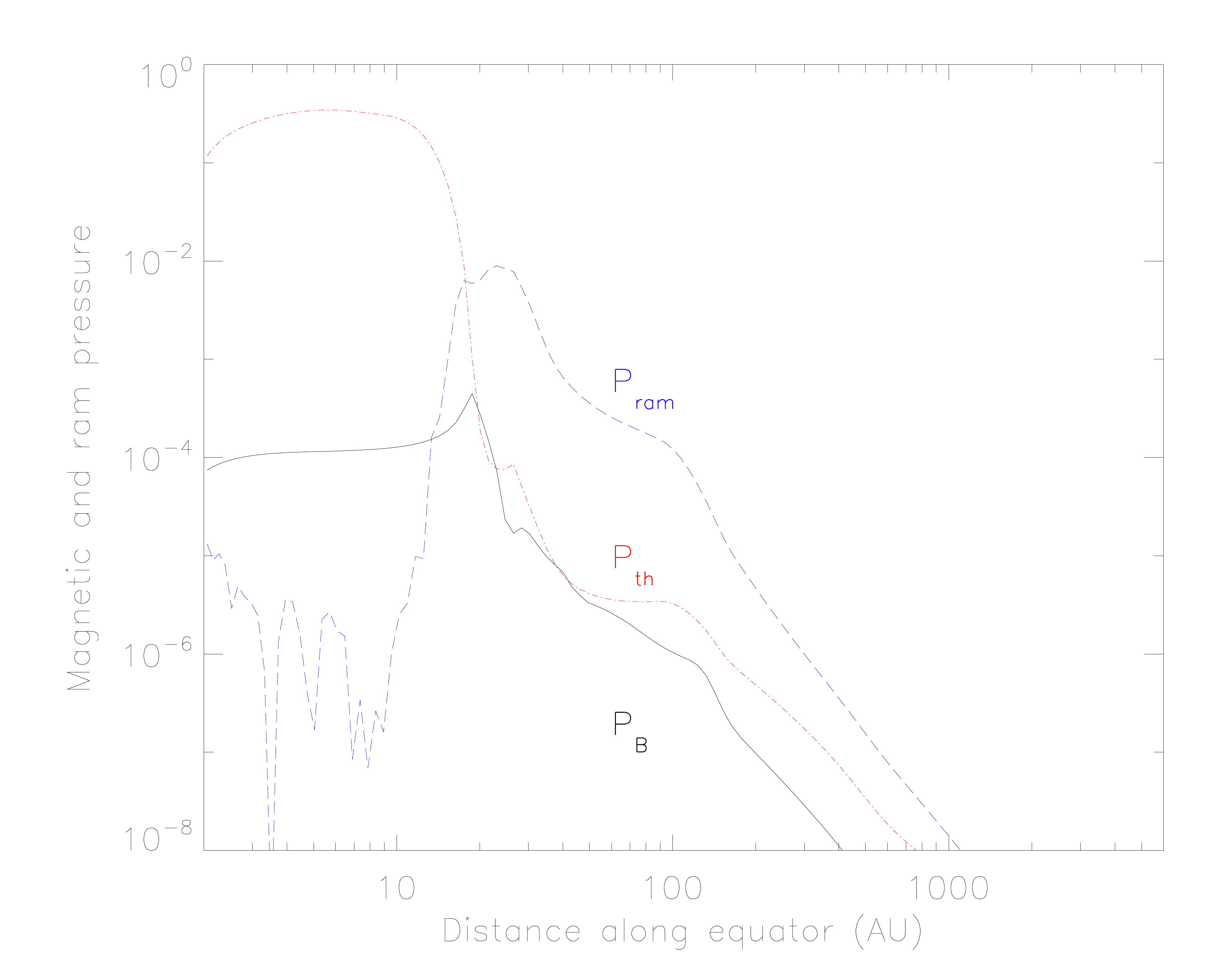}
\end{tabular}
\caption{Left panel: profile of equatorial infall and rotation speed 
in the $\lambda2.4$ Slw-trMRN1 model at $t \approx 148.0$~kyr
(same time frame as Fig.~\ref{Fig:2.4Slw-trMRN1}). 
The Keplerian speed is also plotted based on the central mass. 
Right panel: profile of thermal $P_{\rm th}$, magnetic $P_{\rm B}$, 
and ram $P_{\rm ram}$ pressures along the equator.}
\label{Fig:VP2.4SlwtrMRN1}
\end{figure*}

We present in Fig.~\ref{Fig:VP2.4SlwtrMRN1} the typical centrifugal shock 
formed in model $\lambda2.4$ Slw1-trMRN at a representative time 
$t \approx 148.0$~kyr. 
The centrifugal barrier locates near the disk outer edge at $\sim$20~AU 
from the central star, where gas density is high and 
magnetic field strengthens to $\sim$0.1~G (Fig.~\ref{Fig:2.4Slw-trMRN1}).
In the pre-shock region ($\gtrsim$20~AU), 
there is a rapid increase in magnetic pressure $P_{\rm B}$ 
and a corresponding drop in the infall ram pressure $P_{\rm ram}$. 
The rotational velocity ${\rm v}_\phi$ also rises quickly as the free-falling 
gas starts to become centrifugally supported and spins up when moving to 
smaller radii. 
In the post-shock region ($\lesssim$15~AU), gas infall motion along 
the equator halts almost completely (${\rm v_r} \rightarrow 0$), 
leading to the drastic decrease in ram pressure. 
It implies that the disk gas has high enough angular momentum to orbit 
without falling further inward. 
The lower rotation speed in the middle section of the disk is simply 
because of the small total mass and gravitational potential inside. 
In the innermost 6~AU where gravity is mainly dominated by the central star, 
gas is able to rotate with its supposed Keplerian speed again. 
Notice that the post-shock region is dominated by thermal pressure, and 
magnetic pressure flattens to $\sim$10$^{-4}$~g~cm$^{-1}$~s$^{-2}$; 
hence the plasma $\beta$ can reach values above 1000.

The abrupt change of magnetic field geometry across the centrifugal radius 
causes a very large difference in magnetic braking efficiency. 
We now show that the straight field lines inside the centrifugal radius 
leads to inefficient magnetic braking, while the pinched field lines 
outside the centrifugal radius leads to very strong braking.
The strength of magnetic braking can be quantified by the magnetic torque 
integrated over a finite volume $V$, 
\begin{equation}
\label{Eq:MagnetTorq}
\mathcal{N}_m(V)={1 \over 4\pi} \int_V \left\{\bmath{r} \times [(\nabla \times \bmath{B}) \times \bmath{B}]\right\} {\rm d}V~.
\end{equation}
Typically, the magnetic torque comes mainly from magnetic tension force 
rather than magnetic pressure force (the latter requires a magnetic pressure 
gradient in the azimuthal direction, which is zero in an axisymmetric 
simulation). 
Thus, we consider only the magnetic tension term which can be simplified to 
a surface integral \citep{MatsumotoTomisaka2004}, 
\begin{equation}
\mathcal{N}_t(S)={1 \over 4\pi} \int_S (\bmath{r} \times \bmath{B})(\bmath{B} \cdot {\rm d}\bmath{S})~,
\end{equation}
over the surface {\bf S} of volume V. For a spherical shell at given radius 
$r_{\rm sh}$, the net magnetic torque $\mathcal{N}_t(r_{\rm sh})$ 
exerting on it equals the difference between the two integrals 
over its outer and inner surfaces ($S_{\rm out}$ and $S_{\rm in}$), 
\begin{equation}
\Delta \mathcal{N}_t(r_{\rm sh})=\mathcal{N}_t(S_{\rm out})-\mathcal{N}_t(S_{\rm in})~.
\end{equation}
Therefore, the timescale to magnetically torque down the total 
angular momentum in this shell $\mathcal{L}(r_{\rm sh})$ can be estimated as, 
\begin{equation}
\label{Eq:Tbrake}
t_{\rm brake} (r_{\rm sh}) \approx {\mathcal{L}(r_{\rm sh}) \over \Delta \mathcal{N}_t(r_{\rm sh})}
\end{equation}
which directly measures the magnetic braking efficiency in that shell.

\begin{figure}
\includegraphics[width=\columnwidth]{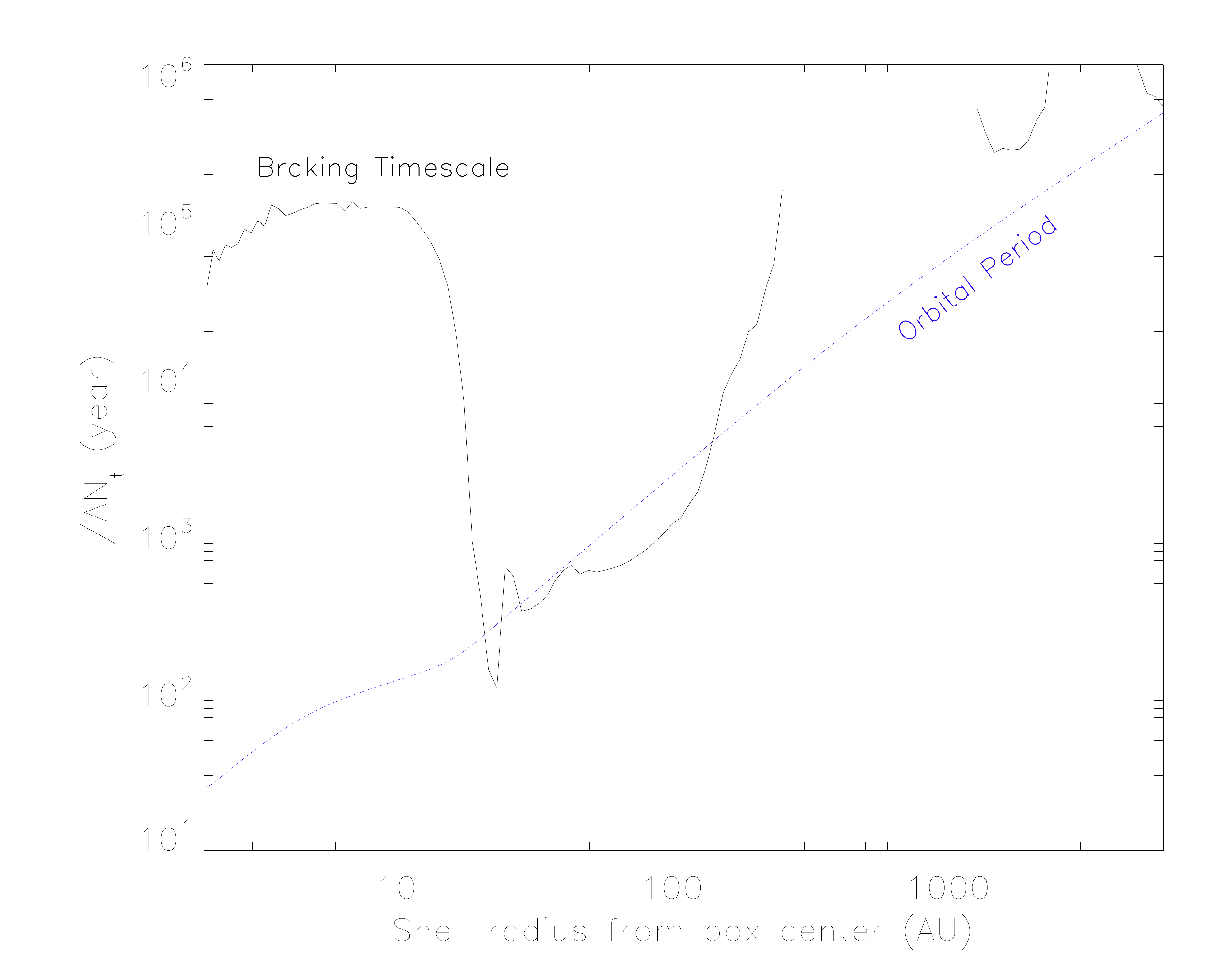}
\caption{Estimated magnetic braking timescale for model 
$\lambda2.4$ Slw-trMRN1 at $t \approx 148.0$~kyr. The orbital timescale 
is derived from the centrifugal velocity originated by all mass inside 
the given shell radius, different than that derived from stellar mass 
alone (Keplerian) by a factor of a few. The gap at few 10$^2$~AU indicates
that magnetic torque there is transporting angular momentum inward instead of 
outward.}
\label{Fig:Braking}
\end{figure}

Fig.~\ref{Fig:Braking} shows the distribution of magnetic braking timescale 
$t_{\rm brake}$ for shells at different radii, in the $\lambda2.4$ Slw-trMRN1
model at $t \approx 148.0$~kyr. The braking timescale is the shortest 
just outside the centrifugal radius $\sim$20~AU, with 
$t_{\rm brake}\approx 100$~years that is about 1/2 of the orbital period 
at that location. The strong magnetic braking matches with the field 
geometry outside the centrifugal radius, where strongly pinched field lines 
are piled up. In contrast, the braking timescale is as high as 10$^5$~years 
inside the disk, guaranteeing $\sim$1000 orbits for materials on the disk.
It is a direct outcome of the straight field lines inside the centrifugal
radius because magnetic torque $N_t$ is $\propto B_{\rm r} B_{\phi}$ 
and $B_{\rm r} \rightarrow 0$ due to negligible infall motion. 

We expect in 3D a larger infall motion along the disk due to 
gravitational and possibly MHD instabilities, which would slightly pinch 
magnetic field lines again inside the centrifugal radius. However, the general 
result should still hold because of the large ambipolar diffusivity in 
the disk, which will greatly weaken the magnetic braking effect even when 
the field lines are pinched and winded-up slightly \citep{Krasnopolsky+2010}.

\subsection{Compound Disk: Effect of Initial Rotation and Magnetic Field}
\label{S.CompoundDisk}

We now investigate the RSDs formed in models with faster initial rotation 
($\lambda2.4$ Fst-trMRN1) and/or weaker initial magnetic field 
($\lambda4.8$ Fst-trMRN1); both help the 
formation of RSDs by increasing the amount of angular momentum 
in circumstellar region. 
The two models $\lambda2.4$ Fst-trMRN1 and $\lambda4.8$ Fst-trMRN1 
demonstrate formation of a ``compound disk'' consisting 
a small inner disk (ID) and a more massive outer ring (OR). 
The compound disk is an evolutionary outcome of a large single disk 
($\gtrsim$30~AU) that forms first and separates into the two substructures 
(ID and OR). 
The separation is mainly caused by the competition of gravitational potential 
at the origin and that near the massive centrifugal barrier. 
Gas in the disk middle section feels the least gravitational force; 
therefore it either falls towards the origin or it is pulled backward 
by the mass accumulated near the centrifugal barrier depending on 
its location. 
Gradually, a gap opens up in the disk middle section, which separates 
the single disk into ID and OR; 
\footnote{For the previous model $\lambda2.4$ Slw-trMRN1 with slower 
rotation and stronger magnetic field, the single disk does not separate 
into a ``compound disk'' because the centrifugal radius is small, 
so that the gravitational potential peaked at the centrifugal barrier is 
not well separated from the primary gravitational potential at the origin.} 
both are rotationally supported. 
For a much wider gap in model $\lambda=4.8$ Fst-trMRN1, the ID can eventually 
disappear, leaving only a massive self-gravitating OR across the 
centrifugal barrier. 

\subsubsection{Effect of Faster Initial Rotation}

\begin{figure*}
\includegraphics[width=\textwidth]{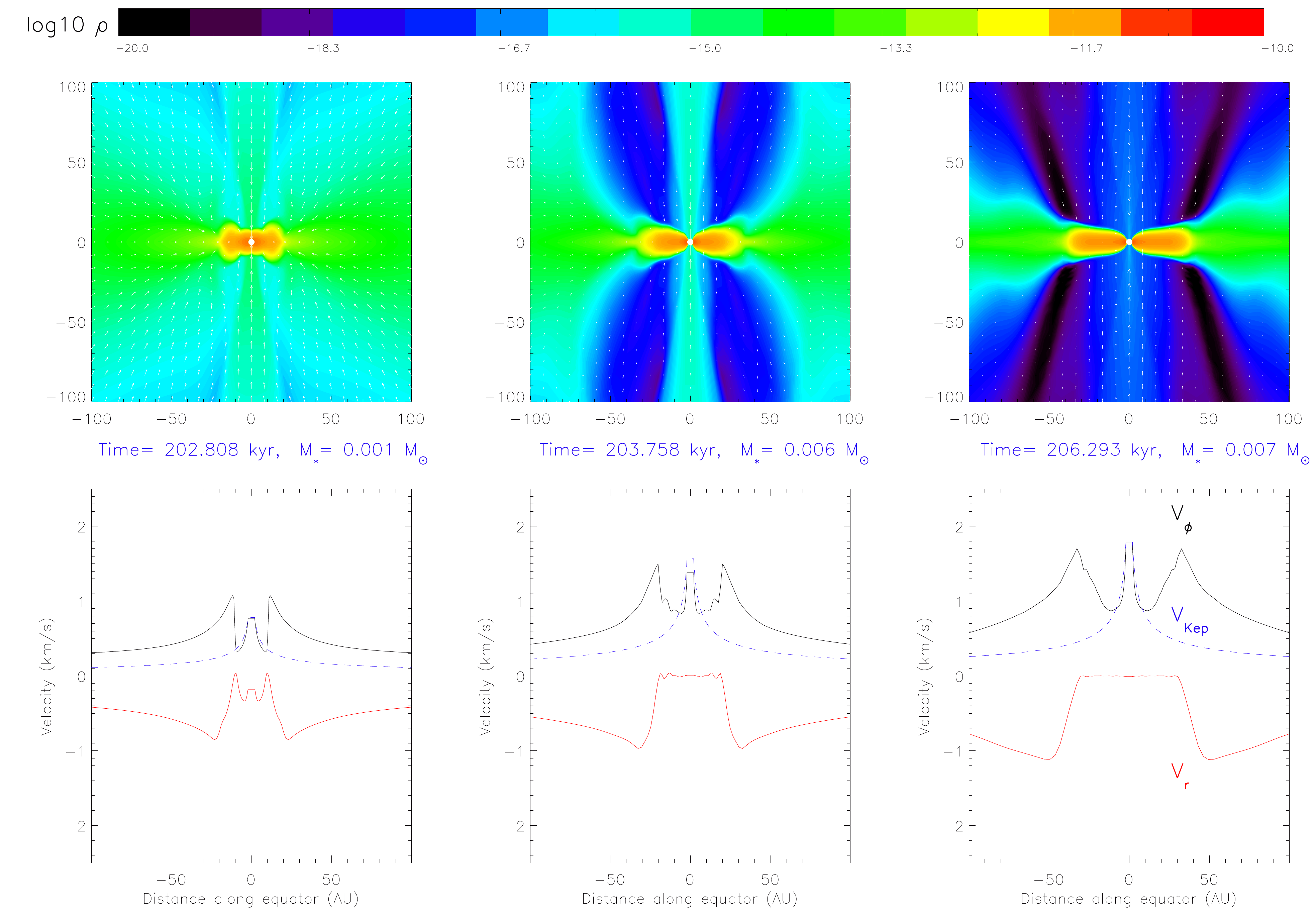}
\caption{Early evolution of density distribution (color map) and 
velocity field (white arrows) for the $\lambda2.4$ Fst-trMRN1 model 
inside 100~AU radius. The corresponding velocity profiles along the equator 
are plotted in the bottom panels.}
\label{Fig:earlyFst-trMRN1}
\end{figure*}

Fig.~\ref{Fig:earlyFst-trMRN1} shows the early evolution of the RSD 
formed in model $\lambda2.4$ Fst-trMRN1. 
Within 3.5~kyr after the ``first core'' stage, the disk grows quickly from 
$\sim$15~AU to $\sim$35~AU --- twice of the disk size in model 
$\lambda2.4$ Slw-trMRN1 --- owing to the higher specific angular momentum 
of infalling gas in this faster rotating core. The growth of disk size 
indicates that envelope materials at larger radii have larger 
centrifugal radii (solid-body rotation profile). 
The rotation motion along the equatorial region is also well above
Keplerian throughout the evolution, except for the disk middle section 
(few AU size) where cancellation of gravitational force occurs. 

\begin{figure*}
\includegraphics[width=\textwidth]{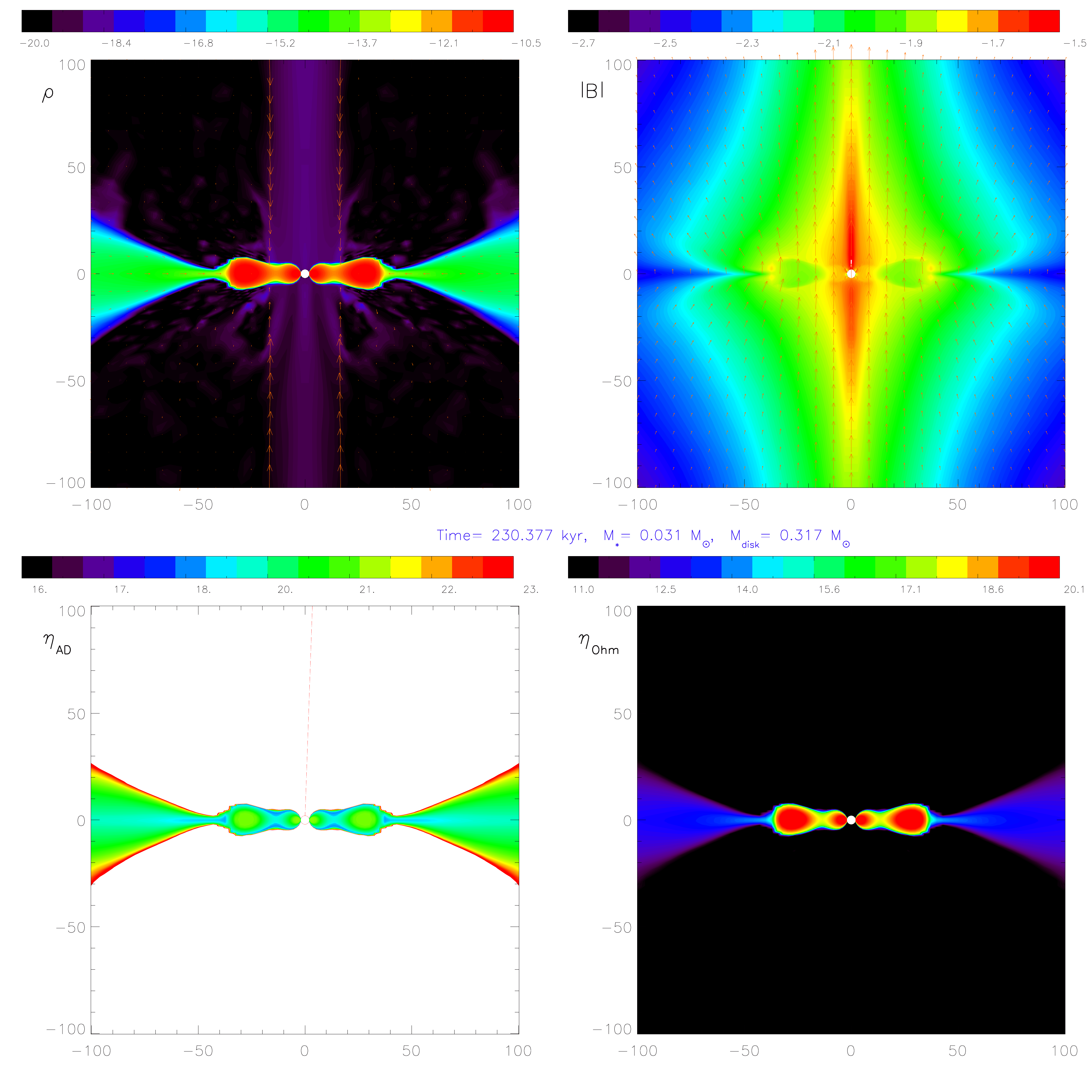}
\caption{Distributions of density $\rho$, magnetic field strength |$B$|, 
ambipolar diffusivity $\eta_{\rm AD}$ and Ohmic diffusivity $\eta_{\rm Ohm}$, 
all in logarithmic scale, for the $\lambda=2.4$ Fst-trMRN1 model 
at a later time $t \approx 230.4$~kyr. 
The poloidal velocity field (top-left) and magnetic field (top-right) 
are shown as orange arrows. The white bipolar regions in the bottom-left 
panel are of high $\eta_{\rm AD}$ above 10$^{23}$~cm$^2$~s$^{-1}$. 
Length unit of the axes is in AU.}
\label{Fig:2.4Fst-trMRN1}
\end{figure*}

The compound disk for model $\lambda2.4$ Fst-trMRN1 is clearly 
presented in Fig.~\ref{Fig:2.4Fst-trMRN1} at a later time 
$t \approx 230.4$~kyr. The whole disk is roughly twice the size of 
that in the slower rotating Slw-trMRN1 model. 
The small inner disk (ID) has a radius of 
$\sim$5~AU and the outer ring (OR) extends from 20~AU to 40~AU. 
The total mass of compound disk is about $\sim$0.317~M$_{\sun}$ at this time, 
with around $\sim$95$\%$ of mass in the OR. Such a massive ring in 3D will 
become gravitationally unstable, driving spiral waves that would 
redistribute angular momentum, as well as potentially fragmenting 
into companion objects. 
In contrast, the stellar mass is only $\sim$0.031~M$_{\sun}$, 10$\%$ 
of the whole compound disk, but still twice of the ID mass. Again, 
we expect in 3D the mass ratio of star to disk be higher 
as a result of efficient gravitational instability in the disk which 
promotes infall towards the star.

\begin{figure*}
\begin{tabular}{ll}
\includegraphics[width=\columnwidth]{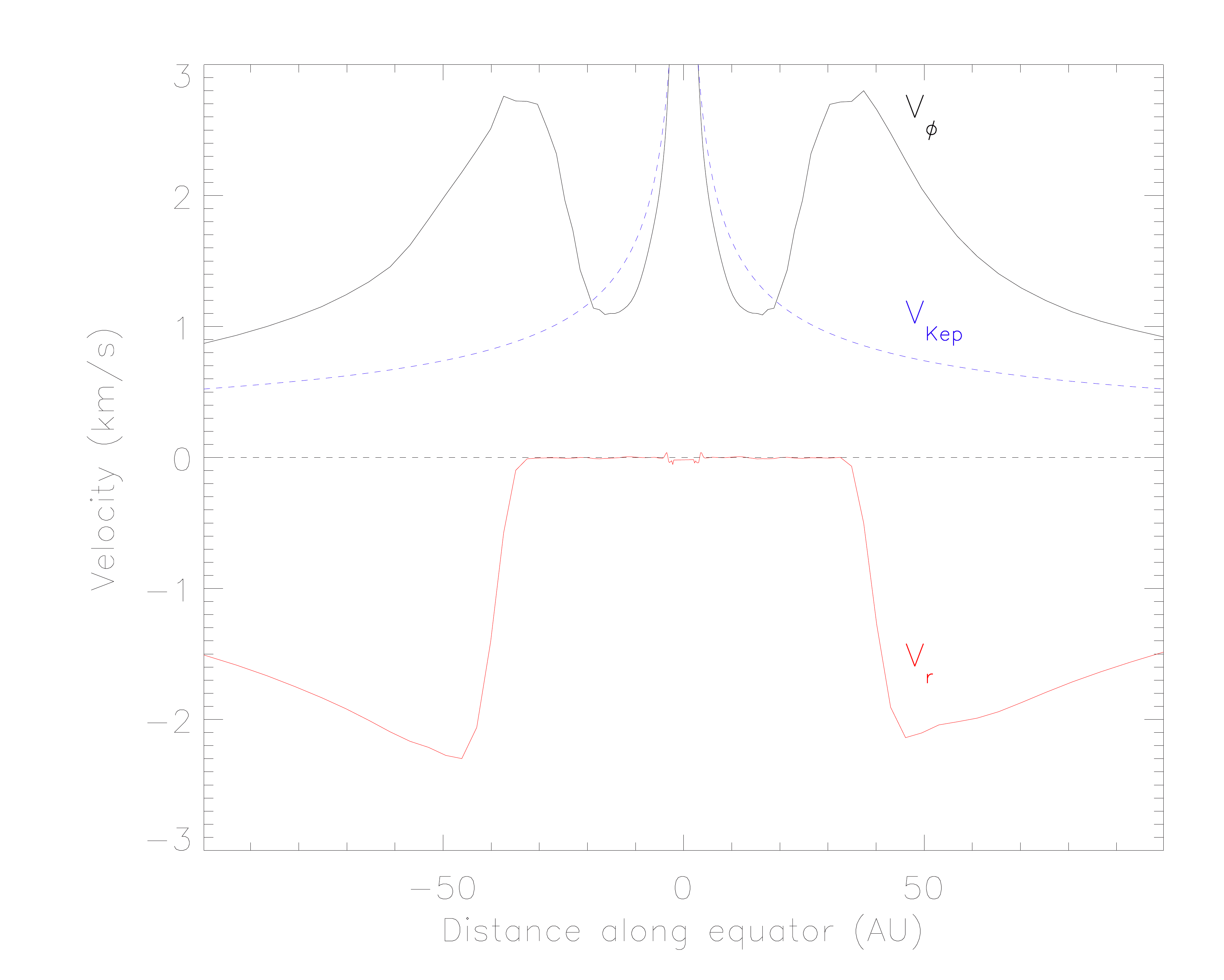}
\includegraphics[width=\columnwidth]{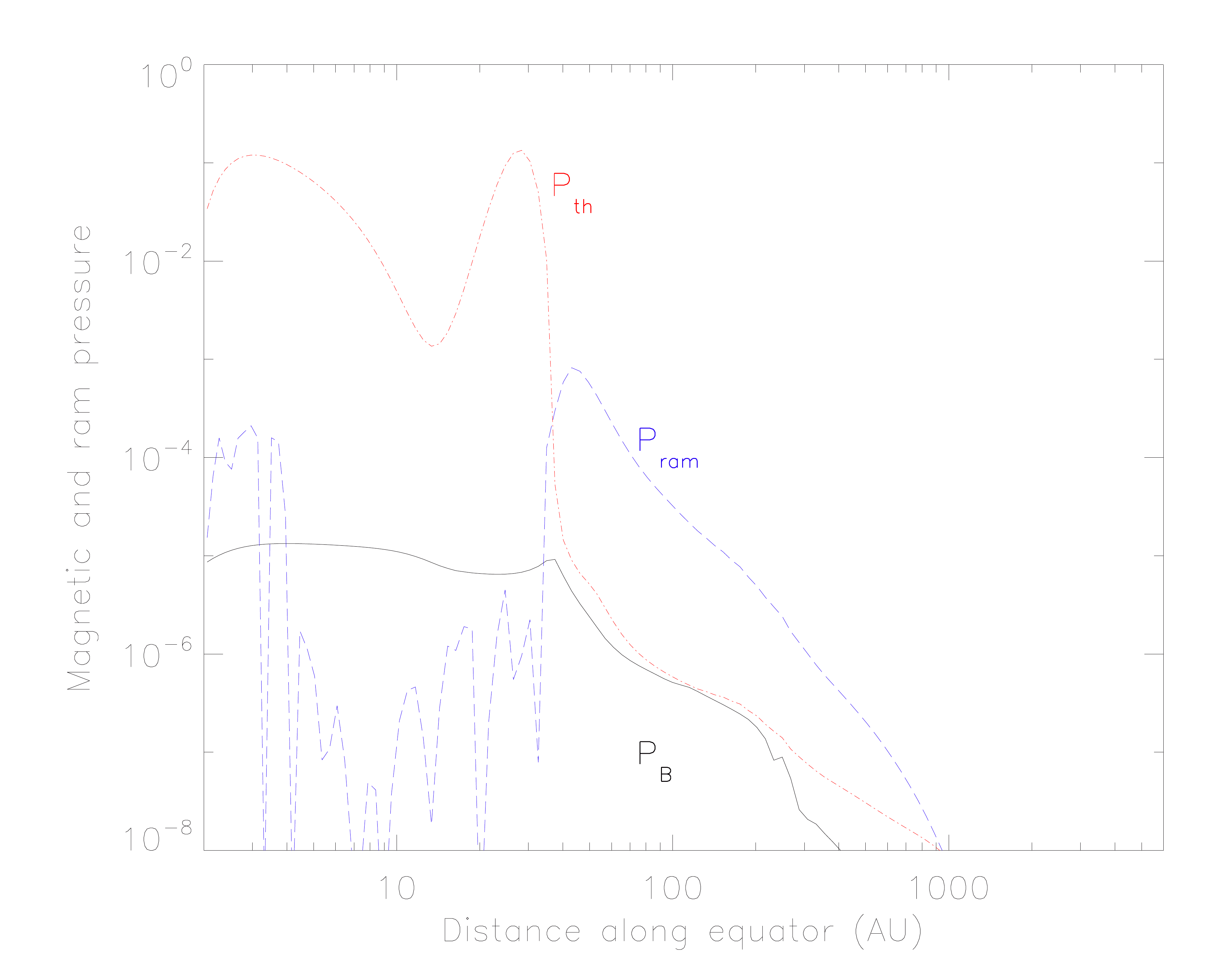}
\end{tabular}
\caption{Left panel: profile of equatorial infall and rotation speed 
in the $\lambda2.4$ Fst-trMRN1 model at $t \approx 230.4$~kyr
(same time frame as Fig.~\ref{Fig:2.4Fst-trMRN1}). 
The Keplerian speed is also plotted based on the central mass. 
Right panel: profile of thermal $P_{\rm th}$, magnetic $P_{\rm B}$, 
and ram $P_{\rm ram}$ pressures along the equator.}
\label{Fig:VP2.4Fst-trMRN1}. 
\end{figure*}

The centrifugal shock locates at $\sim$35--40~AU away from the center. 
Magnetic field lines are pinched outside the centrifugal radius 
but straighten up inside. 
Across the centrifugal shock, magnetic field lines pile up and field strength 
enhances to $\sim$0.015~G.
The values of ambipolar and Ohmic diffusivity along the equatorial region 
are similar to those in the Slw-trMRN1 model. In both ID and OR, 
diffusivities are high, and AD dominates over Ohmic dissipation. 
The strong AD in ID and OR, as well as an inefficient magnetic braking 
there due to straight field lines, helps gas to maintain a super-Keplerian 
rotation (Fig.~\ref{Fig:VP2.4Fst-trMRN1}). 
The disk gap is rotating slightly slower than Keplerian speed; 
yet it is doing so in the positive direction, which results from a 
competition between thermal pressure gradient and gravitational force 
(not shown). Besides, both ID and OR are dominated by thermal pressure, 
with plasma $\beta$ of few 10$^3$ up to 10$^4$.

\subsubsection{Effect of Weaker Initial B-field}

\begin{figure*}
\includegraphics[width=\textwidth]{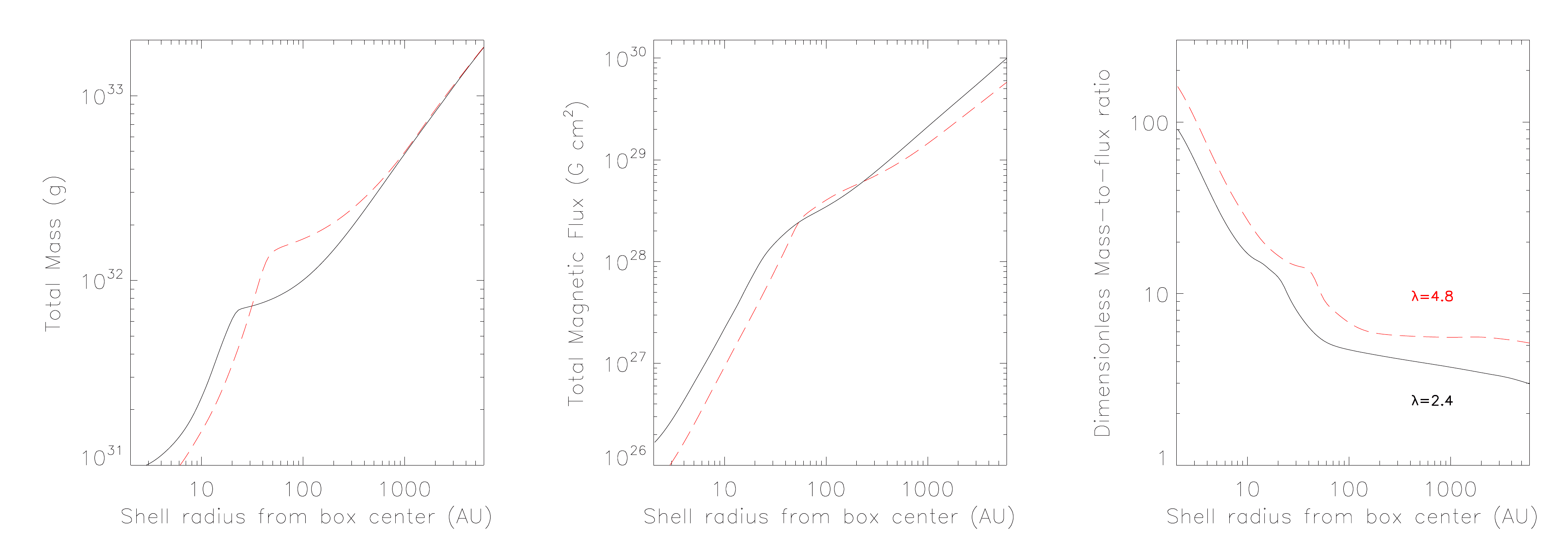}
\caption{Distribution of total mass (left panel), 
total magnetic flux (middle panel) and the corresponding 
mass-to-flux ratio (right panel) inside spheres of different radii for 
the strong field model $\lambda2.4$ Fst-trMRN1 (black solid) at 
$t \approx 203.1$~kyr and the weak field model $\lambda4.8$ Fst-trMRN1 
(red dashed) at $t \approx 164.8$~kyr. 
The time frames are chosen at roughly the 
``first-core'' phase when physical structures are relatively simple.}
\label{Fig:2.4Fst1kmtf}
\end{figure*}

We investigate the effect of initial magnetization by comparing 
model $\lambda4.8$ Fst-trMRN1 to the previous model $\lambda2.4$ Fst-trMRN1. 
A weaker magnetic field results in a lower total magnetic flux. 
We again compare the ``first core'' stage for the two models, when 
structures in the central region are relatively simple. 
As shown in middle panel of Fig.~\ref{Fig:2.4Fst1kmtf}, 
the magnetic flux in $\lambda=4.8$ model is about half of that in the 
$\lambda=2.4$ model almost everywhere. The high magnetic flux in the 
$\lambda=4.8$ model between $\sim$50--200~AU owes to the large amount of 
mass piled up at the centrifugal barrier (left panel). 
Nevertheless, the resulting mass-to-flux ratio 
clearly shows the $\sim$2 times difference between the two models 
(right panel) at all radii.

\begin{figure*}
\begin{tabular}{ll}
\includegraphics[width=\columnwidth]{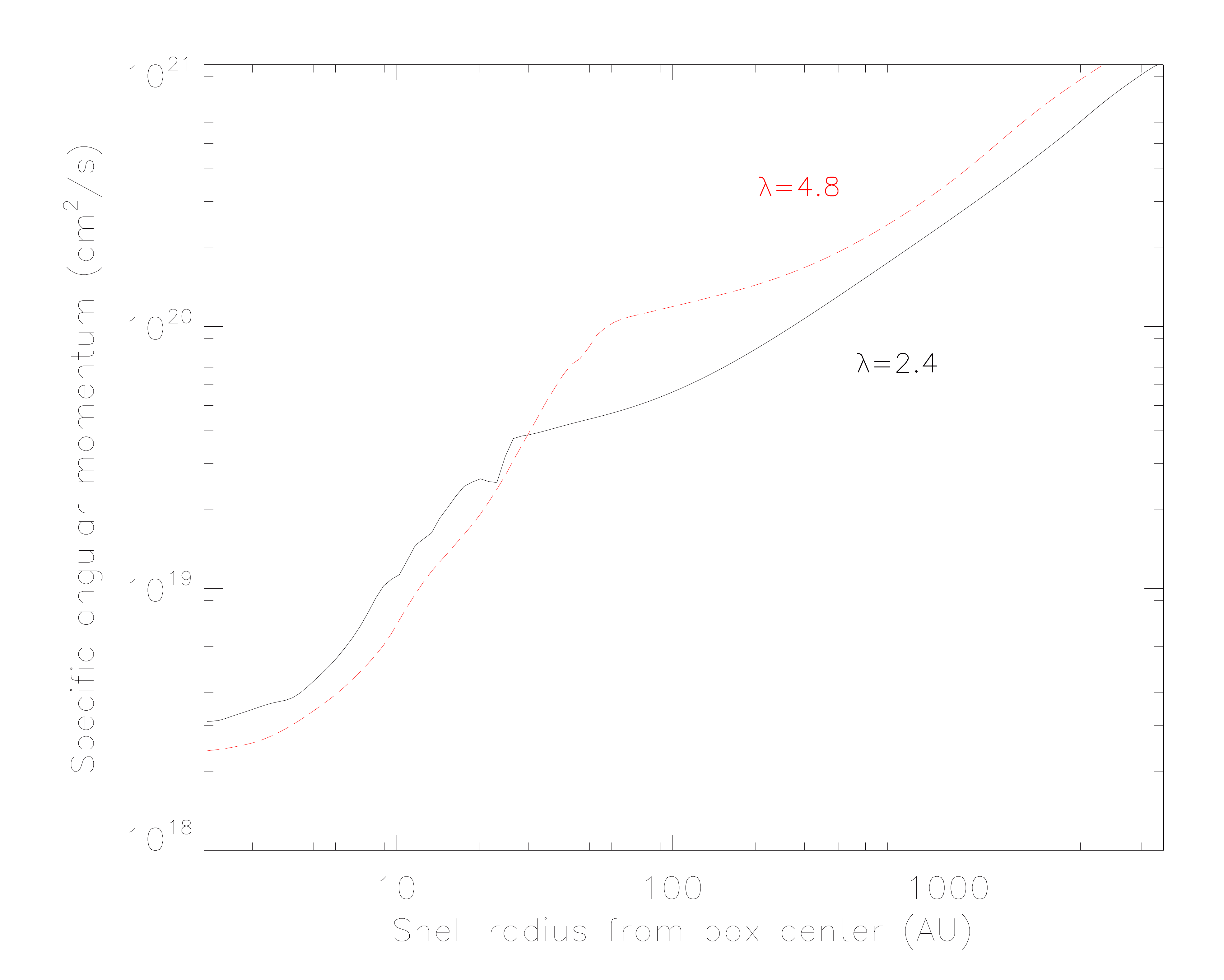}
\includegraphics[width=\columnwidth]{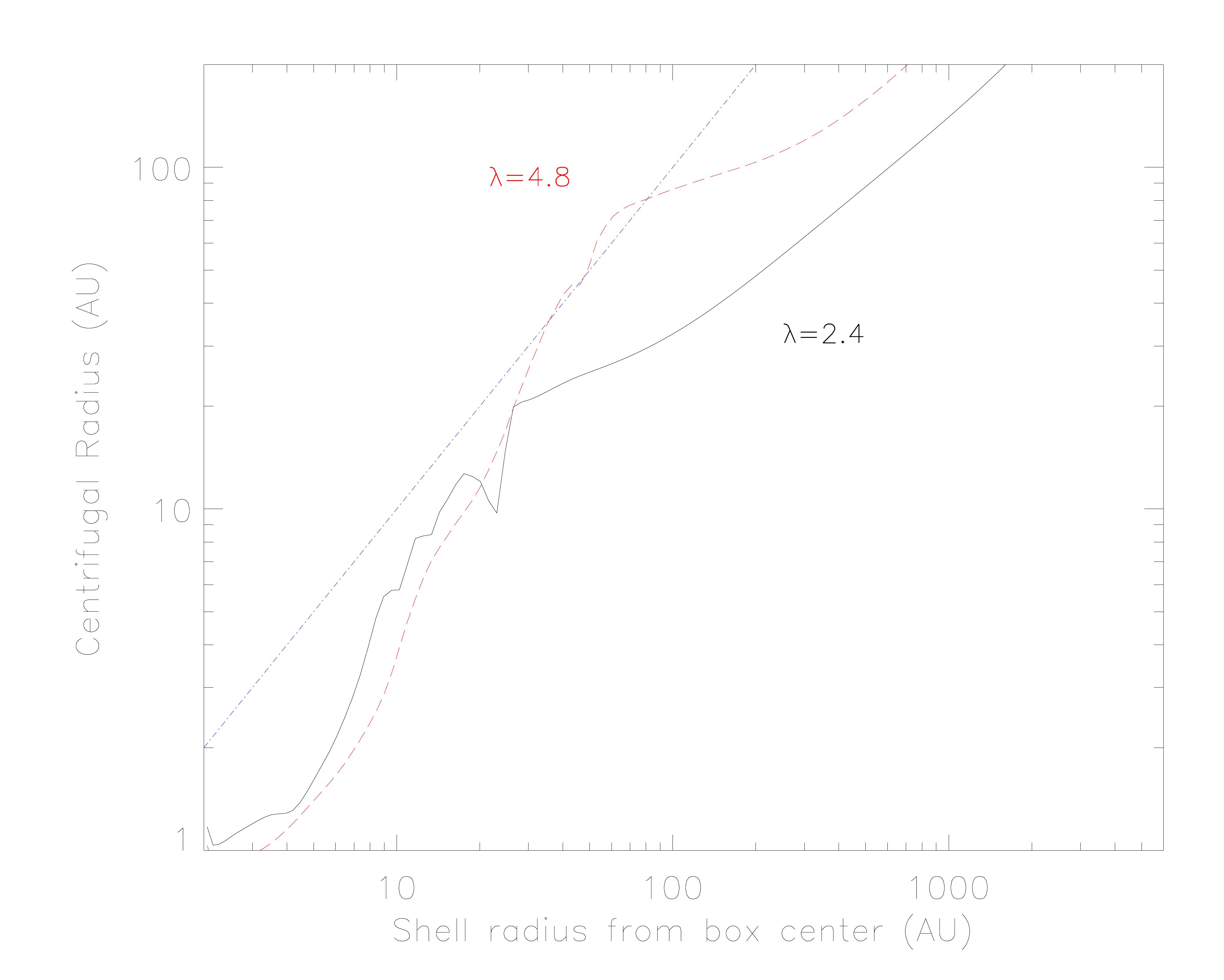}
\end{tabular}
\caption{Left panel: distribution of specific angular momentum inside shells 
at different radii for the strong field model $\lambda2.4$ Fst-trMRN1 
at $t \approx 203.1$~kyr (black solid) and the weak field model 
$\lambda4.8$ Fst-trMRN1 at $t \approx 164.8$~kyr (red dashed). 
Right panel: the estimated centrifugal radius from Eq.~\ref{Eq:Rcent} 
for the two models in the left panel. Blue dash-dotted line represents 
a ``break-even'' relation $R_{\rm cent}(r_{\rm sh})=r_{\rm sh}$.}
\label{Fig:1kBfAMRcent}
\end{figure*}

The reduced magnetic flux and higher mass-to-flux ratio in the $\lambda=4.8$ 
model enable the infalling gas to retain even more angular momentum. 
The angular momentum on the $\sim$50AU scale is so high that 
little matter reaches closer to the center, leaving the specific 
angular momentum there paradoxically lower than that of the stronger 
field case (left panel of Fig.~\ref{Fig:1kBfAMRcent}). 
This is consistent with the delay in mass inflow shown in 
Fig.~\ref{Fig:2.4Fst1kmtf}. 
The ``plateau'' on the specific angular momentum curve again indicates 
the current centrifugal barrier. Difference in specific angular momentum 
is the largest near the plateau, where the $\lambda=4.8$ curve is 2--3 times 
higher than the $\lambda=2.4$ curve. 
We can also estimate the expected centrifugal radius for gas in any given 
shell from Eq.~\ref{Eq:Rcent}, plotted in the right panel of 
Fig.~\ref{Fig:1kBfAMRcent}. In the $\lambda=2.4$ model, circumstellar 
gas between $\sim$25--200 AU are likely to land within 
$\sim$20--50~AU; while the bulk of envelope gas in the $\lambda=4.8$ model 
are expected to gather beyond $\gtrsim$70~AU. Furthermore, part of 
the $\lambda=4.8$ curve even runs above the ``break-even'' line 
$R_{\rm cent}(r_{\rm sh})=r_{\rm sh}$ at $\sim$40--80~AU, 
predicting that materials in these locations will move outward 
instead of falling in, because the total mass inside is unable to hold 
the current orbits of these materials. 
The large centrifugal radius in the $\lambda=4.8$ model, 
though an approximation, is direct evidence of high angular momentum 
in the infalling gas; it again implies weak magnetic braking 
as a result of the reduced magnetic flux. 

\begin{figure*}
\includegraphics[width=\textwidth]{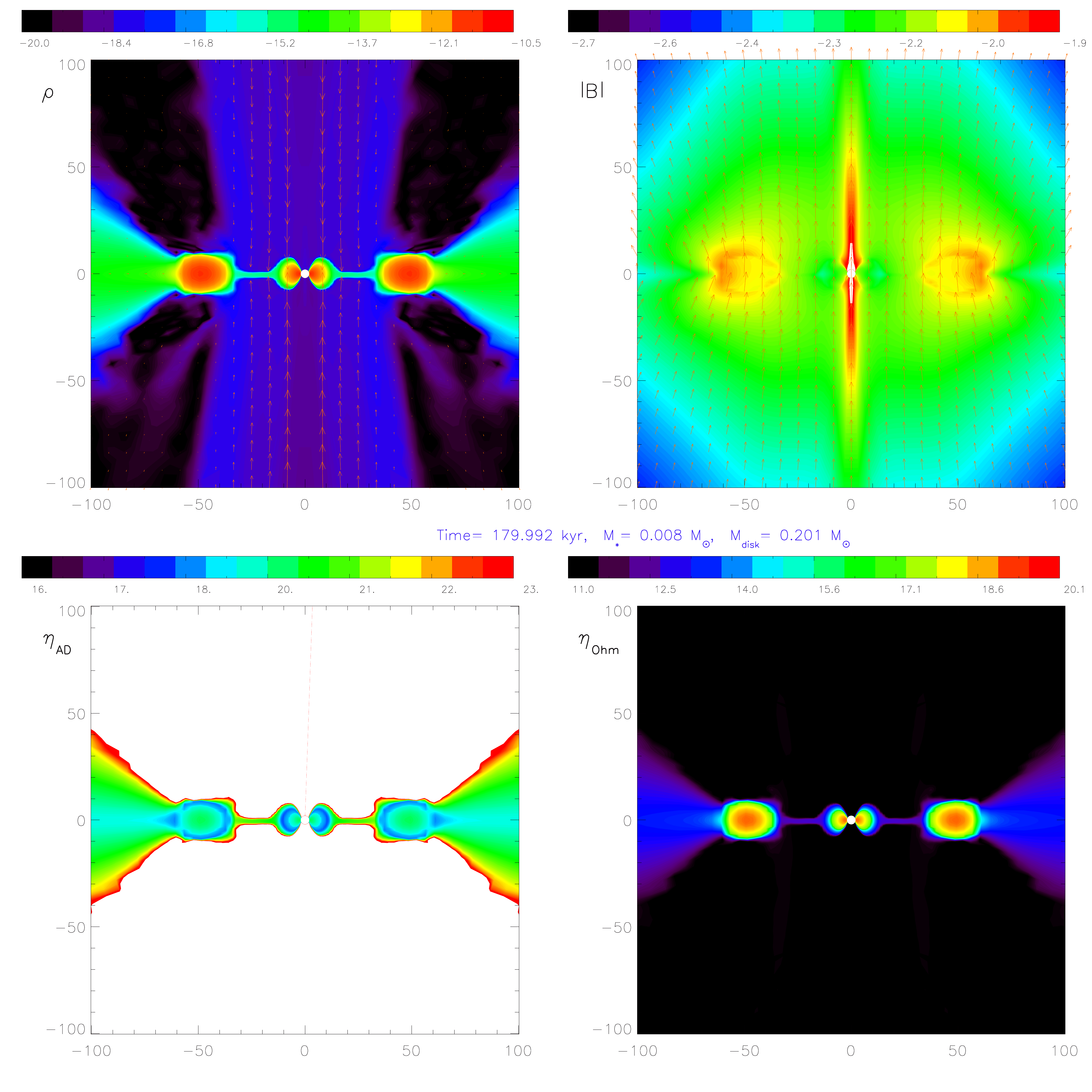}
\caption{Distributions of density $\rho$, magnetic field strength |$B$|, 
ambipolar diffusivity $\eta_{\rm AD}$ and Ohmic diffusivity $\eta_{\rm Ohm}$, 
all in logarithmic scale, for the $\lambda=4.8$ Fst-trMRN1 model 
at a later time $t \approx 180.0$~kyr. 
The poloidal velocity field (top-left) and magnetic field (top-right) 
are shown as orange arrows. The white bipolar regions in the bottom-left 
panel are of high $\eta_{\rm AD}$ above 10$^{23}$~cm$^2$~s$^{-1}$. 
Length unit of the axes is in AU.}
\label{Fig:4.8Fst-trMRN1}
\end{figure*}

The compound disk in this weaker field model ($\lambda4.8$ Fst-trMRN1) 
loses its ID as evolution continues, leaving only the massive OR located at 
$\sim$35--60~AU away from the central star. In Fig.~\ref{Fig:4.8Fst-trMRN1}, 
we present a representative moment at $t \approx 180.0$~kyr, 
about 7~kyrs before the ID's disappearance. 
At the moment, the total mass of compound disk is $\sim$0.201~M$_{\sun}$, 
with $\sim$98.7$\%$ of mass in the OR. 
However, the central star only has $\sim$0.008~M$_{\sun}$, about 
$4\%$ of the compound disk. Even adding up the mass of star and ID 
(0.0026~M$_{\sun}$), the central part only weighs $5.5\%$ of the OR.
Therefore, the massive self-gravitating OR not only will likely fragment as 
in the previous model, but indeed dominates the gravitational potential 
in circumstellar region. Materials in between the star and OR 
are more likely to fall back towards the OR, as the OR builds up 
more mass from the collapsing envelope. 

\begin{figure*}
\begin{tabular}{ll}
\includegraphics[width=\columnwidth]{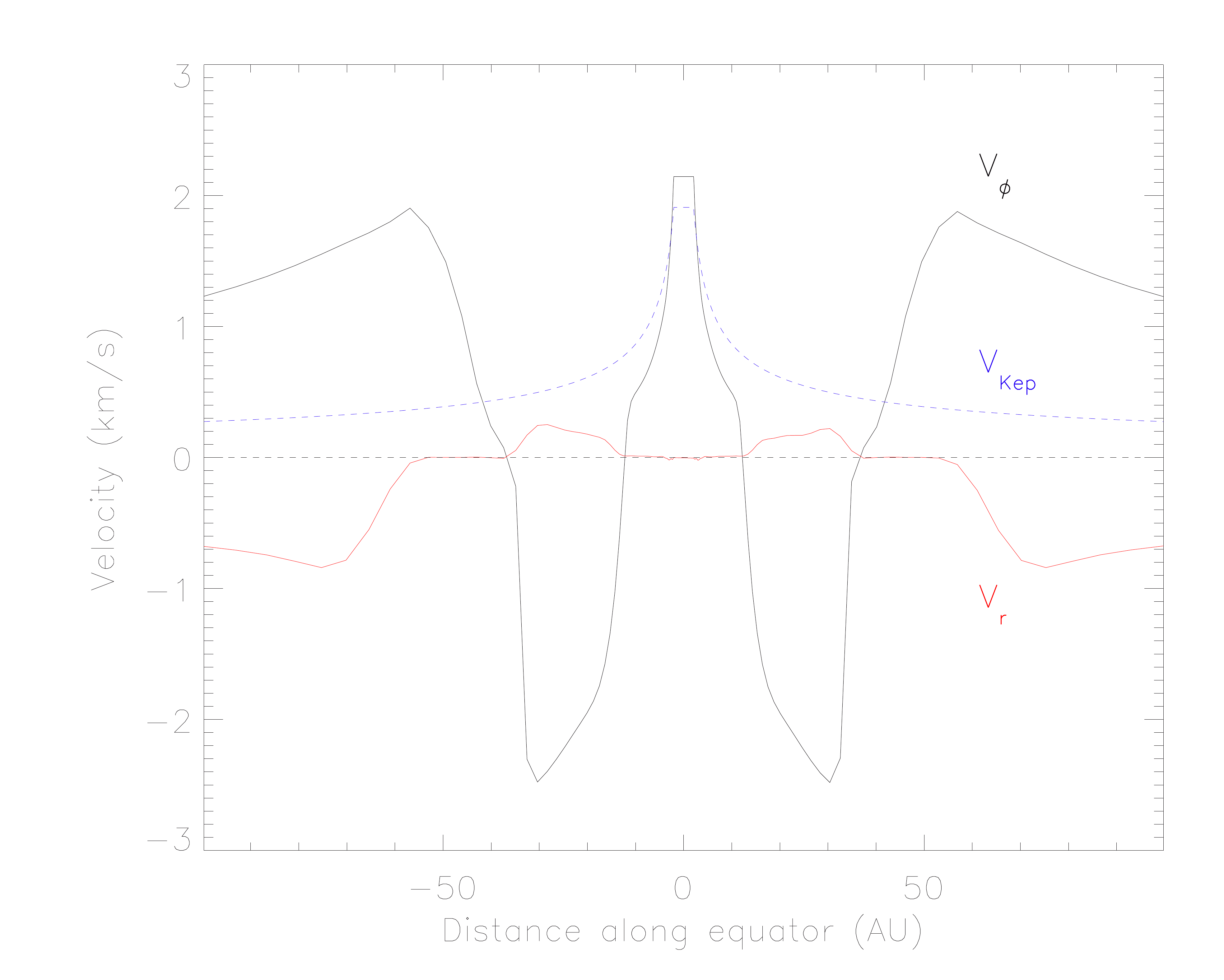}
\includegraphics[width=\columnwidth]{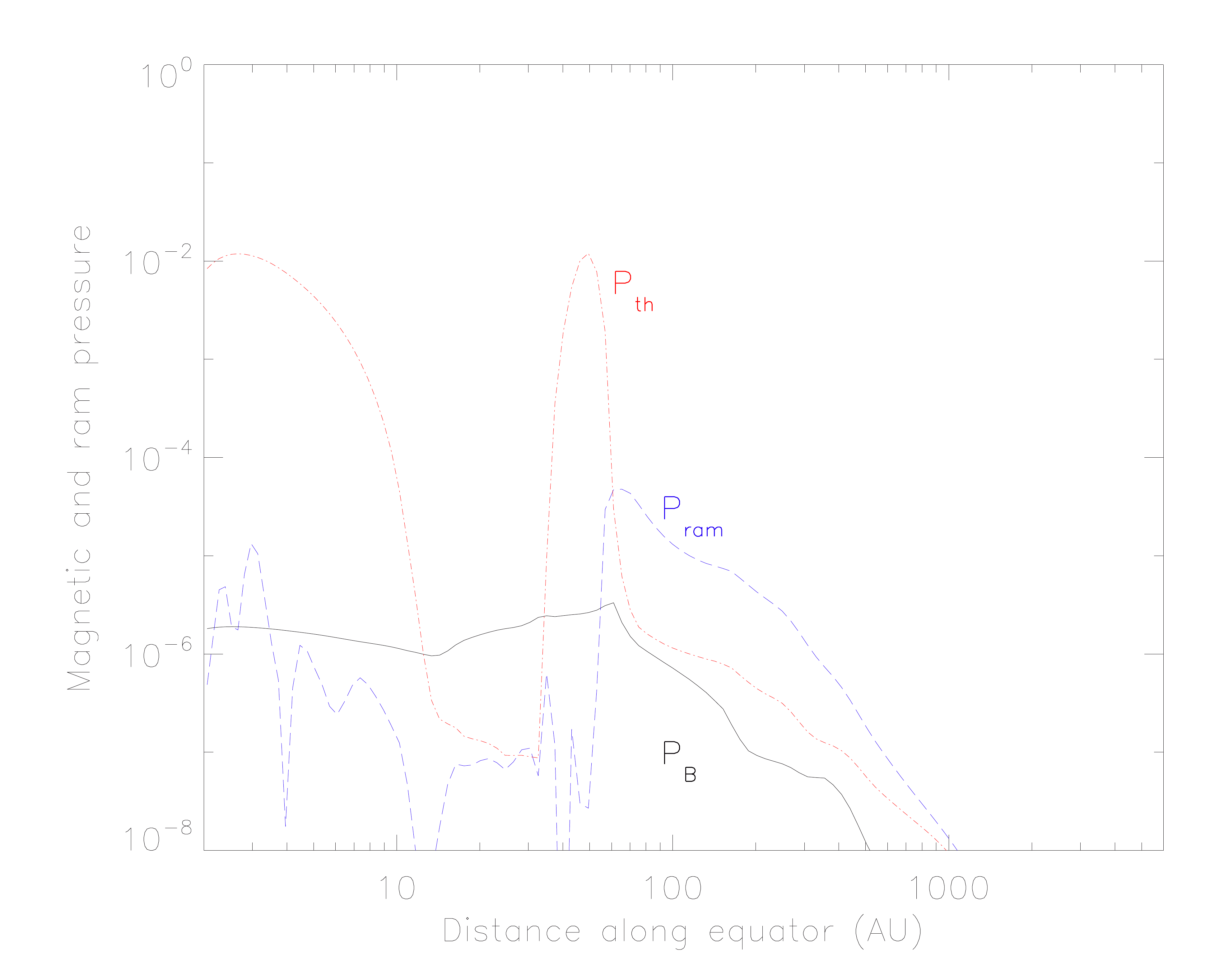}
\end{tabular}
\caption{Left panel: profile of equatorial infall and rotation speed 
in the $\lambda4.8$ Fst-trMRN1 model at $t \approx 180.0$~kyr
(same time frame as Fig.~\ref{Fig:4.8Fst-trMRN1}). 
The Keplerian speed is also plotted based on the central mass. 
Right panel: profile of thermal $P_{\rm th}$, magnetic $P_{\rm B}$, 
and ram $P_{\rm ram}$ pressures along the equator.}
\label{Fig:VP4.8trMRN1}
\end{figure*}

The backward accretion onto the OR is more clearly demonstrated in 
Fig.~\ref{Fig:VP4.8trMRN1}. Materials in the disk gap between $\sim$15--35~AU
gain positive ${\rm v_r}$ and negative ${\rm v}_\phi$, 
which indicates they are moving 
outward and rotating backwardly around the OR! In this weaker field model, 
gravitational and non-magnetic effects dominate the disk dynamics. 
Magnetic effects are only strong outside the centrifugal radius at 
$\sim$60--70~AU, where field lines are pinched and field strength is 
enhanced (to less than 10~mG). 
Inside both ID and OR, ambipolar and Ohmic diffusivities are high; and 
thermal pressure again dominates which results in a plasma $\beta$ 
of several 10$^3$. 

\subsection{Shrinking Disk, LG models, and Cosmic-Ray Ionization Rate}
\label{S.ShrinkDisk}

Finally, we briefly describe the type (ii) shrinking disk formed 
in our simulations. This type of disks mainly form in models with 
fast initial rotation and/or weaker initial B-field, 
but the other two key parameters being more demanding 
(e.g., MRN or LG grain, or high $\zeta_0^{\rm H_2}$). 
RSDs do appear in early times because the fast rotating 
materials have not yet been torqued down by magnetic braking; 
they even grow in size early on, depending on the angular momentum 
reservoir in the circumstellar region in different models. 
However, because of a low ambipolar diffusivity in the low density envelope, 
large amount of magnetic flux is being dragged by the collapsing flow 
into the high density circumstellar region. 
The direct consequence is a strong magnetic braking 
as soon as field lines become pinched along the equator. 
Therefore, when gas eventually lands onto the disk, 
it supplies less specific angular momentum to the disk than what is 
required to sustain the disk rotation. 
Gradually, these RSDs shrink in size, and some may disappear 
in about 10$^4$ years (see Table~\ref{Tab:model1}--\ref{Tab:model2}). 
Therefore, it is the strong magnetic braking and insufficient 
angular momentum influx that fail to maintain a long-lived 
reasonably sized RSD. 

\begin{figure*}
\includegraphics[width=\textwidth]{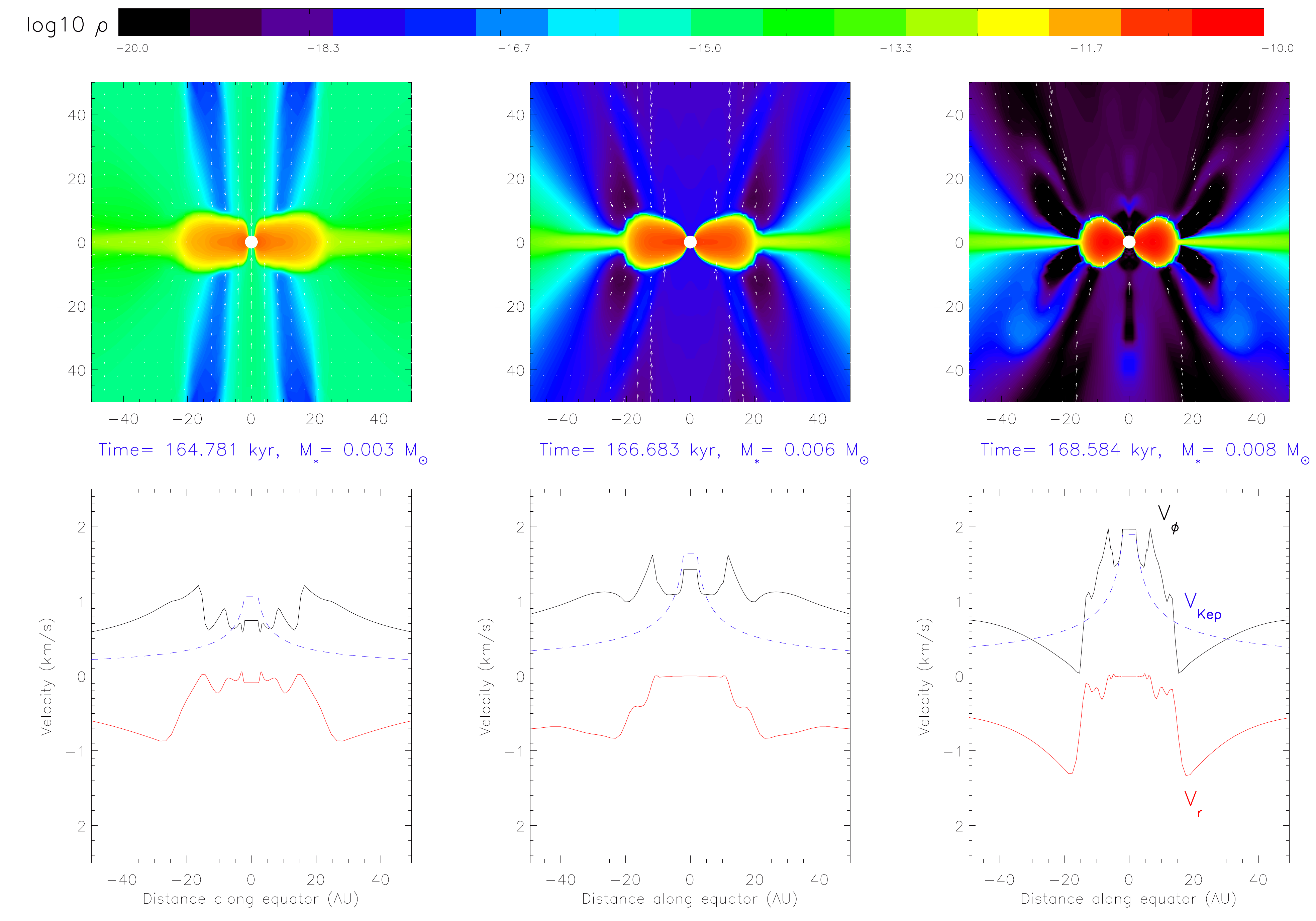}
\caption{Early evolution of density distribution (color map) and 
velocity field (white arrows) for the $\lambda4.8$ Fst-MRN1 model 
inside 50~AU radius. The corresponding velocity profiles along the equator 
are plotted in the bottom panels.}
\label{Fig:early4.8Fst-MRN1}
\end{figure*}

We take model $\lambda4.8$ Fst-MRN1 for example. 
Fig.~\ref{Fig:early4.8Fst-MRN1} shows the evolution of the shrinking disk. 
The disk radius shrinks roughly by 40$\%$, form 25~AU to 15~AU 
over a period of 3.8~kyrs, and will continue to shrink for another 3~kyrs 
to a radius of 10AU. Unlike other models discussed above, the later phases 
of this model (middle and right panels) show a rapid decrease in 
rotation speed ${\rm v}_\phi$ along the flattened pseudo-disk close to 
the disk edge (centrifugal barrier). 
At time 168.6~kyr (right panel) in particular, ${\rm v}_\phi$ has reduced to 
$\sim$0 between 15--20~AU. It is a direct outcome of strong magnetic 
braking outside the centrifugal radius, where magnetic field lines are 
strongly pinched and piled up, as discussed in \S~\ref{S.CentriShock}. 
Besides, there is a plateau in the infall speed just outside the centrifugal 
barrier (between 10--15~AU), which is also a typical feature of the 
``resumed'' infall motion for magnetically braked materials \citep{Li+2011}. 
Note that the infall motion still halts in the innermost $\sim$7~AU inside 
the centrifugal barrier. 

\begin{figure}
\includegraphics[width=\columnwidth]{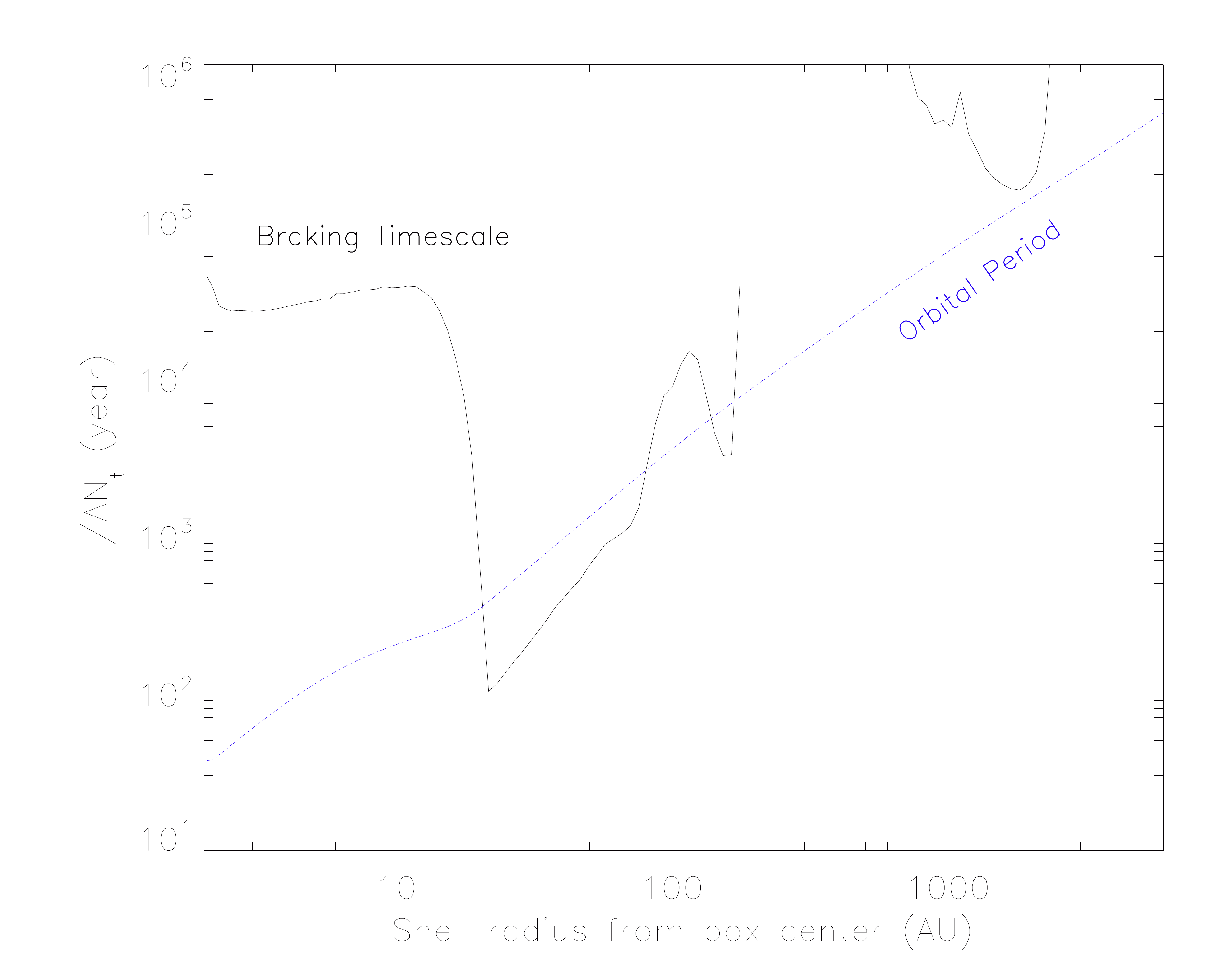}
\caption{The estimated magnetic braking timescale, as in 
Fig.~\ref{Fig:Braking}, but for $\lambda4.8$ Fst-MRN1 case 
at 166.7~kyr (same time frame as the middle panel of 
Fig.~\ref{Fig:early4.8Fst-MRN1}).}
\label{Fig:Braking-Shrk}
\end{figure}

The efficient magnetic braking outside the centrifugal radius 
is shown more clearly in Fig.~\ref{Fig:Braking-Shrk}, 
in which we plot the distribution of magnetic braking timescale 
$t_{\rm brake}$ (Eq.~\ref{Eq:Tbrake}) for the middle panel of 
Fig.~\ref{Fig:early4.8Fst-MRN1} ($t\approx 166.7$~kyr). 
Over a wide circumstellar region between 20-100~AU, 
the braking timescale is already as short as a fraction of the 
estimated orbital period; it reaches a minimum value 
of $\sim$10$^2$ years just outside the centrifugal radius at $\sim$20~AU. 
At such location, the loss of angular momentum via magnetic braking 
takes only 1/4--1/3 orbits, more efficient than the 
$\lambda2.4$ Slw-trMRN1 model discussed above (Fig.~\ref{Fig:Braking}). 

\begin{figure}
\includegraphics[width=\columnwidth]{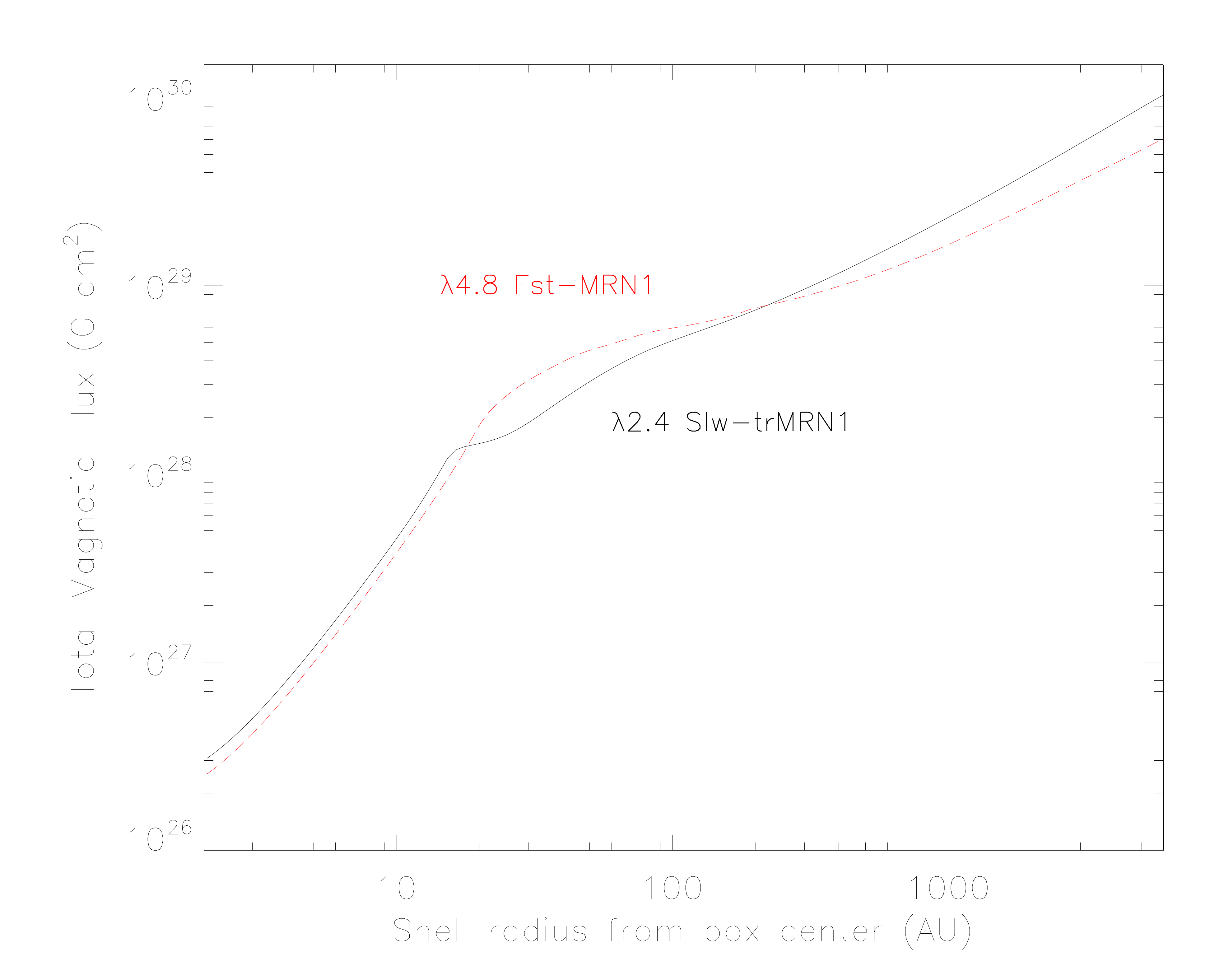}
\caption{Comparison of the distribution of magnetic flux at early times for 
model $\lambda4.8$ Fst-MRN1 at 166.7~kyr (middle panel of 
Fig.~\ref{Fig:early4.8Fst-MRN1}) and $\lambda2.4$ Slw-trMRN1 at 143.2~kyr 
(middle panel of Fig.~\ref{Fig:earlytrMRN1}).}
\label{Fig:compFlux}
\end{figure}

The strong magnetic braking essentially results from a large amount of 
magnetic flux that has brought into the circumstellar region by the 
collapsing flow. AD in this model ($\lambda4.8$ Fst-MRN1) is rather 
inefficient throughout the envelope because of the MRN grains. 
In contrast, the model $\lambda2.4$ Slw-trMRN1 discussed in 
\S~\ref{S.ModeltrMRN}, though starting with a disadvantage in 
the initial magnetic field strength (stronger) and rotation speed (slower), 
has much more efficient AD to help decouple infalling gas from the 
magnetic field because of the tr-MRN grains. 
As a result, less magnetic flux has reached the 
circumstellar region 20--100~AU than in the $\lambda4.8$ Fst-MRN1 model 
at a similar evolution stage (Fig.~\ref{Fig:compFlux}). 
Because the magnetic field lines are strongly pinched along the equatorial 
region outside the centrifugal radius, the larger magnetic flux in 
model $\lambda4.8$ Fst-MRN1 hence yields stronger magnetic braking in 
such region. This is the origin for the shrinkage of the disk.

Shrinking RSDs are also formed in LG models with low cosmic-ray ionization 
rate; faster rotation and/or weaker magnetic field helps as well 
(see Table~\ref{Tab:model1}-\ref{Tab:model2}). 
The performance of LG models in terms of disk formation largely 
resembles that of the MRN models, yet is slightly better in 
terms of disk lifetime and size. 
This is expected from the analytical result above 
(Fig.~\ref{Fig:MHDcoefs}). Compared to the MRN case, 
the LG case has a larger AD diffusivity by $\sim$1 order of magnitude 
at low densities ($\lesssim$10$^{9}$~cm$^{-3}$), which enables 
more angular momentum to be retained in the rotating envelope; 
however, such angular momentum surplus is mostly removed by a 
stronger magnetic braking along the equatorial region, because 
the AD diffusivity at high densities ($\gtrsim$10$^{9}$~cm$^{-3}$) 
in the LG case is indeed lower by a factor of few than that in the MRN case. 
Therefore, LG models can form larger disks than MRN models due to 
larger centrifugal radius of envelope gas. However, the weaker AD 
along the equatorial region leads to a stronger magnetic braking, so that 
these RSDs still shrink over time. 

Although the effect of cosmic-ray ionization rate is not the main 
focus of this study, we rank it as the second most important parameter 
for formation of RSDs after the grain size distribution. 
When we increase $\zeta_0^{\rm H_2}$ to $5.0 \times 10^{-17}$~s$^{-1}$, 
most of long-lived RSDs (Y or Y$^{\rm Shrink}$) formed in 
$\zeta_0^{\rm H_2}=1.0 \times 10^{-17}$~s$^{-1}$ 
either become shrinking disks or shorten their lifetime so much 
that can only be categorized as transient type(i) disks. 
Changing initial rotation or magnetization does not have such a 
strong effect in general.
We have also tested other values of $\zeta_0^{\rm H_2}$, and found 
that $2.0-3.0 \times 10^{-17}$~s$^{-1}$ has much less destructive effect 
than $5.0 \times 10^{-17}$~s$^{-1}$ in terms of forming RSDs in a wider 
parameter space. For instance, with 
$\zeta_0^{\rm H_2}=2.5\times10^{-17}$~s$^{-1}$, long-lived RSD of 20--30~AU 
can still form when grain size sets to tr-MRN, magnetic field to 
$\lambda=2.4$, and rotation to $\beta=0.1$. 
This may provide clues for potentially constraining cosmic-ray 
ionization rate of dense cores using disk observations. 
We will leave the detailed discussion to future studies.

\section{Summary and Discussion}
\label{Chap.Discuss}

We have implemented an equilibrium chemical network to study the effect of 
changing grain size on magnetic diffusivities, 
and coupled the chemistry into 2D axisymmetric MHD simulations to revisit 
the formation of RSDs through non-ideal MHD effects. 
Our main conclusions are summarized as follows. 
\begin{description}
\item 1. Removing VSGs ($\sim$10~$\AA$ to few 100~$\AA$) from the 
grain size distribution can increase the ambipolar diffusivity by 
$\sim$1--2 orders of magnitude at densities below 10$^{10}$~cm$^{-3}$, 
because VSGs are both well-coupled to the magnetic field and 
able to exert strong drag to their surrounding neutral molecules, 
and the large number of VSGs in the size distribution greatly 
increase the fluid conductivity and decrease the ambipolar diffusivity. 
\item 2. Truncating the standard MRN distribution (while keeping the 
dust-to-gas mass ratio at 0.01) at $a_{\rm min}\approx 0.1~\mu$m 
(tr-MRN case) 
has the optimal effect on enhancing the ambipolar diffusivity 
$\eta_{\rm AD}$ in the low density regime. In this case, both 
Pedersen ($\sigma_{\rm P}$) and Hall ($\sigma_{\rm H}$) conductivities 
of the fluid are much smaller than that with the full MRN distribution; 
they together raise the $\eta_{\rm AD}$ by $\sim$1--2 orders of magnitude. 
However, the reduction of total grain surface area only increases the 
ionization fraction by a factor of a few, because the effect scales 
with $\propto a_{\rm min}^{-1}$ and is partially offset by gas-phase 
ion-electron recombinations. 
Note that when $a_{\rm max}$ is set to 1~$\mu$m, the optimal $a_{\rm min}$ 
occurs at $\approx$0.055~$\mu$m. 
\item 3. Pedersen conductivity $\sigma_{\rm P}$ is normally determined by 
ions; however, a large number of VSGs can dominate $\sigma_{\rm P}$ instead, 
especially when $a_{\rm min}$ is between $\sim$10~$\AA$ and $\sim$200~$\AA$ 
(0.02~$\mu$m). For the MRN distribution in particular, the 
$\sigma_{\rm P}(g^-)$ is over 1--2 orders of magnitude larger than the 
contribution of ions. For the optimal tr-MRN distribution, 
the $\sigma_{\rm P}(g^-)$ is reduced by $\sim$10$^4$ times 
and the total $\sigma_{\rm P}$ by $\sim$1--2 orders of magnitude than 
those of the MRN case. 
\item 4. Hall conductivity $\sigma_{\rm H}$ is generally dominated by 
negatively charged grains as long as grain size is below $\lesssim$0.5~$\mu$m 
and above $\gtrsim$10~$\AA$. By increasing $a_{\rm min}$ from the MRN case to 
the tr-MRN case, the $\sigma_{\rm H}$ decreases by up to $\sim$100 times 
at densities below $\sim$10$^{10}$~cm$^{-3}$, mainly because of a reduction 
($\sim$500 times) in the fractional abundance of negatively charged grains 
$x(g^-)$ and an offsetting effect from the grain's Hall parameter 
$\beta_{\rm g^-,H_2}$. The low $\sigma_{\rm H}$ combines with the 
low $\sigma_{\rm P}$ in the tr-MRN case together enhances AD in the 
low density regime. 
\item 5. The enhanced AD by truncated MRN grains 
can enable massive ($\sim$few 10$^{-1}$~M$_{\sun}$) 
long-lived (>few 10$^4$--10$^5$~years) RSDs to form in 2D simulations. 
The efficient ambipolar diffusion of the magnetic field in the envelope 
reduces the amount of magnetic flux being dragged in by the collapsing flow, 
which weakens magnetic braking in the circumstellar region. 
Therefore, the infalling gas can retain enough angular momentum 
required for sustaining a large RSD of tens of AU. 
\item 5. Large gas angular momentum naturally leads to a large 
centrifugal radius that also determines the disk size. 
The centrifugal barrier hosts a hydrodynamic shock as infalling materials 
slow down and pile up. Across the shock, magnetic field geometry changes 
abruptly from pinched field lines outside the centrifugal radius into 
straight field lines inside. As a result, field lines pile up 
near the centrifugal radius, field strength amplifies, and magnetic 
braking becomes the strongest. However, inside the centrifugal radius 
where field lines straighten up (due to lack of infall motion), 
the magnetic braking timescale is much longer than the orbital 
period, by up to a factor of 10$^3$. 
\item 6. Magneto-centrifugal outflows are launched at the centrifugal 
radius (disk edge), which, although are partly affected by the 
numerical technique, continuously carve the bipolar regions.
\item 7. Faster initial rotation speed and/or weaker initial magnetic field 
strength further aid the formation of RSDs and increase disk 
size and lifetime, because of an increased amount of angular momentum 
available in the circumstellar region. In cases with truncated MRN grains 
and low cosmic-ray ionization rate, a large disk ($\gtrsim$30~AU) 
forms first and later evolves into a ``compound disk'' that consists of 
a small Keplerian ID and a massive rotationally supported self-gravitating OR. 
The disk gap that separates the two substructures appears gradually 
as a result of competition of gravitational force from the origin 
and that exerted by the massive OR, i.e., gas either falls towards 
the origin or is pulled backward to the OR. 
\item 8. Cosmic-ray ionization rate $\zeta^{\rm H_2}$ also plays 
an important role, besides grain size, on the formation of RSDs. 
In order for the truncated MRN grains to avert the 
``magnetic braking catastrophe'' and produce RSDs, 
$\zeta^{\rm H_2}$ cannot be too high. 
We find that $\zeta_0^{\rm H_2}\lesssim2.0-3.0\times10^{-17}$~s$^{-1}$ 
on the corescale can still allow RSDs of 20--30~AU to form in strongly 
magnetized ($\lambda=2.4$) cores. 
However, $\zeta_0^{\rm H_2}=5.0\times10^{-17}$~s$^{-1}$ may be 
too high; even when other parameters are less stringent, disk radius 
is at most 10--20~AU and may further shrink somewhat. 
\end{description}

Most of the results presented here should still hold in 3D calculations. 
The main drawback of 2D axisymmetry is the lack of 3D gravitational 
and magnetic instabilities, which can redistribute angular momentum 
inside the disk and drive inward accretion and outward expansion 
(to conserve angular momentum). 
Therefore, in 3D, the disk would be less massive (but larger) and the star 
would be more massive than the ones in 2D. 
Disk fragmentation may also occur in 3D if massive RSDs become 
gravitational unstable \citep{Kratter+2010}. 
Hence tight binary and multiple systems with 50~AU separation or smaller 
may form naturally even without tightening wide binaries 
through magnetic braking \citep{ZhaoLi2013,Zhao+2013}. 
Magnetic interchange instability in the stellar vicinity has been shown 
to strongly hinder disk rotation and suppress disk formation, 
owing to the decoupled magnetic flux from accretion flow that is trapped 
in the circumstellar region \citep[DEMS: decoupling-enabled magnetic 
structure;][]{Zhao+2011,Krasnopolsky+2012}. 
However, the DEMS is likely substantially weakened, if not eliminated, 
by a high AD diffusivity, which makes it difficult for the collapsing 
flow to drag magnetic flux close to the central object. 

Although the numerical study here includes both AD and Ohmic dissipation, 
we have also carried out simulations with AD alone for comparison (not shown) 
and find the same results. The Ohmic dissipation has negligible effect 
on disk formation, at least within the density range and resolution limit 
of this study. It is easily explicable from \S~\ref{Chap.ChemResult} and 
Fig.~\ref{Fig:MHDcoefs}. 
In the MRN case, even though Ohmic diffusivity dominates over 
ambipolar diffusivity at high densities ($\gtrsim$10$^{11}$~cm$^{-3}$), 
it has little opportunity to take effect due to a lack of dense 
long-lived RSD in the first place. 
The low AD diffusivity at low densities is simply unable to preserve 
enough angular momentum to sustain a well defined RSD, if any. 
In the tr-MRN case, Ohmic diffusivity is much smaller than 
ambipolar diffusivity over the entire density range in this study 
(even without the cap on the Ohmic diffusivity), 
and hence makes little difference to the result. 
Future studies regarding the subsequent evolution of protoplanetary disk 
may still consider Ohmic dissipation at a much higher density. 

We have also explored different slopes of the size distribution 
than -3.5 and found the corresponding changes on AD diffusivity 
are much less significant than by removing VSGs. 
Again, we fix the total grain mass fraction at 0.01 and 
$a_{\rm max}=0.25~\mu$m. 
When the slope is in range (-4.0, -3.0), there is only a slight change 
(less than a factor of $\sim$2) to the AD curve. 
When the slope is in range (-3.0, -2.0), AD diffusivity 
slightly enhances to the level of $a_{\rm min}\approx 0.035~\mu$m in the 
-3.5 case, which is still more than 1 order of magnitude lower than 
that in the optimal $a_{\rm min}\approx 0.1~\mu$m case. 
Such tests, though very rough, reinforce the pivotal role of removing 
VSGs on enhancing the ambipolar diffusivity, which is unlikely to be 
replaced by a reasonable change in the slope of size distribution. 
We will leave the detailed analysis in a more refined chemical 
framework in future study. 

Although grain coagulation provides natural means of removing VSGs, 
grain growth to 1~$\mu$m or larger has been shown to be difficult 
\citep{HirashitaLi2013} in dense cores unless they are relatively long lived. 
However, the AD enhancement discussed in this study only require moderate 
growth of grains. The optimal cut-off size is just around 0.1~$\mu$m for 
$a_{\rm max}=0.25~\mu$m, and around 0.055$~\mu$m for $a_{\rm max}=1~\mu$m.
Therefore, the removal of VSGs less than few 100~$\AA$ can be 
efficiently achieved in a few 10$^6$ years through grain coagulation 
\citep{Rossi+1991,Ossenkopf1993,Ormel+2009,Hirashita2012}. 
Such a process is also likely to take place during the quiescent 
prestellar phase which can last for several free-fall times 
\citep[e.g.,][]{Ward-Thompson+2007, KetoCaselli2010}. 

Another possible approach to removing VSGs from the collapsing flow 
is proposed by \citet{CiolekMouschovias1996}; they suggest that 
different degrees of grain coupling to the magnetic field can cause 
small grains to be left behind in the low density envelope, 
while allow only larger grains to follow the collapse into dense part 
of the core. In reality, both mechanisms can operate to help removing VSGs 
during the prestellar and collapse phase. 
More complete chemistry models including advection of gas-phase and grain 
species as well as grain evolution are needed in non-ideal MHD simulations 
to obtain a more realistic picture of disk formation in particular and 
star formation in general. 
Although much work remains to be done, our calculations have strengthened 
the case for the non-ideal MHD effects, especially ambipolar diffusion, 
as a viable mechanism for enabling the formation of rotationally supported 
disks in even strongly magnetized cloud cores and the growth of such disks 
to at least tens of AUs in size.

\section*{Acknowledgements}

We thank Daniele Galli, Marco Padovani, Shu-ichiro Inutsuka, 
and Charles Malcolm Walmsley for inspiring discussions. 
BZ and PC acknowledge support from the European Resarch Council 
(ERC; project PALs 320620). 
Z.-Y. L. is supported in part by NASA NNX14AB38G and 
NSF AST-1313083. 
Numerical simulations are carried out on the MPG supercomputer 
HYDRA and our CAS group cluster at MPE.


\bsp	
\label{lastpage}
\end{document}